# Civic Technologies
## Research, Practice, and Open Challenges

Proposal, outcome and position papers of the 23rd ACM
Conference on Computer-Supported Cooperative Work
and Social Computing (CSCW 2020) workshop

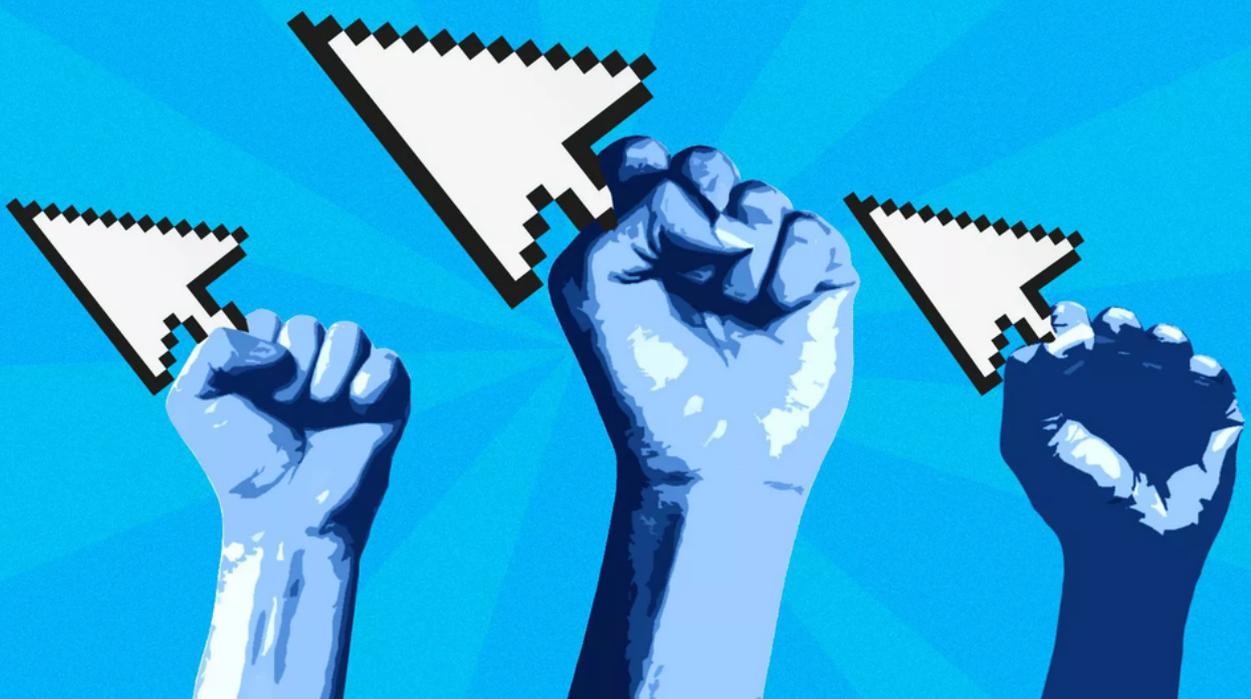

# Civic Technologies

## Research, Practice, and Open Challenges

Proposal, outcome and position papers of the 23rd ACM Conference on Computer-Supported Cooperative Work and Social Computing (CSCW 2020) workshop, held virtually on October 17, 2020.

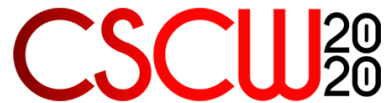

## Credits

| | |
|---|---|
| **Editors:** | Pablo Aragón, Adriana Alvarado Garcia, Christopher A. Le Dantec, Claudia Flores-Saviaga, Jorge Saldivar. |
| **Workshop organizers:** | Pablo Aragón, Adriana Alvarado Garcia, Christopher A. Le Dantec, Claudia Flores-Saviaga, Jorge Saldivar. |
| **Workshop participants:** | Abhishek Srivastava, Anna De Liddo, Annika Wolff, Anqi Cao, Antti Knutas, Belén Agüero, Beto Saavedra, Briane Paul V. Samson, Cristhian Parra, Curtis McCord, Darío Federico Uribe, Delsy Denis, Ebtihaj, Edgar Martinez, Eric Gordon, Esteban Pelaez, Gionnieve Lim, John Harlow, Mahmood Jasim, M.A Ibraheem, Marta Poblet, Mojin Yu, Muhammad Masood, Myeong Lee, Narges Mahyar, Pablo Martín Muñoz, Patriya Wiesmann, Saúl Esparza, Samantha Dalal, Shawn Janzen, Shazia Ansari, Sofia Rivas, Stephanie Blucker, Tarik Nesh-Nash, Victoria Palacin, Weiyu Zhang, XueYan Chen. |
| **Position paper authors:** | Ali Sarvghad, Amy X. Zhang, Anam Zakaria, Anna De Liddo, Annika Wolff, Anqi Cao, Antonio Aranda-Eggermont, Antti Knutas, Belén Agüero, Brian C. Keegan, Briane Paul V. Samson, Cristhian Parra, Curtis McCord, Delsy Denis, Diego Flores, Ebtihaj, Edgar Martínez, Eric Gordon, Erik Johnston, Esteban Pelaez, Fabian Medina, Fatima-Ezzahra Denial, Gionnieve Lim, Irene Martín, Jieshu Wang, John Harlow, Jon E. Froehlich, Jorge Saldivar, Julio Paciello, Liliana Savage, Luca Cernuzzi, Luis F. Cervantes, Luis Godoy, Mahmood Jasim, M.A Ibraheem, Marta Poblet, Michael Saugstad, Mojin Yu, Myeong Lee, Narges Mahyar, Norma Elva Chavez, Pablo Martín Muñoz, Patriya Wiesmann, Philipp Grunewald, Pompeu Casanovas, Rebeca de Buen Kalman, Ricardo Granados, Saiph Savage, Saúl Esparza, Samantha Dalal, Sami Hyrynsalmi, Shawn Janzen, Shazia Ansari, Simon T. Perrault, Sofia Rivas, Stephanie Blucker, Susan Winter, Tarik Nesh-Nash, Víctor Rodríguez Doncel, Victoria Giraldo, Victoria Palacin, Weiyu Zhang, XueYan Chen, Ziyi Wang. |
| **Cover author:** | Aïda Amer (Axios) |

Workshop proposal published in [11]

# Abstract

Over the last years, civic technology projects have emerged around the world to advance open government and community action. Although Computer-Supported Cooperative Work (CSCW) and Human-Computer Interaction (HCI) communities have shown a growing interest in researching issues around civic technologies, yet most research still focuses on projects from the Global North. The goal of this workshop is, therefore, to advance CSCW research by raising awareness for the ongoing challenges and open questions around civic technology by bridging the gap between researchers and practitioners from different regions.

The workshop was organized around three central topics:

1. discuss how the local context and infrastructure affect the design, implementation, adoption, and maintenance of civic technology;

2. identify key elements of the configuration of trust among government, citizenry, and local organizations and how these elements change depending on the sociopolitical context where community engagement takes place;

3. discover what methods and strategies are best suited for conducting research on civic technologies in different contexts.

# Contents

# 1

# Workshop proposal

**Pablo Aragón**
Eurecat, Centre Tecnològic de Catalunya
Universitat Pompeu Fabra
Barcelona, Spain
*elaragon@gmail.com*

**Adriana Alvarado Garcia**
**Christopher A. Le Dantec**
Georgia Institute of Technology
Atlanta, USA
*adriana.ag@gatech.edu*
*ledantec@gatech.edu*

**Claudia Flores-Saviaga**
West Virginia University
West Virginia, USA
*cif0001@mix.wvu.edu*

**Jorge Saldivar**
Barcelona Supercomputing Center
Barcelona, Spain
*jorgesaldivar@gmail.com*

## 1.1. Introduction

The Internet was heralded for its democratic potential empowering citizens and challenging existing power structures by diversifying the relationship between governments and citizens [30, 139]. In the last two decades, a large number of political innovations [207], powered by digital technologies, have emerged to scale up citizen participation and to promote new forms of governance. As noted by Linders [124], there is a plethora of competing labels for these initiatives: *collaborative government* [145], *citizen sourcing* [217], *wiki government* [160], *government as a platform* [164], *do-it-yourself government* [138], *participatory civics* [230], *digital civics* [162], etc. Among them, the term *civic technologies* (or simply **civic tech**), proposed in a report by the Knight Foundation [172] and motivated by the expected civic outcome of such technological approaches, has gained popularity in recent years.

The phenomenon of civic technologies has resulted in increasing research on different projects around the world. The first works, inspired by initiatives in the United States and Europe, focused on operationalizing the notion of civic tech and mapping existing projects into component areas [42, 54, 56, 152, 197, 201, 210]. This early literature—originated primarily in the business and social innovation sectors—was followed by academic works to develop knowledge on civic tech and its relation to public libraries [15], digital data analytics [10, 133], hackathons [94, 199], and urban collaborative governance [80]. Recent research has started to offer a broader perspective of the civic tech movement by covering case studies from geographical regions of the Global South, including Latin America [174, 188, 189], Africa [36, 174, 185], Asia [95, 213] and Oceania [187].

Although most works about civic technologies have come from social and political sciences, there has been an increase in the scholarship within the CSCW research community that examines the role of the Internet, social media, and ICTs on supporting civic engagement [14], mobilizing communities [193], and examining civic data practices [6, 24, 118, 148] and software development processes in civic projects [114, 204]. Nevertheless, there is still a tension in community technologies between novelty contributions and sustained engagement. As explained by Liu et al. [126], the broader HCI and CSCW literature has traditionally emphasized technological innovation rather than social impact. Similarly, previous work has suggested considering not only the results of civic technologies but also community practices [81, 106, 141]. Thus, we observe the disconnection between research and practice as an opportunity for future CSCW research [192]. By bringing practitioners and members from different disciplines, we aim to bridge experiences about civic technologies from both sides.

Civic technologies are constrained by their context [82], such as infrastructure [224], history of the communities [53], local practices [156], and perceived trust [45]. Therefore, it is important to identify how these elements affect the design, implementation, adoption, and maintenance of civic tech in the targeted region. Up to now most of the CSCW research on civic technologies focused on projects from the Global North. This difference between Global North and South was measured in a recent systematic review literature of more than 100 papers about civic technologies: over 85% were designed and implemented in the Global North [192]. This inequality motivates the need to promote dialogue and collaboration with key players in civic technologies from the Global South. During our workshop, rather than erasing particularities, our goal is to identify common patterns, intersections on the approaches, and similarities in practices to address open challenges.

## 1.2. Goal of the Workshop

The goal of the workshop is twofold. First, to exchange knowledge and experiences when designing, implementing, deploying and maintaining civic technologies across regions with different infrastructures, needs, and local histories. Second, to bridge the gap between researchers and civic tech practitioners (e.g., policymakers, public officers, social innovators, developers, designers, activists, etc.). To this end, our activities will focus on discussing similarities, nuances and differences among civic technologies from different regions and unpacking ongoing research challenges such as:

- **Civics, Infrastructure, and Local Context**
    - Local conditions that favour the development and deployment of civic technologies
    - Challenges when adopting existing technologies in new socio-geographic environments
    - Hybridization of online and offline participation in civic technologies

- **Civics, Trust and Government**
    - Methods for building trust among civic tech participants and with government bodies
    - Challenges in making government data available to the public

- **Sharing Methods and Strategies**
    - Governance models of civic technologies based on participatory principles
    - Approaches to ensure project sustainability and the community engagement
    - Indicators for measuring community health and democratic quality online

Lastly, due to the exceptional virtual nature of *CSCW 2020* as a response to the global crisis of COVID-19, this will be a unique occasion to attract participants from the non-academic sectors and different regions to the venue. We expect to leverage the benefits of the virtual edition to foster the participation of communities that have historically lacked visibility in top-tier academic conferences. Therefore, we intend to give priority voice to civic technology initiatives developed in the Global South.

## 1.3. Call For Participation

We seek participants who engage with research and/or practice focused on developing technologies, supporting civic engagement, or examine the mechanisms that citizens and organizations follow to influence change and decision-making on issues of concern. We will explicitly seek increased participation from researchers and practitioners from geographical regions that have traditionally been underrepresented in these academic venues, in specific from the Global South.

We will promote the call for participation in our workshop via online channels such as Twitter, Facebook groups, relevant mailing lists, and by contacting researchers and practitioners who are interested in these topics. In particular, we will contact the organizers of the *CHI 2016 Special Interest Group on Digital Civics* [221], the *CSCW 2017 Workshop on Crowdsourcing Law and Policy* [146] and the *CSCW 2019 Workshop on Social Technologies for Digital Wellbeing among Marginalized Communities* [52].

Submissions and Review

Applicants will be asked to submit a proposal including previous or ongoing research or practice that reflects on the process, lessons learned, or emerging challenges while examining, designing, or deploying civic technologies. We will give preferential treatment to applications including a 2-4 pages position paper (ACM Extended Abstract format) on their projects centered on civic technologies. Position papers are not limited to these topics, and broader discussions on digital civics are encouraged. The organizing committee will review the submissions according to their relevance and demonstrated experience with the goals of the workshop. We expect the maximum number of participants to be 25.

## 1.4. Workshop Format

### 1.4.1. Pre-Workshop Activities

Since *CSCW 2020* will take the form of a virtual conference, we will rely on the technological infrastructure provided by the conference chairs to facilitate workshops of this edition. Holding the workshop virtually will allow us to reach a broader type of participants, but this format also imposes several challenges such as reduction of depth on communication, reluctance to actively participate, and increased levels of distraction depending on the particularities of each participant's remote environment. To ameliorate some of these challenges, we are planning to send a survey before the workshop to learn about participants' time zones, identify any particular constraint, and accessibility needs that participants may have. With the results of the survey, we will be able to prepare and respond to any accessibility request and prevent unexpected situations. Additionally, we will make sure of making our workshop materials accessible. Lastly, to help to build community among participants before the workshop, we will create a Slack channel two weeks before the workshop to encourage them to begin a conversation.

### 1.4.2. Agenda

Table 1.1: Agenda of the workshop

| Time | Activity | Outcome |
|---|---|---|
| 45 min. | Introduction and Brief Remarks | - |
| 1 hour | **First Session**<br>Civics, Infrastructure, and Local Context | Collages |
| 20 min. | *Break* | |
| 1 hour | **Second Session**<br>Civics, Trust, and Government | Stakeholders Maps |
| 20 min. | *Break* | |
| 1 hour | **Third session**<br>Sharing Methods<br>and Strategies | Affinity Diagrams |

After the introductory session, our one-day workshop will be organized in three sessions, in each of which participants will brainstorm and reflect on the different challenges to research and practice of civic technologies (see Table 1.1). Participants will work in groups based on the topics that emerge from the position papers received. For the formation of the teams, we will consider the particularities of each position paper, such as target population, methods, the status of the project, and technology used. The organization of the groups will seek a balance between people from different regions and diverse backgrounds to encourage a richer discussion.

- **Introduction and Brief Remarks:** In this introductory session, the workshop's organizers will conduct brief remarks about the goal and motivation of the workshop. Then, each participant will introduce their work.

- **First Session | Civics, Infrastructure, and Local Context:** In this session, we will encourage discussion on infrastructure and local context, and how those two elements affect the design, implementation, adoption, and maintenance of civic technology. To this end, we will ask participants to craft a collage in which they describe the existent or lacking infrastructure in the context where they work. To facilitate this activity, we will provide participants with a collage kit.

- **Second Session | Civics, Trust, and Government:** Participants in this session will focus the discussion on how trust in digital civics depends on the sociopolitical context where community engagement takes place. We will encourage them to identify key elements of the configuration of trust among government, citizenry, and local organizations. To this end, we will ask participants to use an adapted version of stakeholder maps to visually communicate who are the key constituents of their ongoing projects and to define hierarchies and key relationships. To facilitate this process, we will provide participants with digital templates and visual materials on Jamboard. Similarly to the previous session, we will ask each group to present their maps to the rest of the participants.

- **Third Session | Sharing Methods and Strategies:** Building on the discussions of the two previous sessions, we will ask participants in the last session to reflect on how the key elements of infrastructure, local context, and trust of the region where they have been conducting their research have influenced their selection and adaptation of research methods. Through an affinity diagram activity, participants will share and discover what methods and strategies are best suited for conducting research on civic technologies in specific contexts. After the activity, each group will present their affinity diagram to the rest of the participants.

### 1.4.3. Website

We have created a website[1] to provide an overview of the workshop, the agenda, and expected outcomes. The website will also be used to post the call for submissions and to feature accepted position papers, relevant materials and, after the conclusion of the workshop, a summary of the contributions to the CSCW community.

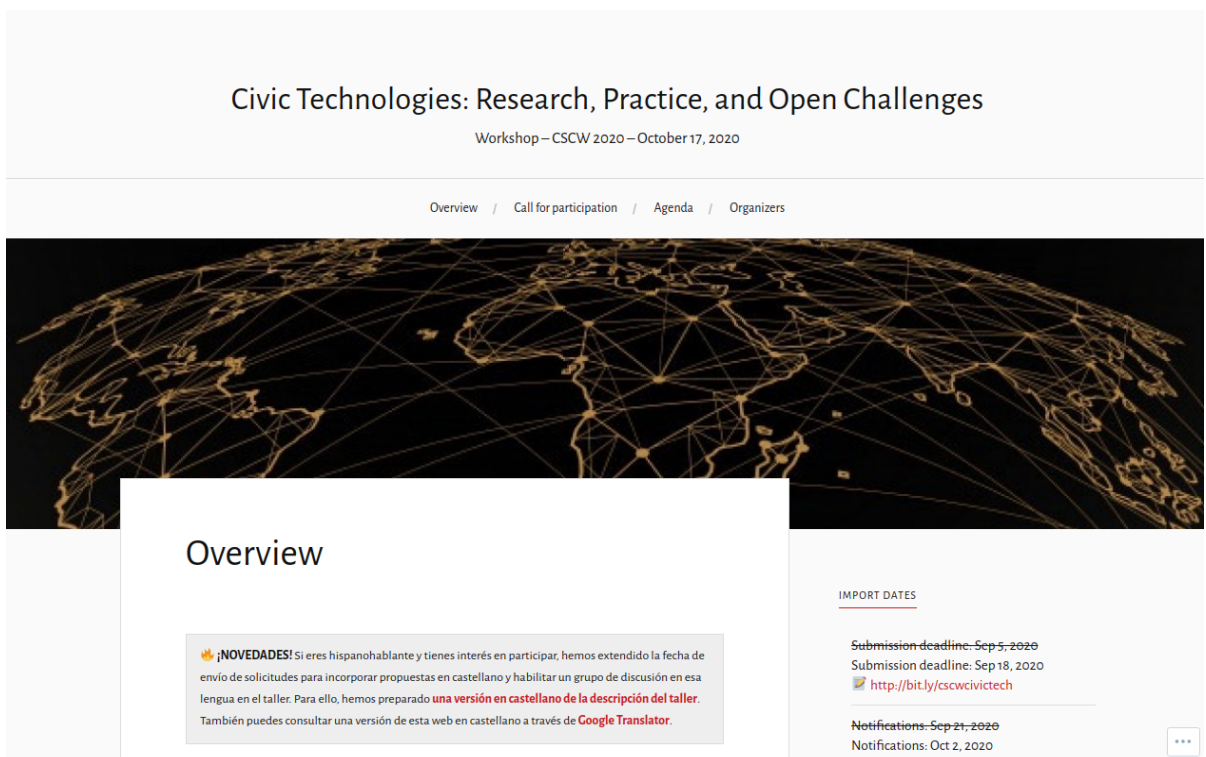

Figure 1.1: Screenshot of the website.

---

[1] cscwcivictechnologies.wordpress.com

# 2

# Outcome

In this section, we present the discussion groups of the workshop and their outcome with the toolkits provided by the organizers.

## 2.1. Groups

The workshop involved participants from institutions covering 15 countries: Australia, Bolivia, Canada, Colombia, Finland, Hong Kong, India, Mexico, Pakistan, Paraguay, Philippines, Singapore, Spain, UK and USA. Given the many time zones, the workshop was held on October 17, 2020 (Saturday) in two different intervals:

- Interval 1 (2.00 - 6.00, Central Time): 2 English groups (1a, 1b).

- Interval 2 (7.00 - 11.00, Central Time): 3 English groups (2a, 2b, 2C) and 1 Spanish group.

### 2.1.1. Participants in group 1a English

- Abhishek Srivastava (Indian Institute of Management Jammu)

- Briane Paul V. Samson (De La Salle University)

- Belén Agüero (CIECODE)

- Beto Saavedra (bolivia.ai)

- Muhammad Ibraheem Saleem (Code for Pakistan)

- Tarik Nesh-Nash (Impact For Development)

**Organizer**: Adriana Alvarado Garcia (Georgia Institute of Technology)

### 2.1.2. Participants in group 1b English

- Marta Poblet (RMIT University, Melbourne)

- Weiyu Zhang (National University of Singapore)

- Muhammad Masood (City University of Hong Kong)

- Ebtihaj (Code for Pakistan)

- Estebán Pelaez (Fundación Corona)

**Organizer**: Pablo Aragón (Eurecat & Universitat Pompeu Fabra)

### 2.1.3. Participants in group 2a English

- Antti Knutas Antti Knutas (LUT University)

- John Harlow (Emerson College)

- Shawn Janzen (University of Maryland)

- Samantha Dalal (University of Colorado Boulder)

- Anqi Cao (HCDE, UW)

- XueYan Chen (HCDE, UW)

**Organizer**: Claudia Flores-Saviaga (West Virginia University)

### 2.1.4. Participants in group 2b English

- Anna De Liddo (Knowledge Media Institute, The Open University)

- Curtis McCord (University of Toronto)

- Victoria Palacin (University of Helsinki & LUT University)

- Gionnieve Lim (Singapore University of Technology and Design)

- Stephanie Blucker (HCDE, UW)

- Mojin Yu (HCDE, UW)

**Organizer**: Pablo Aragón (Eurecat & Universitat Pompeu Fabra)

### 2.1.5. Participants in group 2c English

- Eric Gordon (Emerson College)

- Annika Wolff (LUT University)

- Myeong Lee (George Mason University)

- Mahmood Jasim (University of Massachusetts)

- Shazia Ansari (Civic Innovation Lab, UNAM)

- Patriya Wiesmann (HCDE, UW)

- Saúl Esparza (Civic Innovation Lab, UNAM)

**Organizer**: Christopher A. Le Dantec (Georgia Institute of Technology)

### 2.1.6. Participants in group 2 Spanish

- Cristhian Parra (Universidad Católica Nuestra)

- Delsy Denis (Universidad Nacional de Asunción)

- Sofia Rivas (Universidad Nacional de Asunción)

- Darío Federico Uribe (Movilizatorio)

- Pablo Martín Muñoz (CIECODE)

- Edgar Martínez (Liga Peatonal)

**Organizer**: Jorge Saldivar (Barcelona Supercomputing Center)

## 2.2. Toolkits

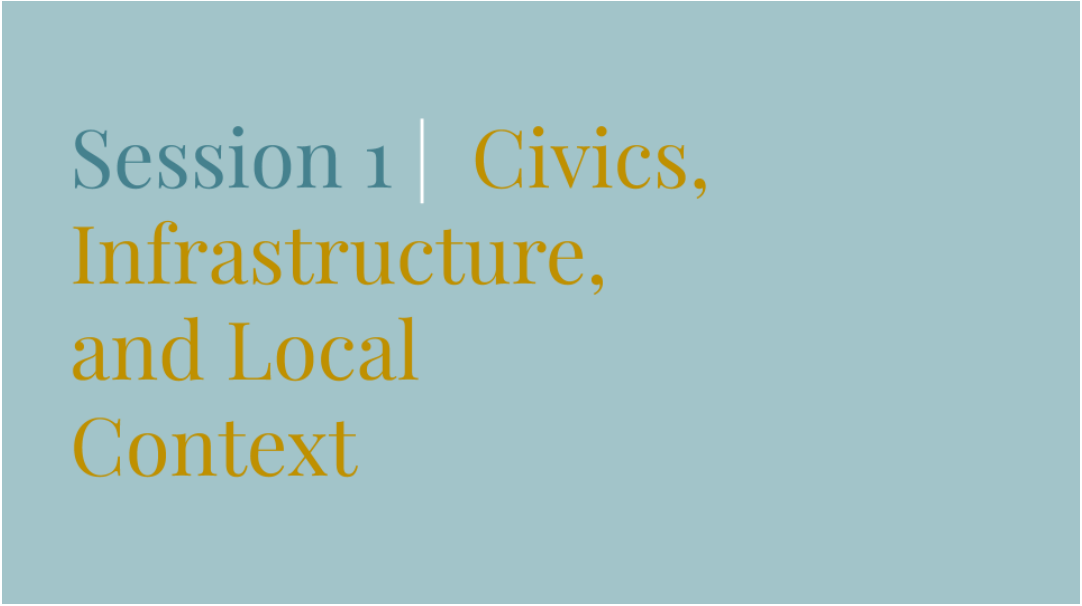

Figure 2.1: Introduction to the first session.

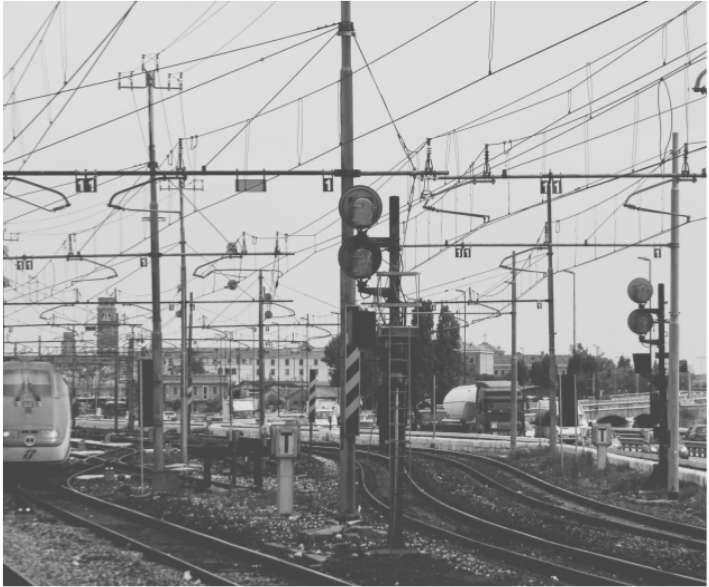

Figure 2.2: Description of the first session.

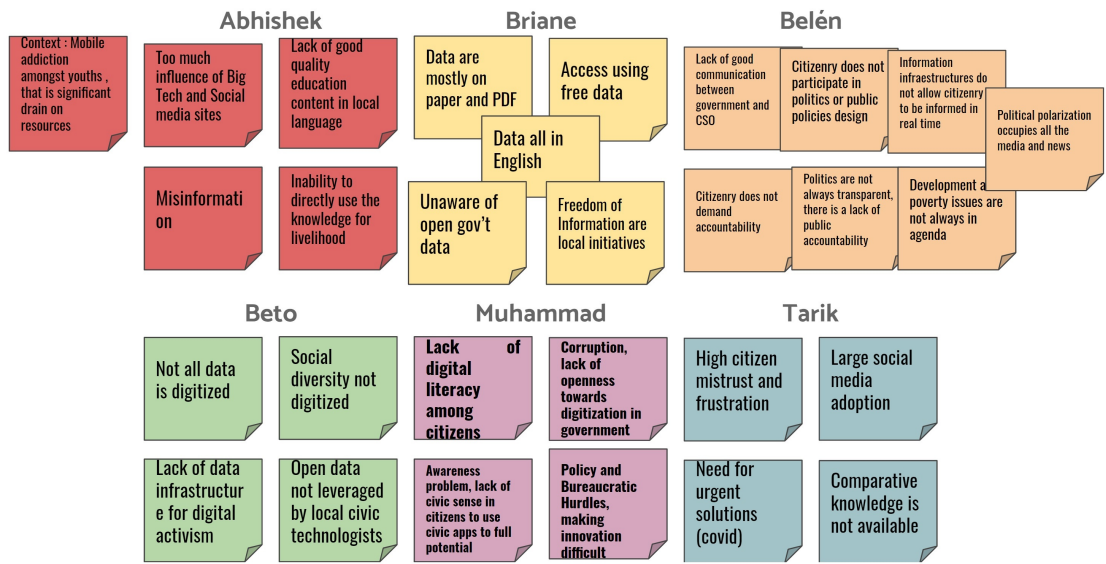

Figure 2.3: Partial outcome of the first session by Group 1a English.

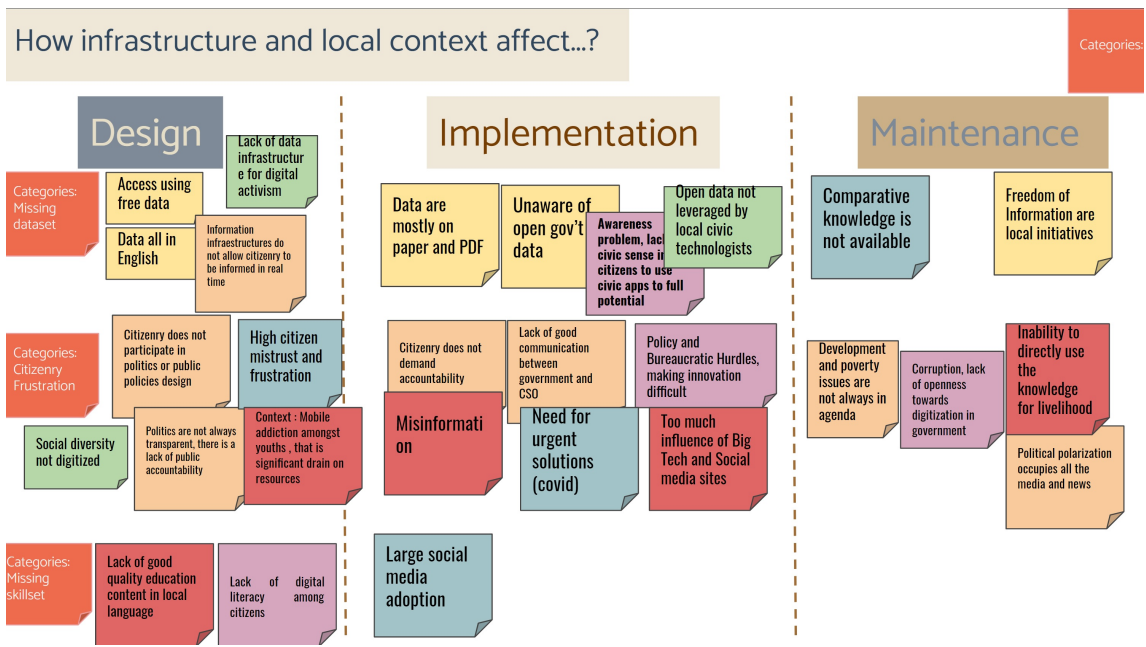

Figure 2.4: Final outcome of the first session by Group 1a English.

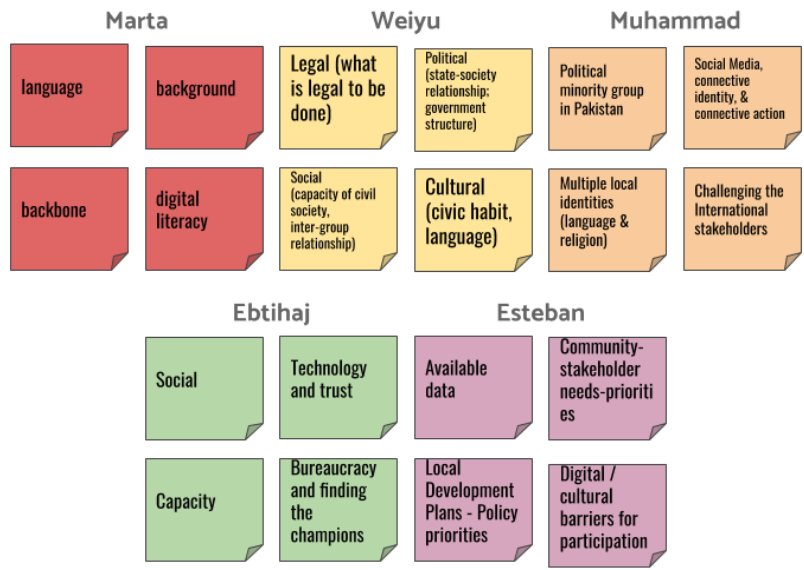

Figure 2.5: Partial outcome of the first session by Group 1b English.

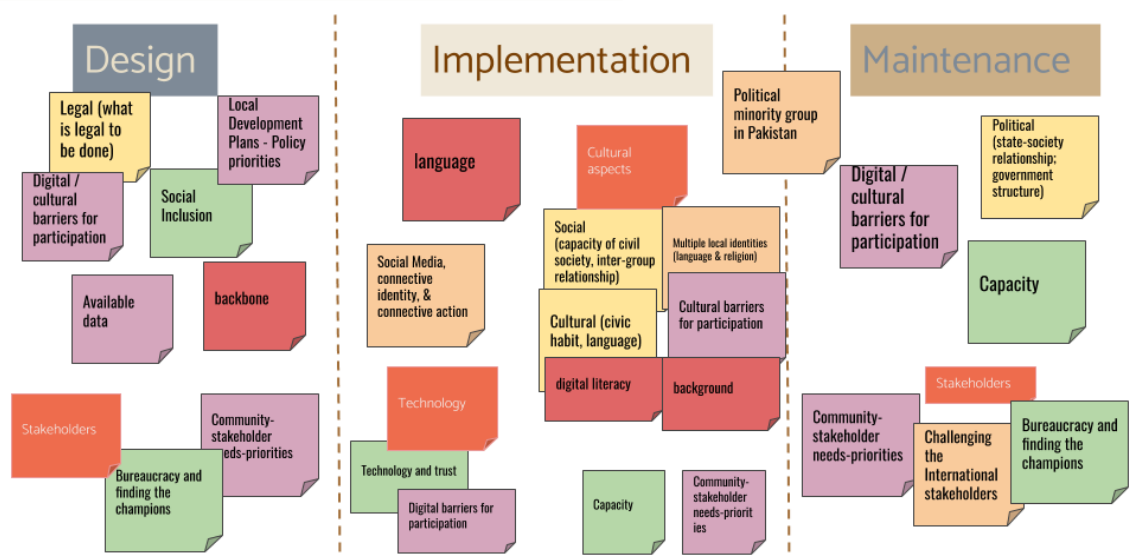

Figure 2.6: Final outcome of the first session by Group 1b English.

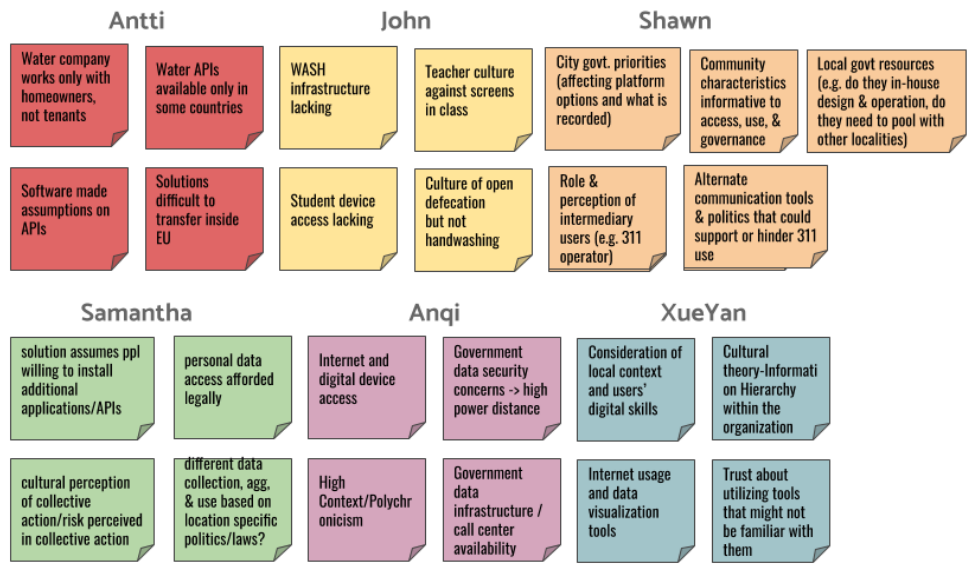

Figure 2.7: Partial outcome of the first session by Group 2a English.

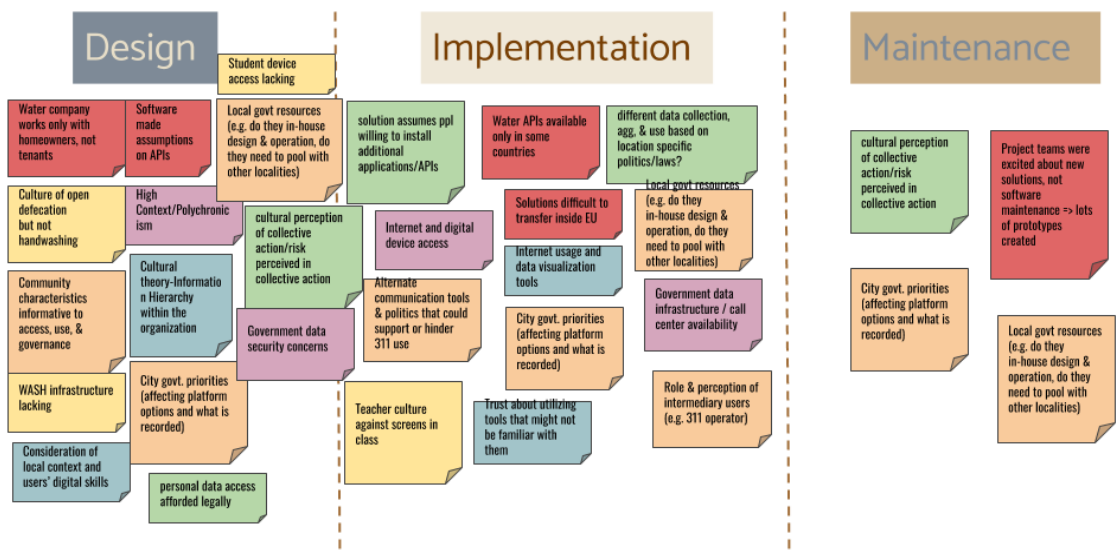

Figure 2.8: Final outcome of the first session by Group 2a English.

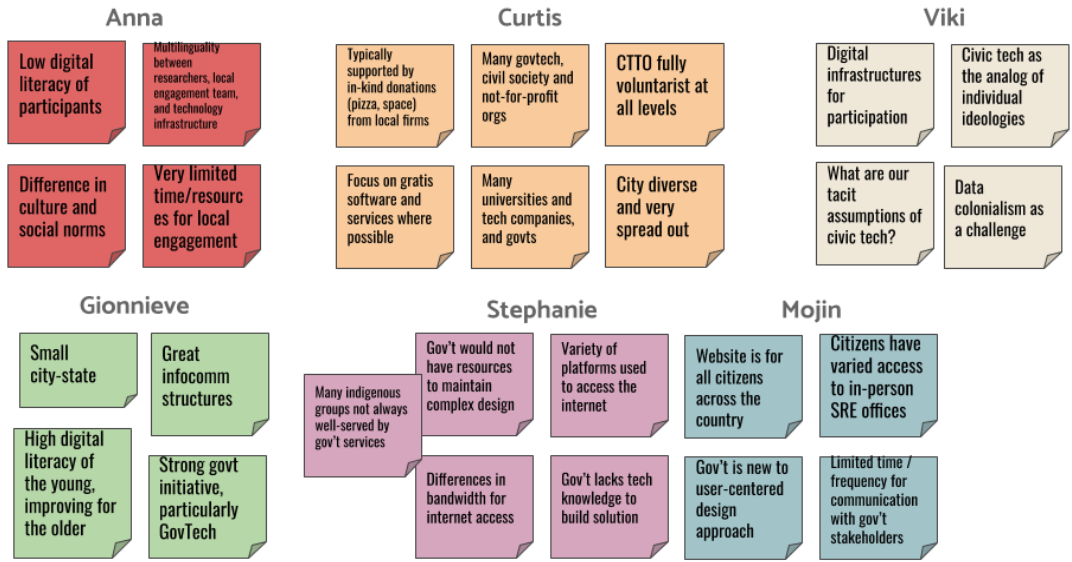

Figure 2.9: Partial outcome of the first session by Group 2b English.

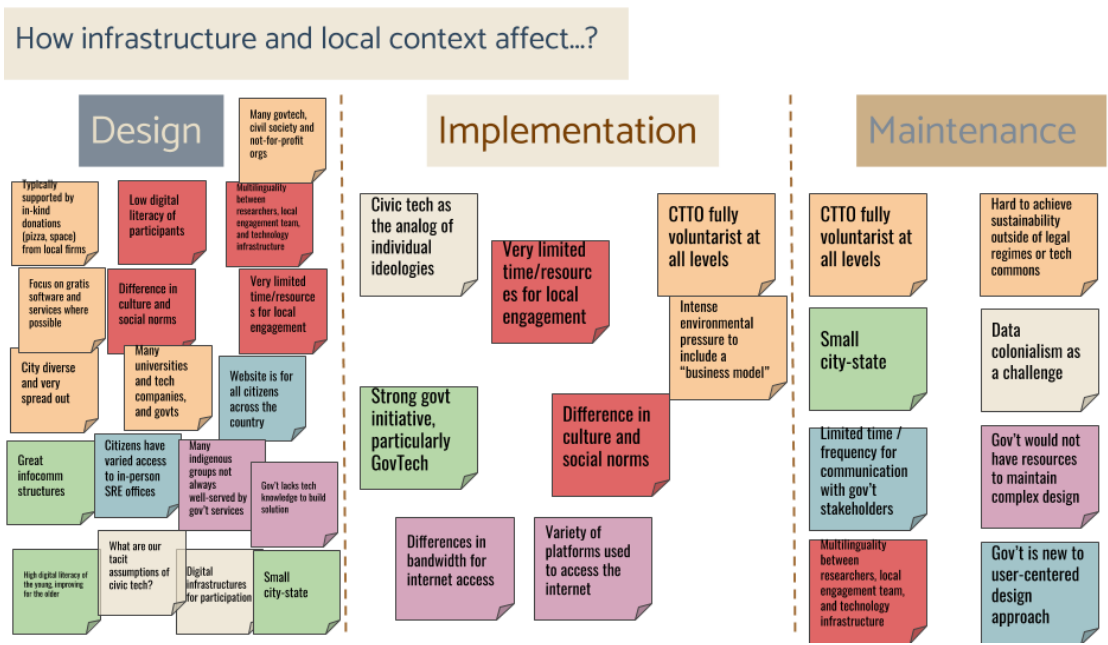

Figure 2.10: Final outcome of the first session by Group 2b English.

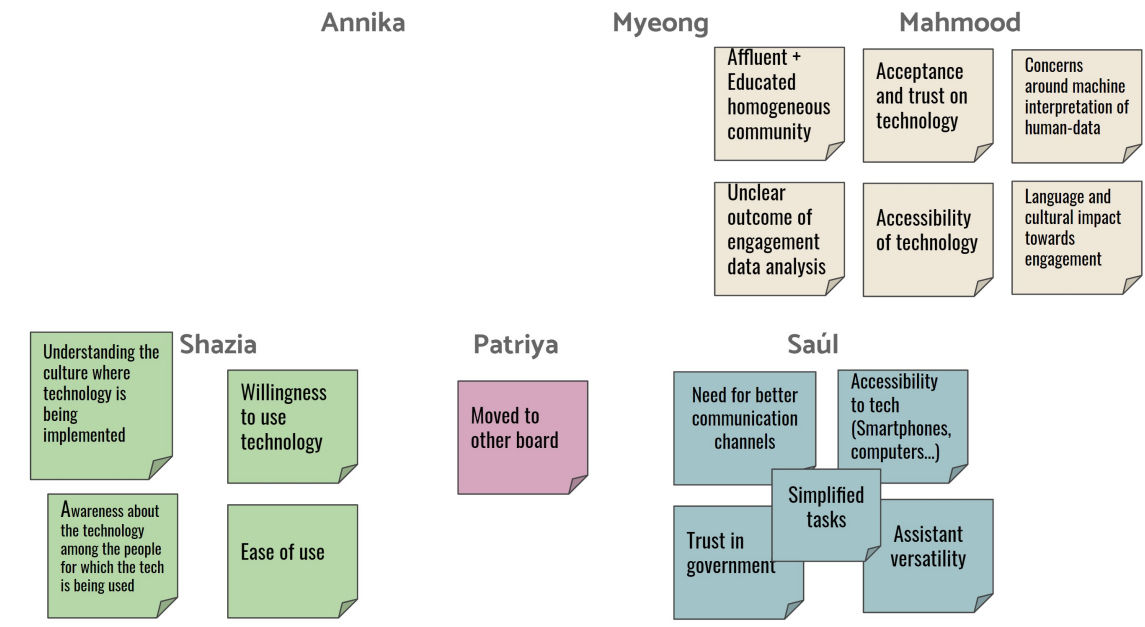

Figure 2.11: Partial outcome of the first session by Group 2c English.

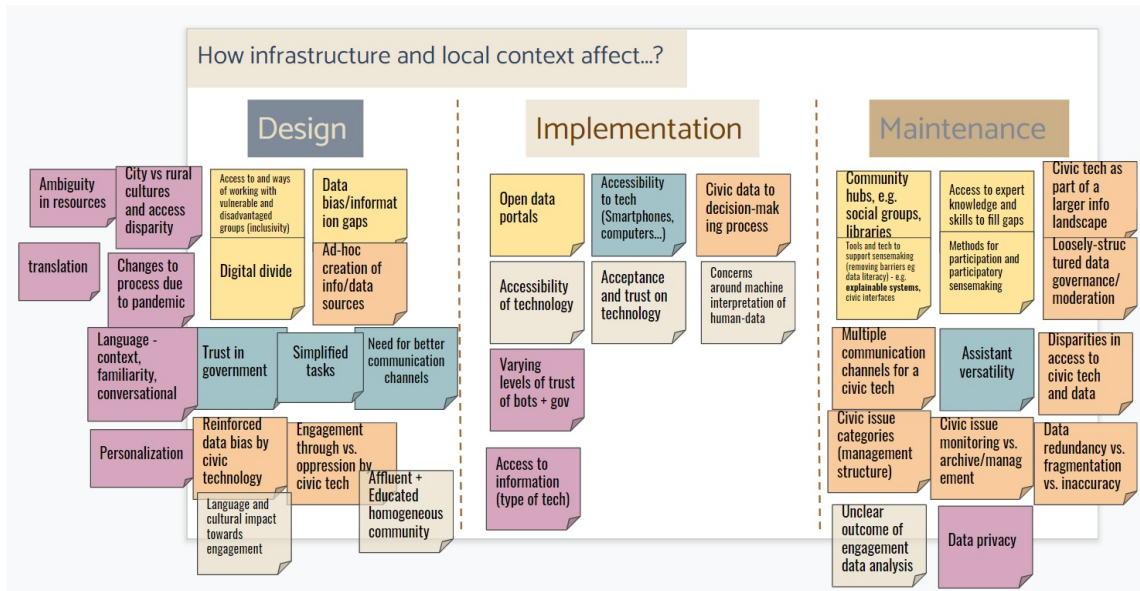

Figure 2.12: Final outcome of the first session by Group 2c English.

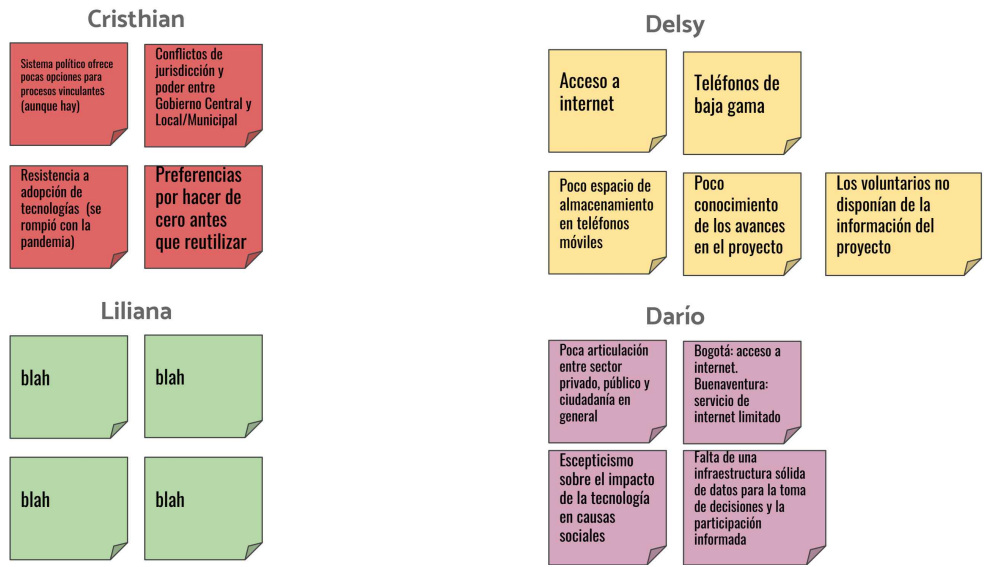

Figure 2.13: Partial outcome of the first session by Group 2d Spanish.

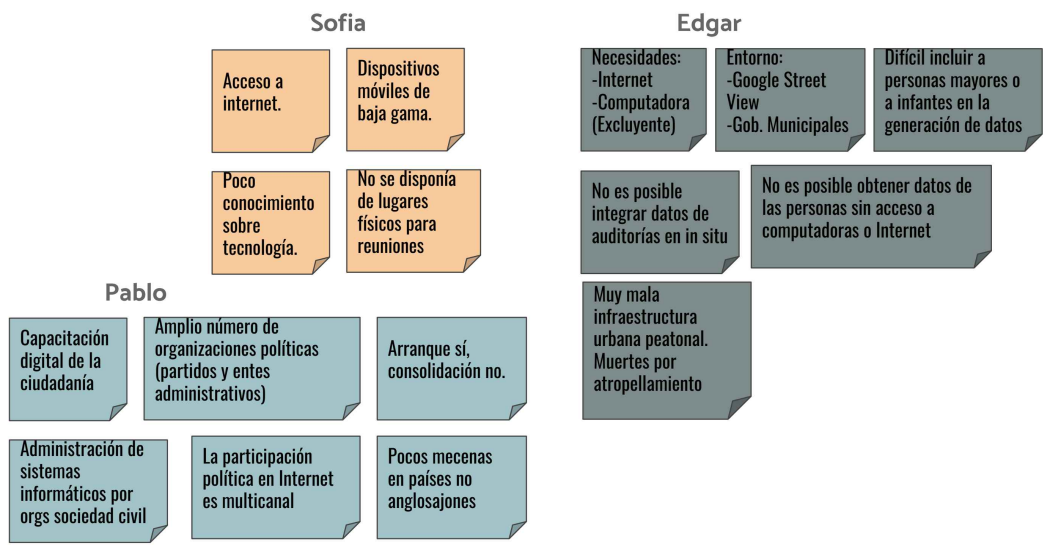

Figure 2.14: Partial outcome of the first session by Group 2d Spanish.

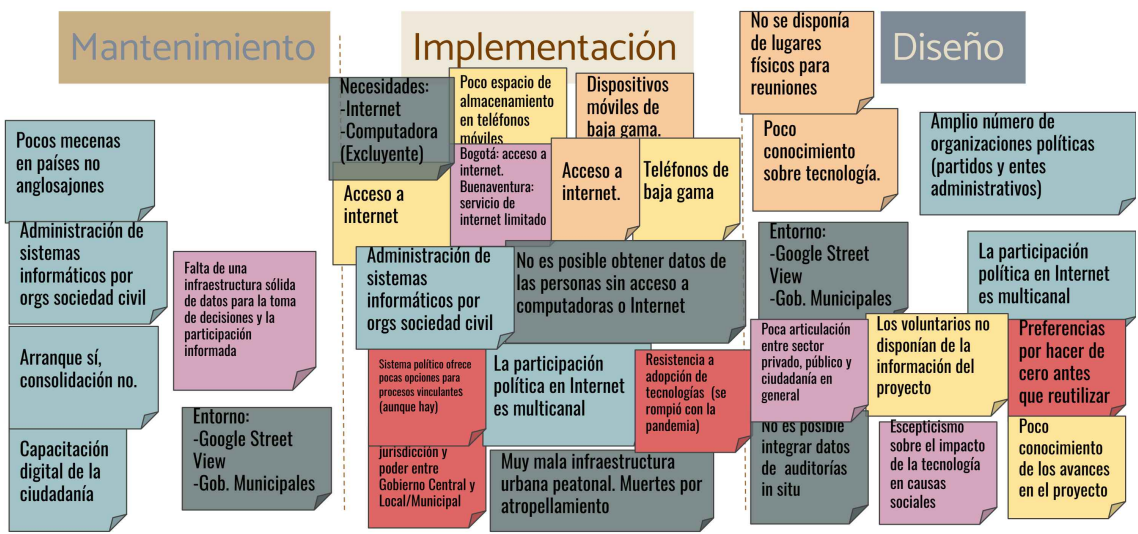

Figure 2.15: Final outcome of the first session by Group 2d Spanish.

Session 2 |
Civics, Trust, and
Government

Figure 2.16: Introduction to the second session.

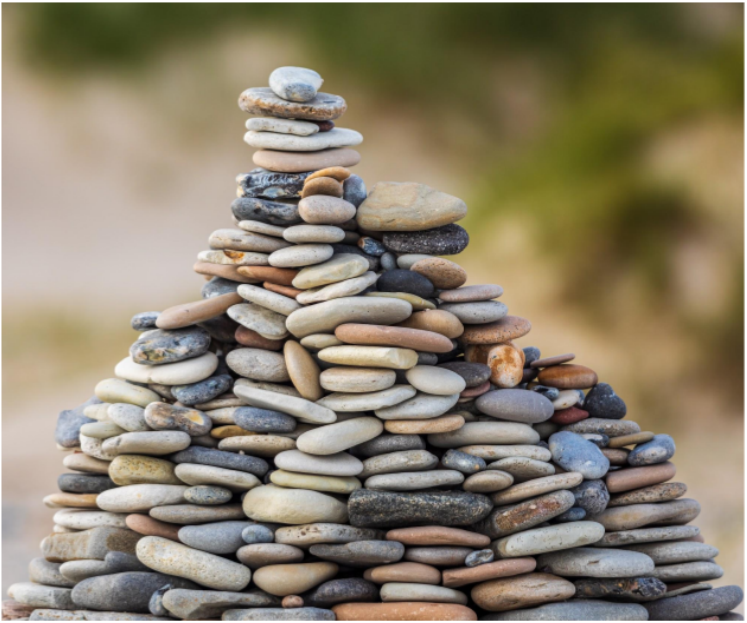

The goal of this session is to discuss how the sociopolitical context where community engagement takes place affects trust in digital civics.

We will identify key elements of the configuration of trust among government, citizenry, and local organizations.

Figure 2.17: Description of the second session.

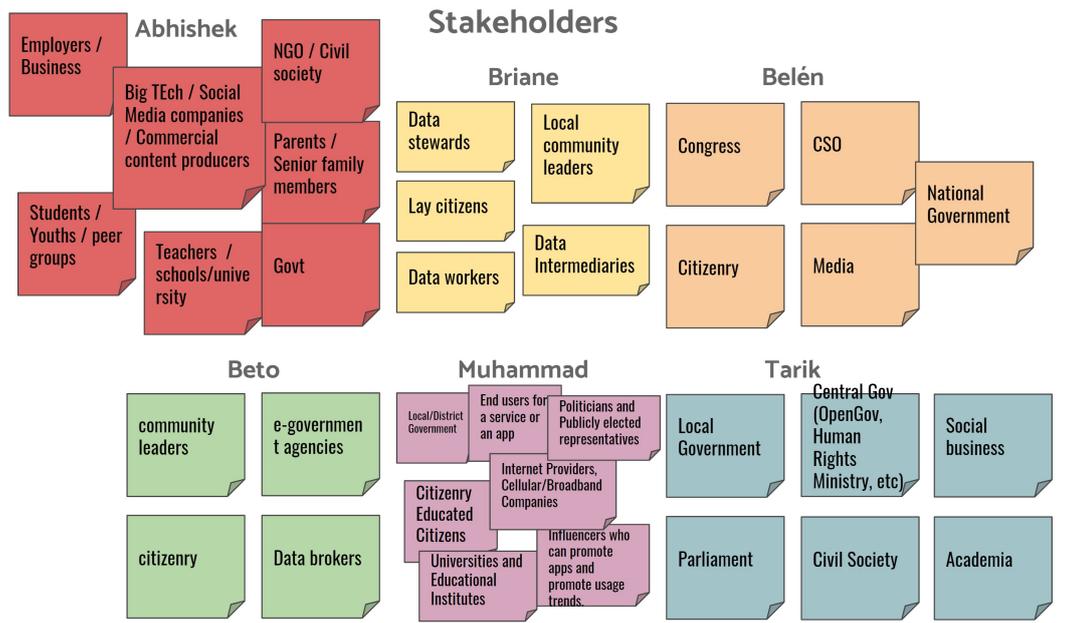

Figure 2.18: Partial outcome of the second session by Group 1a English.

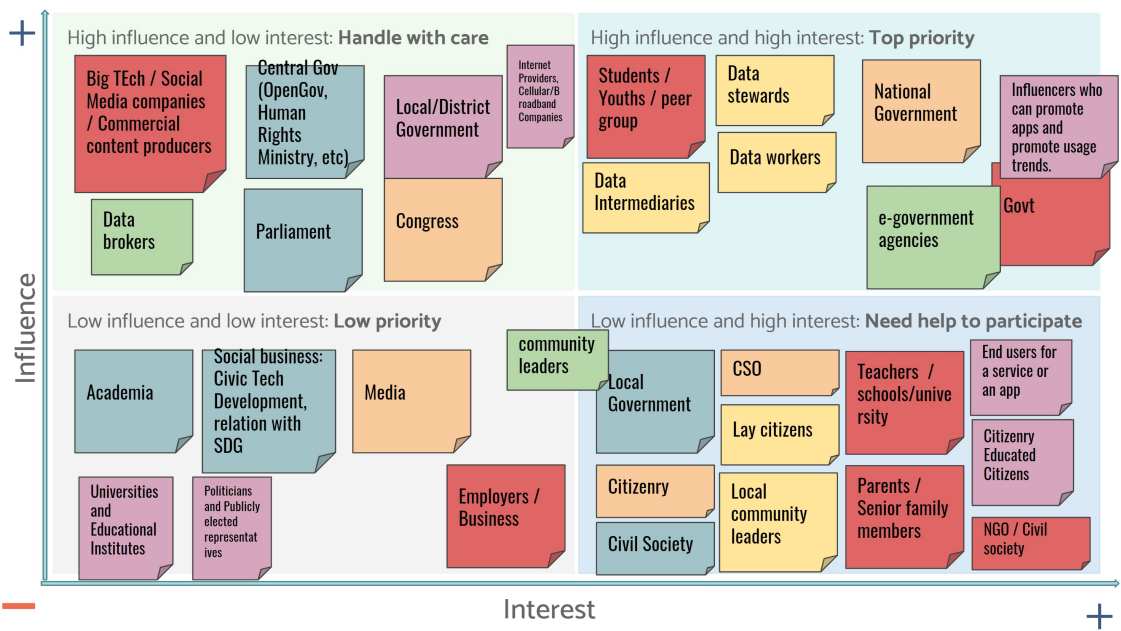

Figure 2.19: Final outcome of the second session by Group 1a English.

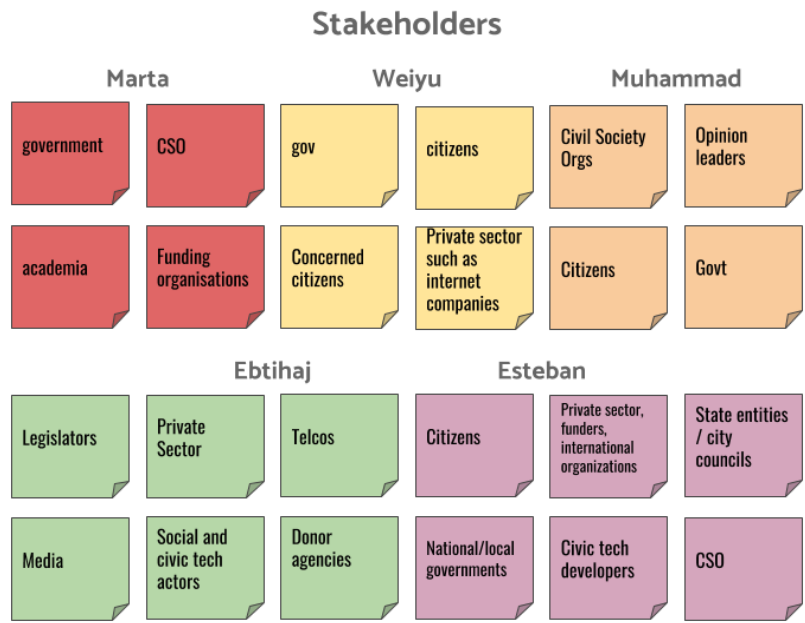

Figure 2.20: Partial outcome of the second session by Group 1b English.

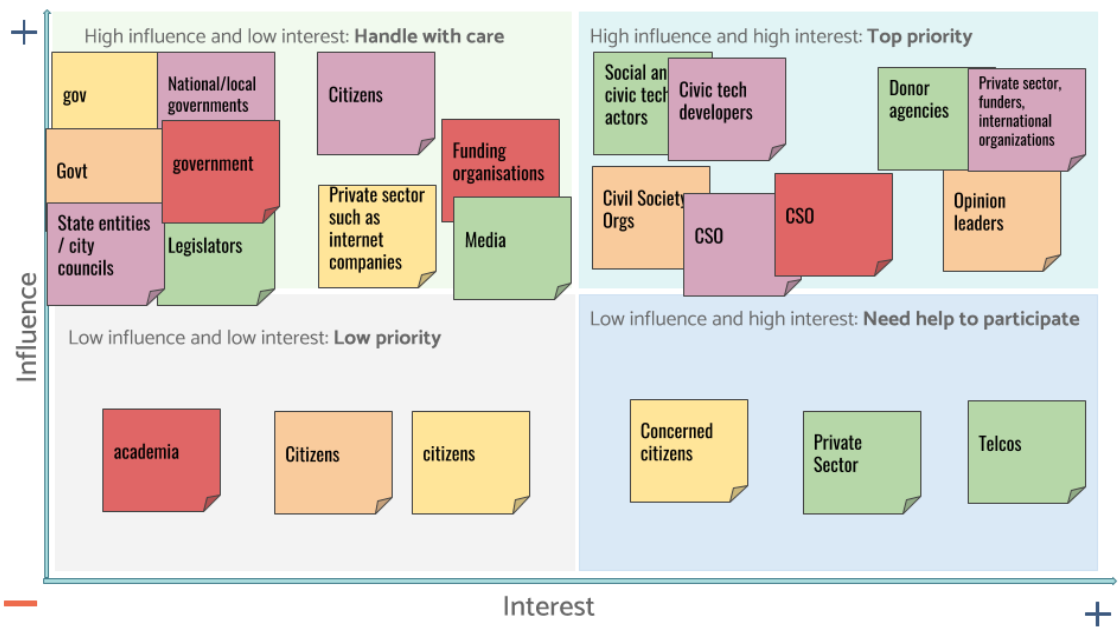

Figure 2.21: Final outcome of the second session by Group 1b English.

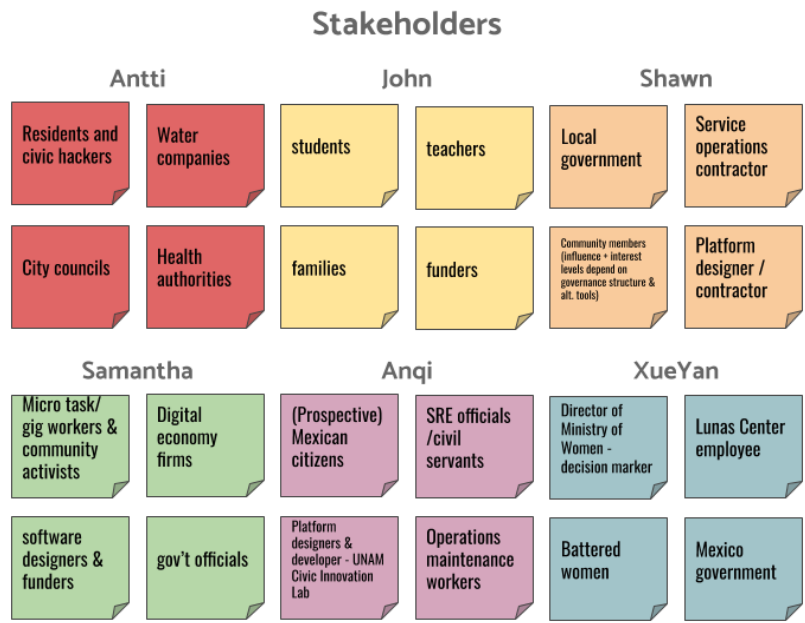

Figure 2.22: Partial outcome of the second session by Group 2a English.

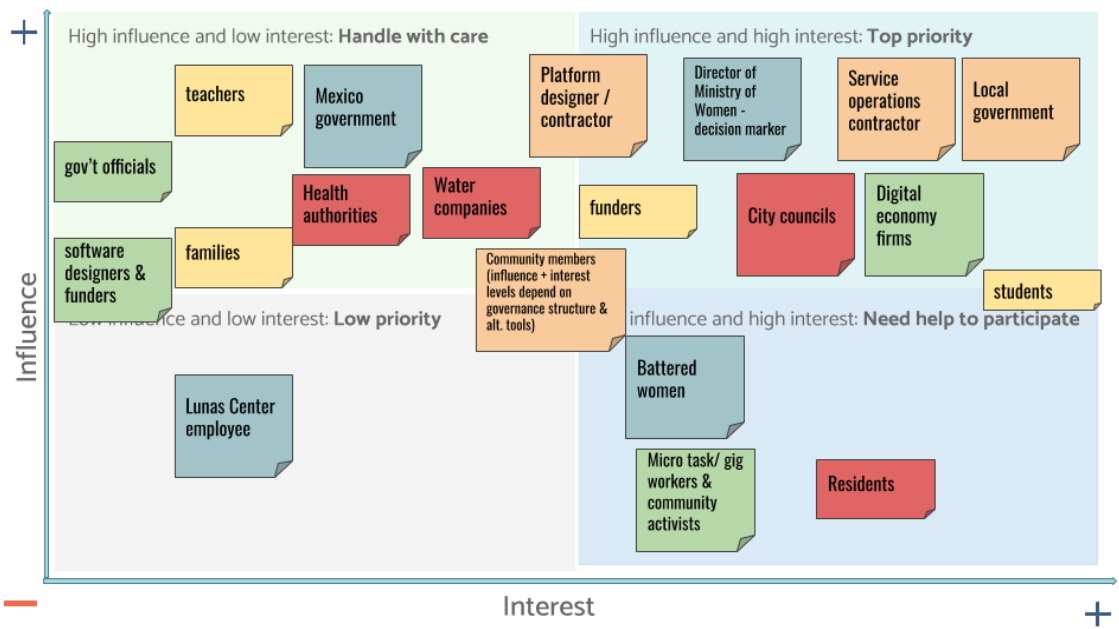

Figure 2.23: Final outcome of the second session by Group 2a English.

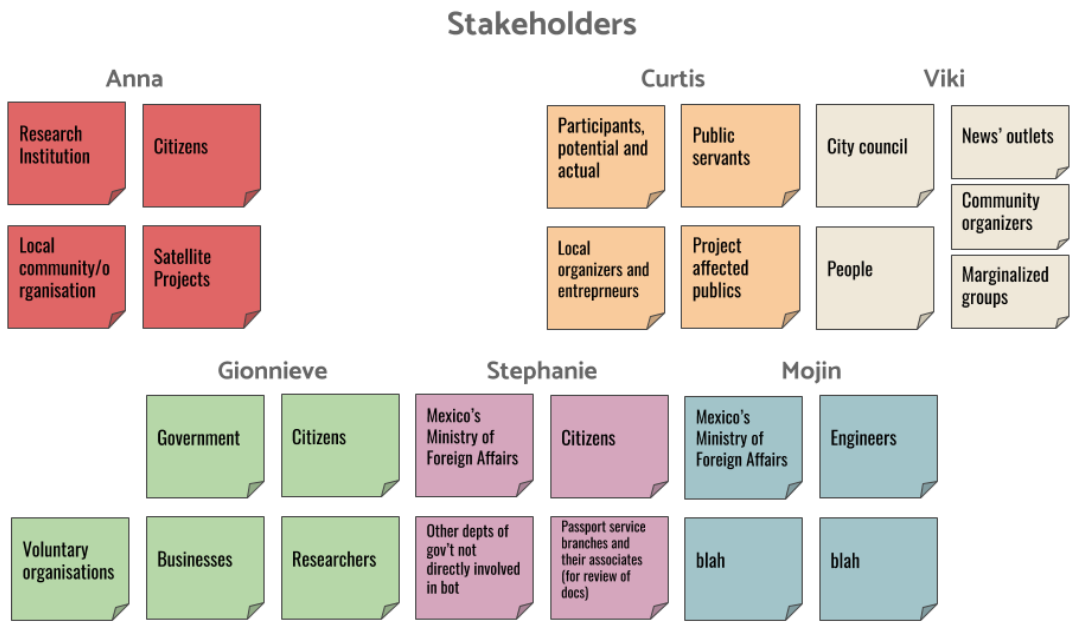

Figure 2.24: Partial outcome of the second session by Group 2b English.

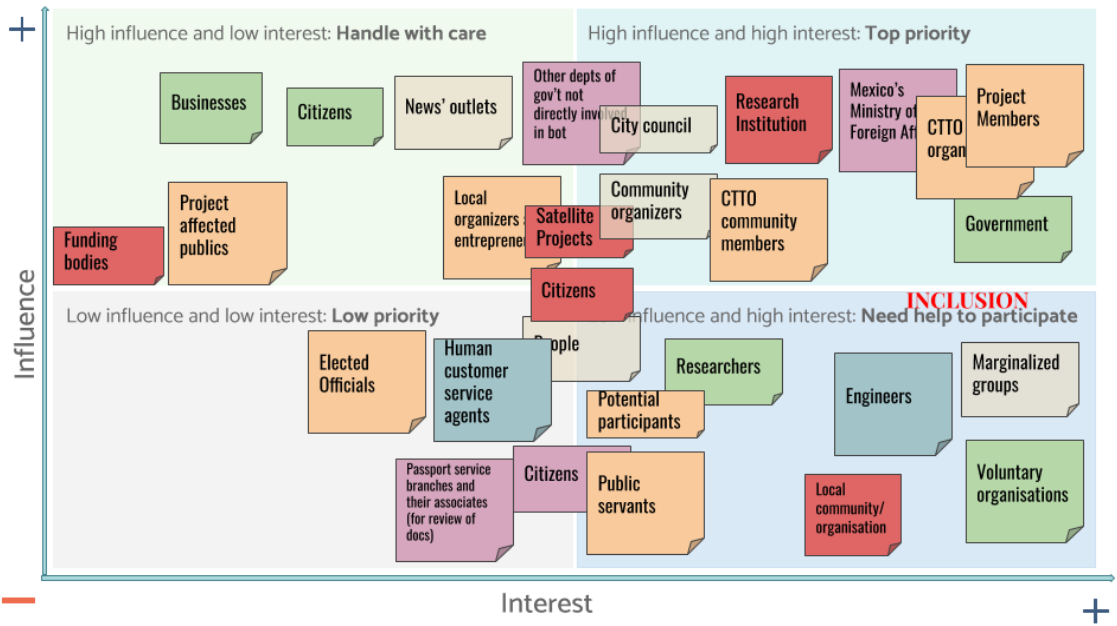

Figure 2.25: Final outcome of the second session by Group 2b English.

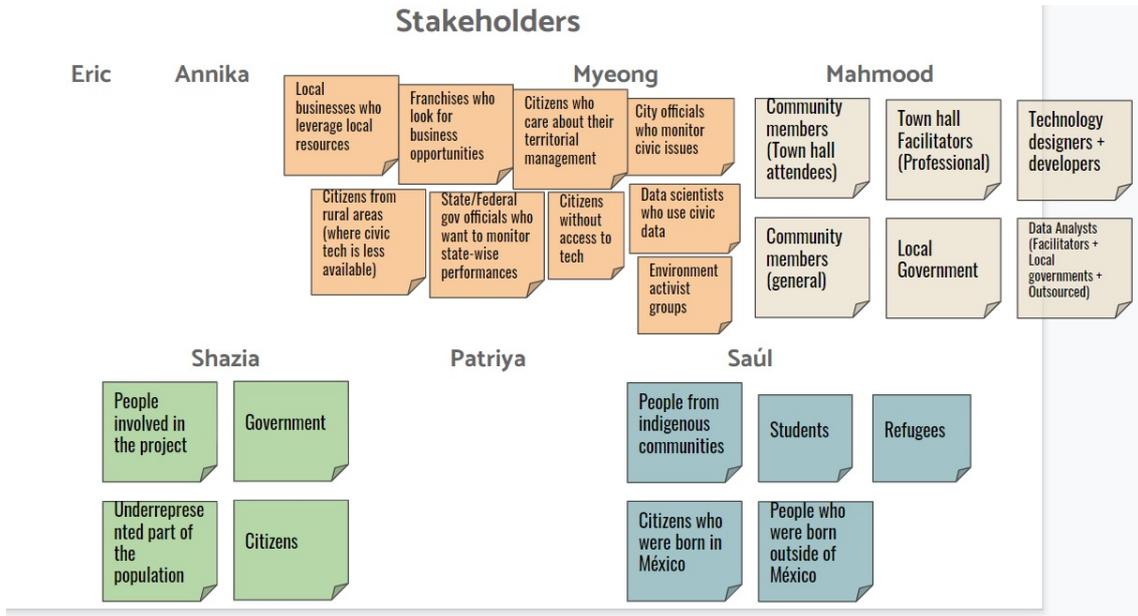

Figure 2.26: Partial outcome of the second session by Group 2c English.

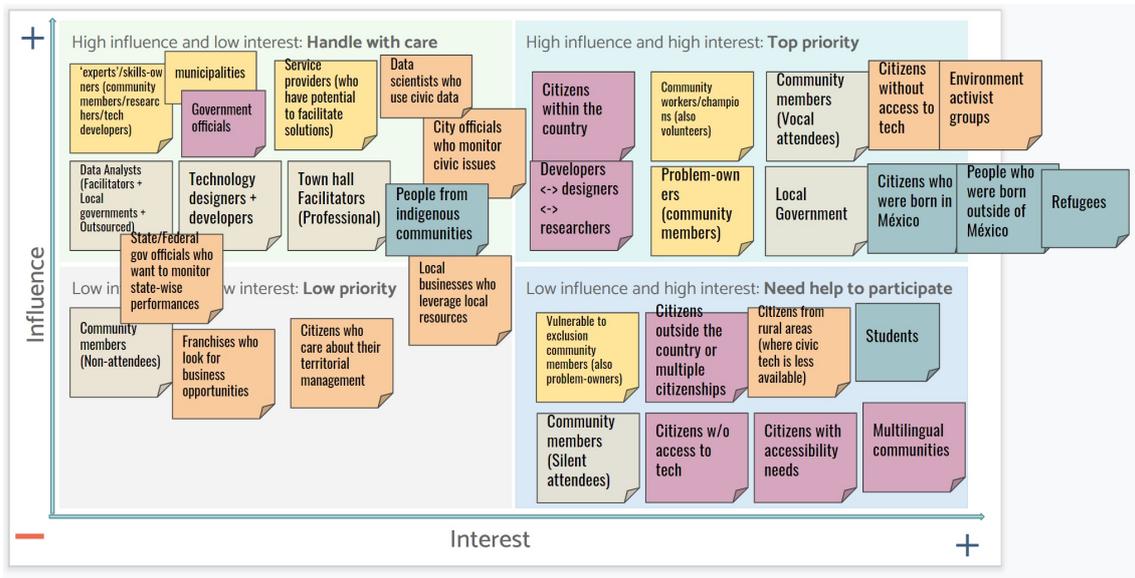

Figure 2.27: Final outcome of the second session by Group 2c English.

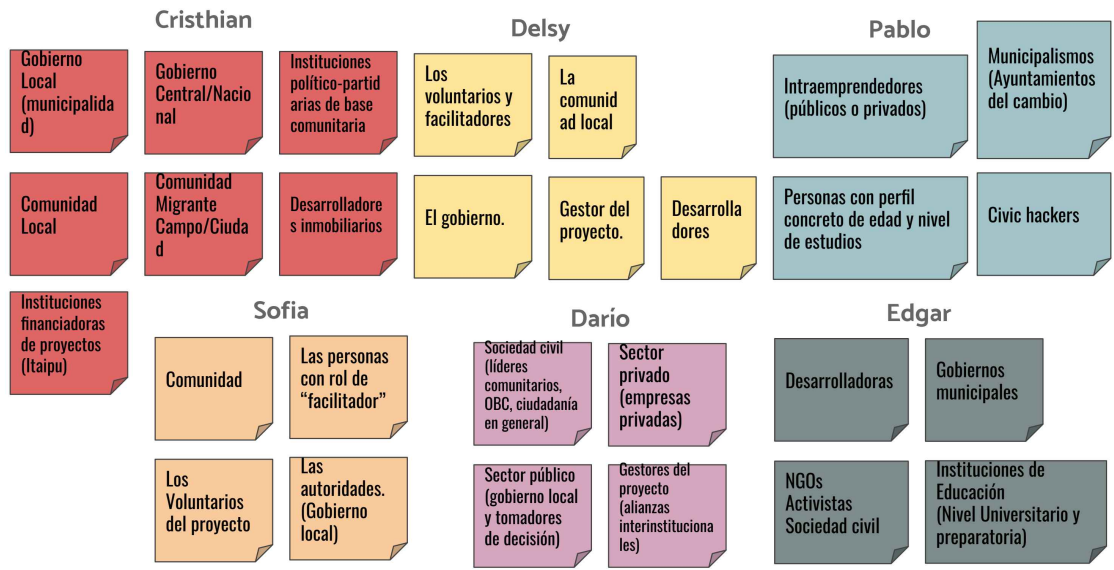

Figure 2.28: Partial outcome of the second session by Group 2d Spanish.

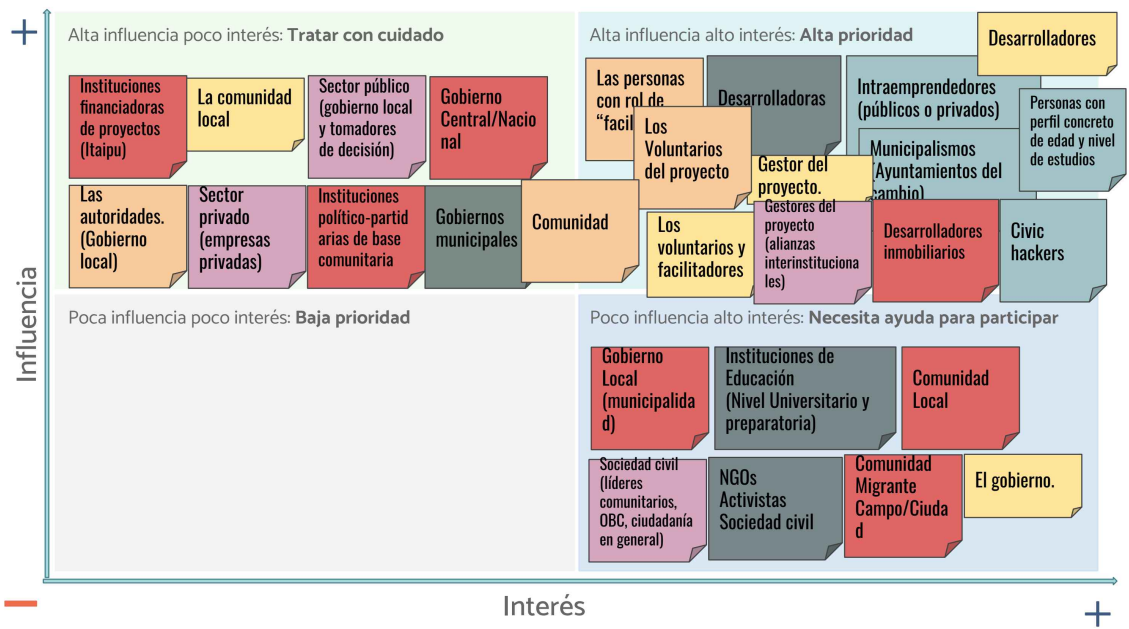

Figure 2.29: Final outcome of the second session by Group 2d Spanish.

# Session 3 |
# Sharing Methods and Strategies

Figure 2.30: Introduction to the third session.

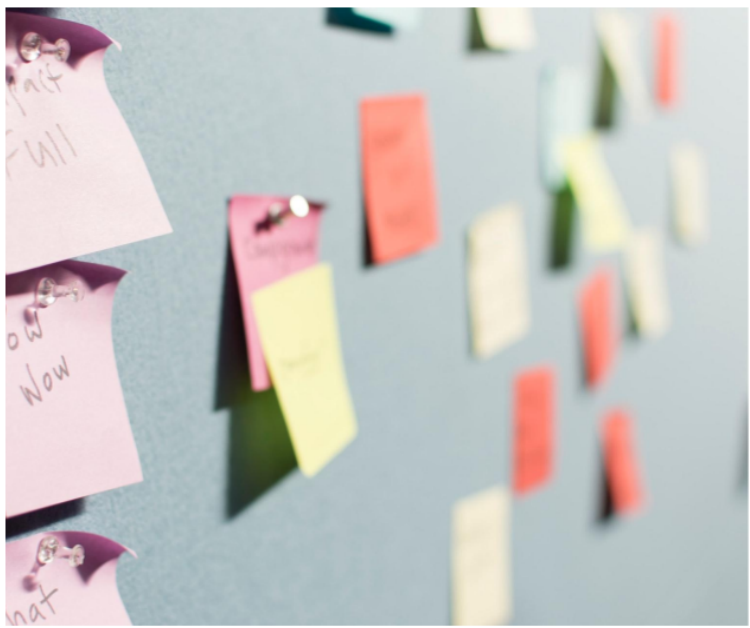

The goal of this session is to reflect on how the primary elements of infrastructure, local context, and trust of the region where we are conducting research have influenced our selection and adaptation of research methods.

Through an affinity diagram activity, we will share and discover what methods and strategies are best suited for conducting research on civic technologies in specific contexts.

Figure 2.31: Description of the third session.

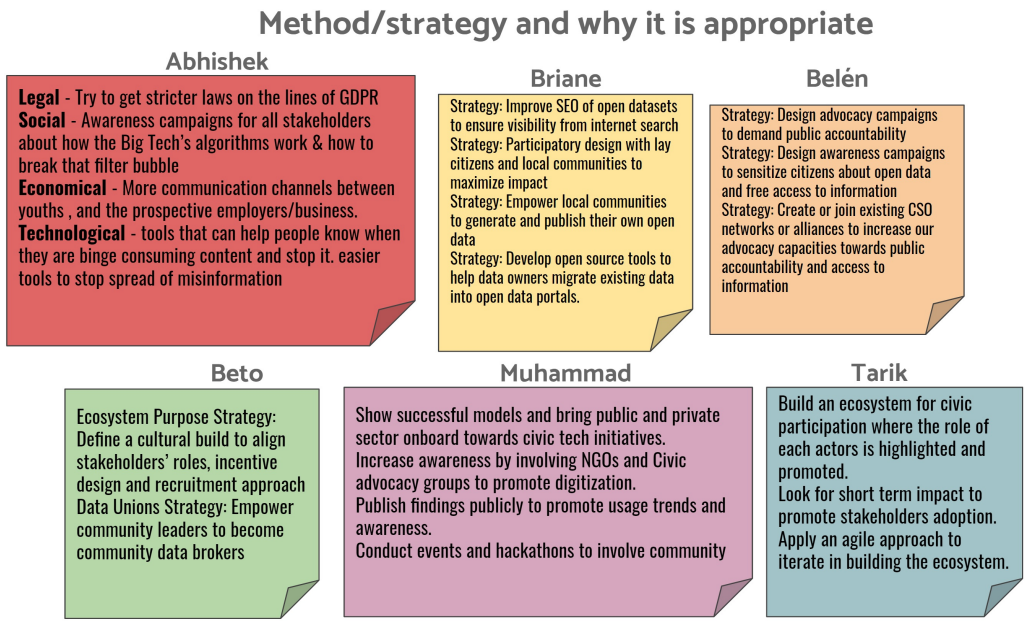

Figure 2.32: Outcome of the third session by Group 1a English.

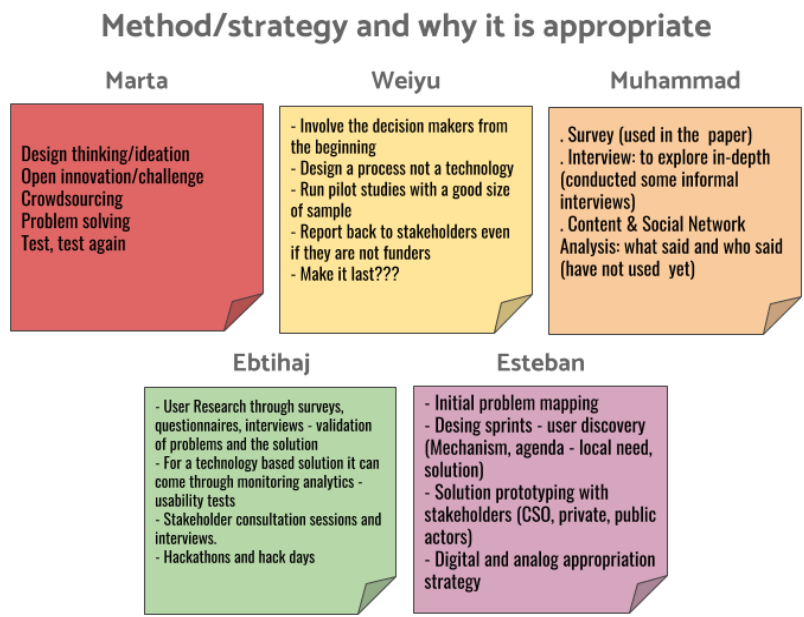

Figure 2.33: Outcome of the third session by Group 1b English.

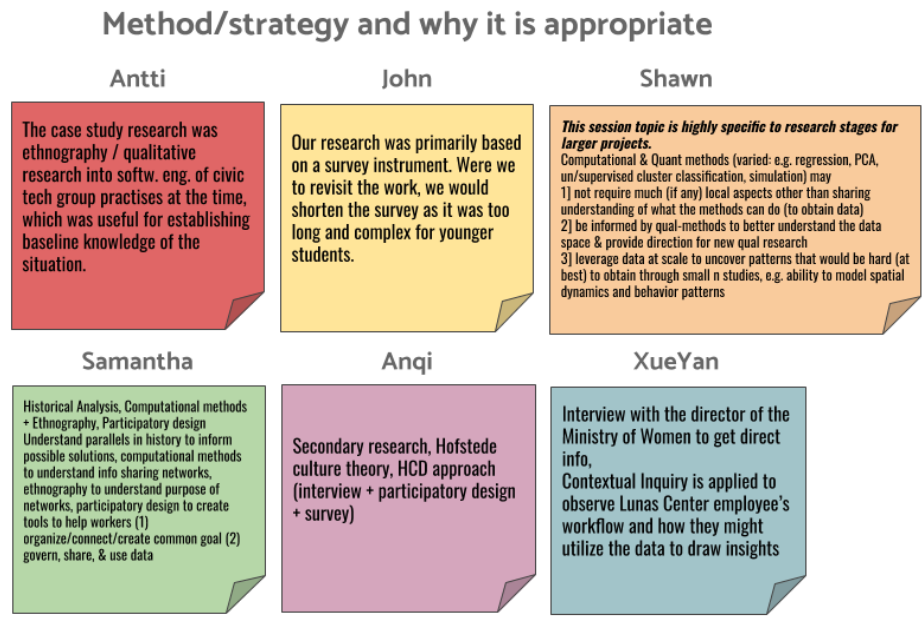

Figure 2.34: Outcome of the third session by Group 2a English.

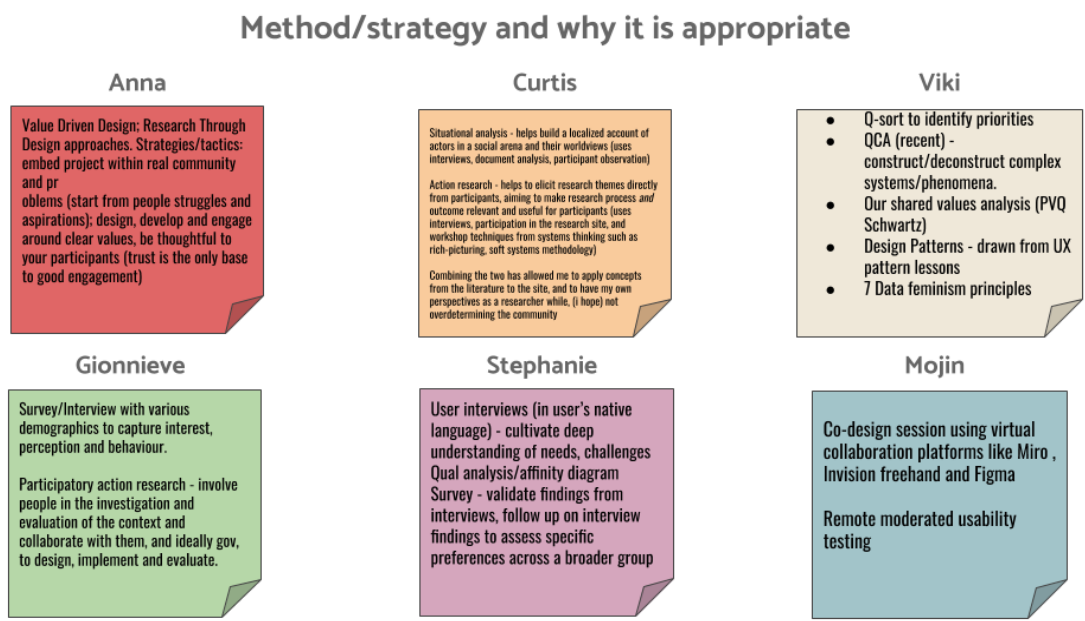

Figure 2.35: Outcome of the third session by Group 2b English.

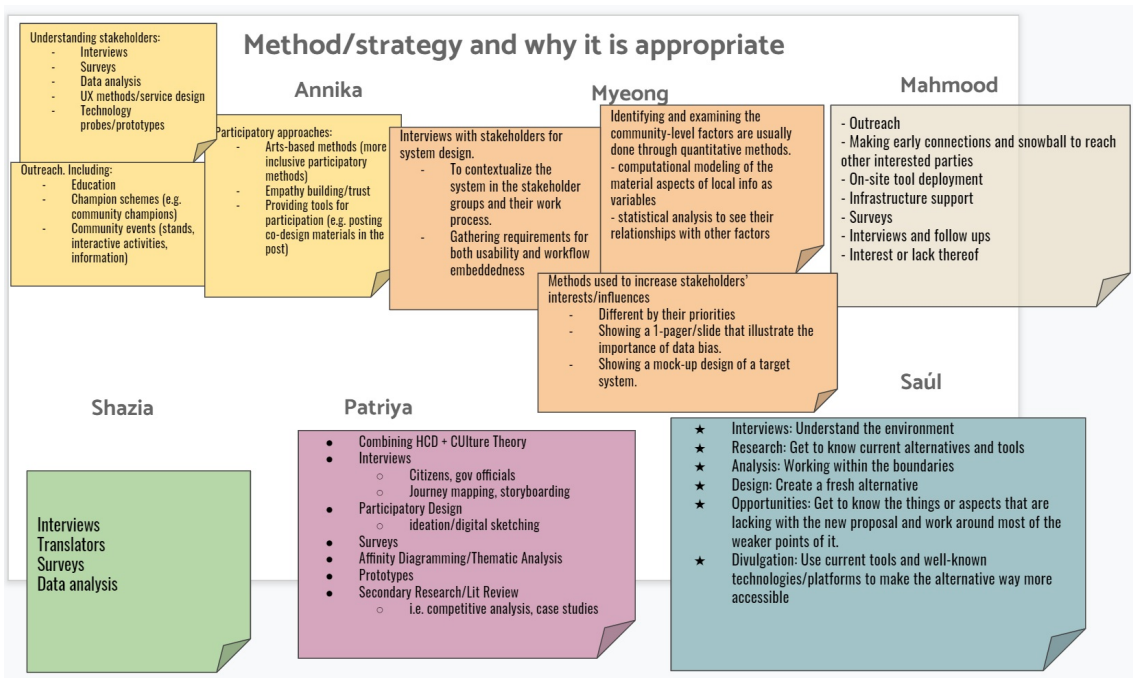

Figure 2.36: Outcome of the third session by Group 2c English.

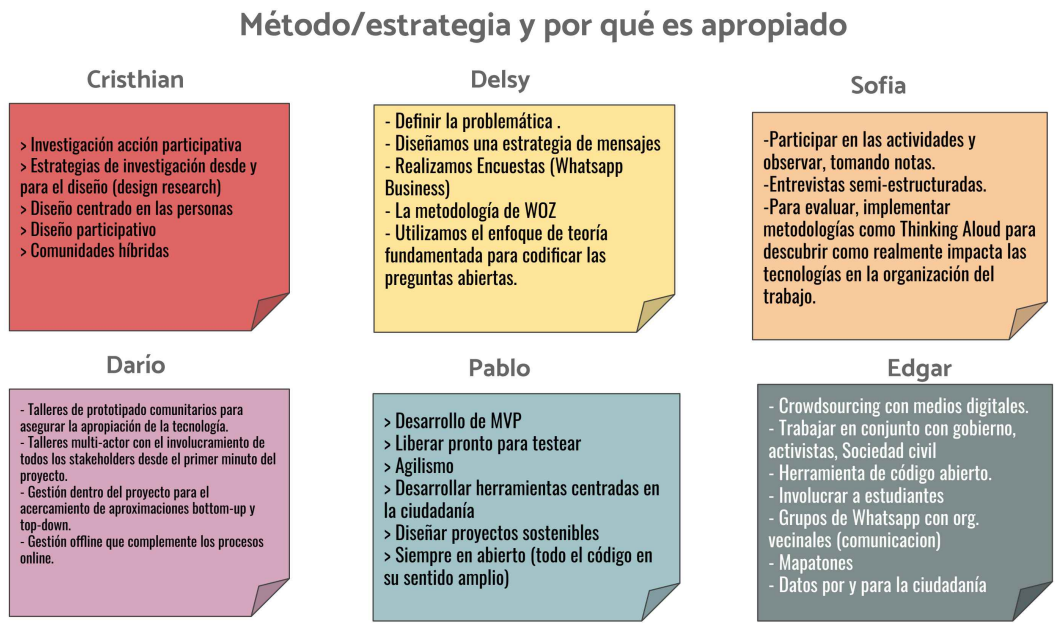

Figure 2.37: Outcome of the third session by Group 2d Spanish.

# 3

# Position papers

## 3.1. #CivicTech For And By Citizens: A Review And A Meta-Evaluation

**Weiyu Zhang**
National University of
Singapore (NUS)
Singapore
*weiyu.zhang@nus.edu.sg*

**Gionnieve Lim**
Singapore University of Technology
and Design (SUTD)
Singapore
*gionnievelim@gmail.com*

**Ziyi Wang**
National University of
Singapore (NUS)
Singapore
*wziyi27@u.nus.edu*

**Simon T. Perrault**
Singapore University of Technology
and Design (SUTD)
Singapore
*perrault.simon@gmail.com*

This paper takes a first step to systematically review and collectively evaluate #CivicTech works done in the computer science discipline, especially the vibrant community of CSCW. Based on 50 full papers published in CSCW, we ran a quantitative content analysis of the works. We found that civic tech is a growing young field with interests from all over the world, across academic, governmental, and commercial sections. While we are progressing well towards the goal of "for the citizens", "by the citizens" remains largely absent. We call for a more balanced approach to civic tech, both in developing cutting edge technologies and in adapting laymen and popular technologies for civic purposes.

### 3.1.1. Introduction

The field of #CivicTech is truly interdisciplinary. One such intersection is found in the overlapping interests in developing technologies for civic purposes between social scientists and computer scientists. "Civic technology" and "digital civics" are terms used to refer to technological innovations aimed "for the public good" [1].

A recent report[2] emphasized "promoting civic outcomes" as an important criterion to identifying civic tech projects. The report listed a wide range of technologies, including e-government and community participation, as forming the spectrum [24].

This paper takes a first step to systematically review and collectively evaluate works done in the computer science discipline, especially the vibrant community of CSCW. Based on 50 full papers published in CSCW, we ran a quantitative content analysis of the works to understand four key questions critical to the development of the #CivicTech field:

- Which civic groups are served, and which civic topics are focused on?

- What technologies are used or designed?

---

[1] https://blogs.microsoft.com/on-the-issues/2016/04/27/towards-taxonomy-civic-technology/
[2] https://knightfoundation.org/wp-content/uploads/2019/06/knight-civic-tech.pdf

| | |
|---|---|
| Civic tech | Saldivar et al [192] defined civic technology as "technology (mainly information technology) that facilitates democratic governance among citizens". This definition is intended to be broad enough to include both government-centric and citizen-centric approaches. Our definition is inclined to the latter approach, which stresses both for and by the citizens. Civic tech needs to not only serve the citizens but also engage them in its design and implementation (e.g., participatory design). However, we are fully aware of the difficulty with such a definition, which may find citizen participation in design easy to say but hard to do. |

Table 3.1: Definitions

- Who are designing or supporting the design of civic tech?

- What are the limitations and challenges for the field?

### 3.1.2. Related Work

Two systematic reviews [192, 205] of civic tech platforms provide us an initial description of the field. Researchers found that:

- Many civic tech initiatives exist. Skarzauskiene & Maciulene [205] found 614 such platforms and Saldivar et al. [192] found 1,246 such papers.

- Government-oriented civic tech is a leading genre of civic tech. Governments used platforms to engage citizens to improve their services and functions. Examples include participatory budgeting, urban planning, policy-making, or public sector innovation.

- Citizen-oriented civic tech is also growing, with focuses on a large range of issues from mundane ones such as improving quality of life, to grand ones such as building a stronger democracy with transparency and accountability in the government.

- Both studies point out a lack of collaboration between stakeholders; and a limited engagement of citizens in academic- and practitioner-led projects.

These two reviews provide a bird's eye overview of the field, but there are some notable missing pieces. First, they lack a meta-evaluation of the field: if the field is growing well, what had been done right? Are there any significant problems with the field? Second, the nudge-type of design is not emphasized in their reviews. Nudge-type design rely on existing platforms and make influence through playing user psychology. Although not so innovative, these nudge-type civic techs have long been studied by scholars who pay close attention to how widely-used information technology makes an impact on citizens [149].

Our paper tries to address the above-mentioned limitations of existing works by emphasizing both whole-sale and piece-meal solutions, and by carefully evaluating the effectiveness and sustainability of civic tech projects.

### 3.1.3. Method

A keyword search on papers published by CSCW up to March 2020 was conducted using the selection criteria: research article, full text, and the keyword "civic" in the abstract. We found 63 papers, with 50 full papers involving civic tech. We then developed a codebook of 22 items and used 10 papers to assess inter-coder reliability between two coders (Scott's Pi: 0.89). Disagreements were resolved through discussion. Five results each from the first and second coders were included with the remaining papers for the final coded dataset.

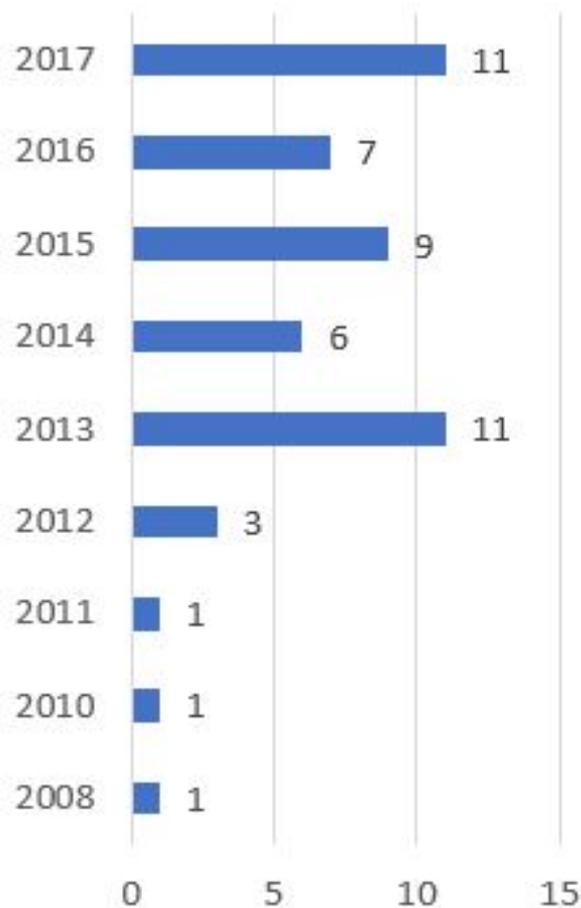

Figure 3.1: Number of publications per year (2008-2017)

### 3.1.4. Results

We found that civic tech papers started to appear in 2008 and first peaked in 2013 (see Figure 1). Since then, the number of full papers published were steady with about 10 each year. However, this trend stopped since 2018, which may be a worrying signal. We note, however, that 8 extended abstracts were published in 2018 and 2019, which may indicate that these studies are still works-in-progress.

Location wise, 38% of the research were conducted in the US, followed by virtual locales without specific identification of a physical locale, and then European countries. There was one paper each from China, Taiwan, South Korea, and Mexico. Despite being mainly developed in the geographical West, a vast majority (92%) either claims that their contribution has no national boundary or does not clearly identify the physical limits.

Civic groups wise, 32% studies targeted general citizens and 26% targeted online citizens. 14% aimed for helping community/organization members and 12% targeted volunteers. Minority groups, e.g. elderly, homeless people, and Native Americans also appeared in the target user groups.

Civic topics wise, 60% of the papers focused on citizen engagement, 18% social capital, 12% political participation, and 10% others. A closer look at the topics shows a highly diverse range of issues being addressed. Those ranked at the top of the long list include urban planning, citizen science, neighborhood issues, disasters, and minority groups.

We found 60% of technologies studied are existing and only 20% are tailor-made. Among those, which are supposed to be design heavy, most of them do not have a clear design approach (84%) nor an empirically based evaluation of the effectiveness of the technology designed (80%). 10% of studies combined both existing technologies and researchers' own inventions. When we look closer to the specific technologies, popular social media platforms such as Twitter (32%), Facebook and websites (24% each), and emails (16%) were the mainstream. It is to our pleasant surprise that most civic techs (72%) are still in use, probably because many such technologies are existing platforms or applications. But there are still some civic techs no longer in use, which suggests the challenge of sustainability.

It is expectable that as we focus on academic research, 98% of the papers involve academics as the main designers of the studies. The pattern is reversed if we look at the designers of the technologies used in the studies: companies such as Facebook and Twitter (62%) became the dominant group and academics only designed 24% of the technologies. Half of the studies (50%) were supported by the governments, followed by foundations and organizations. Companies only provided support to a mere 4% of the studies we analyzed. Research partnership wise, most studies were solely run by academics (68%). If partnership did exist, they were most likely to be schools (16%), followed by governments and organizations (8% each).

### 3.1.5. Discussion

Although the field looks generally healthy, our empirical data suggest some limitations and challenges we need to keep in mind. At least two areas of improvement exist along the line of "for the citizens". Firstly, citizens from non-democratic or non-Western contexts are under-served by civic tech. These contexts differ in aspects like tech development capacity, political systems, and cultures. We should not assume the universality of technological design in these contexts, especially for civic purposes. Secondly, most civic tech designs still target general citizens, which is understandable considering the youth of the field. However, we need to expand our target users to help those who are in high need of help, such as the minorities. Civic tech is not only for the majority of citizens but also for taking care of the minority, in order to enhance social justice and keep social integration.

What seems to be largely absent in current civic tech research is the goal of "by the citizens". Despite prominent design theories such as participatory design, most studies do not involve citizens in the design phase, but instead merely as naive users. Civic tech scholars need to build infrastructure to reach out to civil society partners.

### 3.1.6. Limitations

Our study is only a first step leading towards a larger project. Thus, many limitations exist. Firstly, our search is limited to one keyword "civic". Secondly, the database we use is only the ACM Digital Library. Thirdly, we note that there are many practitioners who design civic tech but do not always document their practices in the form of academic papers. Lastly, we authors are located in a Global South location. Although this rather marginal position has its advantages (e.g., urge us to be open to both the West and East), we recognize the limits of our perspectives.

### 3.1.7. Conclusion and Future Work

In conclusion, we found that civic tech is a growing young field with interests from all over the world. We believe that academics across disciplines but with shared interests need to join forces; both academics and practitioners around the world should come together to address these common civic challenges. We also call for a more balanced approach to civic tech, both developing cutting edge technologies (e.g., AI, VR/AR) and adapting laymen and popular technologies (e.g., social media). The latter is no less challenging than the former and involving citizens in the design of both technologies and processes [227] would be key.

## 3.2. Civic Fellowships: A Framework for Civic Technology Innovation in Pakistan

**Anam Zakaria**
Code for Pakistan
Lahore, Punjab, Pakistan
*anam@codeforpakistan.org*

**Ebtihaj**
Code for Pakistan
Islamabad, ICT, Pakistan
*ebtihaj@codeforpakistan.org*

**M.A Ibraheem**
Code for Pakistan
Peshawar, KP, Pakistan
*ibraheem@codeforpakistan.org*

In the last decade, civic technology has empowered citizens by allowing them to use their voices to change the way governments operate. Open government, participatory governance, and easier access to government services are revolutionizing governments across the globe. However, in large parts of the world, introducing/bringing digitization, transparency, and openness is often faced with resistance, bureaucratic hurdles, and systemic challenges that are often difficult to navigate.

In the last five years, Code for Pakistan [65], a non-profit, has been on a mission to build a non-partisan civic innovation ecosystem in Pakistan. Its flagship program, the KP Government Innovation Fellowship Program [68], a three-way collaboration between the KP Government, Code for Pakistan, and The World Bank Group, has empowered 80 Fellows who have worked with 21 government departments, developing 28 digital solutions with five more under development this year. From reporting systems to open data portals to automated service delivery apps, these applications have enhanced government efficiency and transparency while improving the lives of thousands of citizens who interact with the government every day.

In this paper, we will discuss how being a civic tech organization, we were able to bring meaningful difference in the civic and public sector, sharing both the achievements and our learnings along the way.

### 3.2.1. Introduction

The use of technology to improve governance has been widely used across the world. In doing so, many developed countries have focused on citizen participation in designing laws, policies, and services that impact them the most. Enabling the citizens to take part in the decision-making processes of the state has turned out to be revolutionary for strengthening democracies. Online communities form the basis for civic platforms, which are incubators for new ideas through peer-to-peer networks, stakeholder mobilization, collaboration and partnership engagement.

In the developed part of the world, civic-technology solutions are ubiquitous. Countries like Taiwan, the US, Canada, UK, France, and Singapore have already championed civic technology by transforming how the public interacts with the government. Global movements like Code for All [43] have also been actively involved in promoting civic technology across the world. Code for All is an international network of organizations that believe that digital technology, when used correctly, can both improve governance and open new channels for citizens to more meaningfully engage in the public sphere and have a positive impact on their communities.

Lately, this co-creation concept between the public sector and the citizens through the use of technology has also gained much traction in developing countries like Pakistan. Civic sector organizations like Code for Pakistan have been working to create a civic innovation ecosystem to improve the quality of life in Pakistan. Code for Pakistan is also an active member of the global Code for All community.

Code for Pakistan's goal is to bring together civic-minded software developers and citizens to give back to Pakistan by innovating in public services using technology, primarily by creating open-source solutions to address citizens' needs. Step by step, Code for Pakistan aims to transform the relationship that the government has with its citizens by leveraging technology to bridge the gap between government interfaces for the public and its users.

The KP Government Innovation Fellowship Program is the flagship initiative of Code for Pakistan supported by the Khyber Pakhtunkhwa Information Technology Board and the World Bank, where a team of talented researchers, designers, community organizers, and developers collaborate to build apps, inspire citizen engagement, improve government, and show how to innovate in public services.

This program revolves around three key features:

- Citizens and government working together, hand-in-hand, to solve problems

- Adopting user-centric, lean, and agile development methodologies

- Increasing civic engagement by creating innovative solutions in public services

Our goal with this paper is to provide researchers, designers, and practitioners, a starting point to understand the state of civic technology in Pakistan, the available opportunities, and the challenges in scaling civic tech initiatives in Pakistan.

### 3.2.2. Learnings and Outcomes

Launched in 2014, the KP Government Innovation Fellowship Program has completed five cycles, with the sixth cycle currently in progress. Working closely with government departments, we have gained tremendous insights in the way civic technology solutions can be developed, deployed, and sustained in developing countries where traditionally an environment of hostility and resistance exits towards digitization and automation initiatives. The initial 2 years of the program were challenging, where as a nascent civic tech organization working towards building a sustainable innovation ecosystem, we had to build trust with government partners on one hand and navigate infrastructural challenges on the other. Along the way there were a lot of lessons learnt and through our proactive interventions, we were able to scale the program effectively.

Today through the digital solutions developed under the Fellowship Program, over 3 million queries generated by the public have been handled, over 14,000 citizens are being facilitated each week, and over 90,000 Government hours have been saved thus far [64]. Significantly, we have also been able to bring a cultural shift in government departments, through advocacy and training; almost 160 government officials have already been trained in 21 government departments.

Looking back at these five years from the vantage point of where the program stands today, there are some great learnings which have enabled us to get this far:

- **Building trust with government departments**: When we initially started the program the IT infrastructure in the KP province was almost non-existent, departments were under-equipped both in terms of human resources and technical resources to sustain digital solutions. Secondly, most departments were hesitant to adopt civic tech solutions considering they would introduce more transparency within the department, resulting in government officials being held more accountable. We had to work closely with our program partners to convince the government departments to make structural changes and be more receptive to these civic tech solutions. We also made a concerted effort in building trust with our partnering government departments, through training and involving them more closely in the development process of civic tech solutions. A significant component of this work was convincing government departments that employing civic technology effectively would make their lives simpler, saving hours and creating easier systems and procedures;

- **Finding the right stakeholders**: In our inaugural Fellowship cycle back in 2014, the local utility company in the KP province faced losses due to consumers defaulting in bill payments and rampant theft of electricity. A Fellowship team developed a citizen reporting app [71] to report electricity theft hotspots in their surrounding areas. Despite being a simple civic-tech solution with a well-defined use case, user-adoption and government buy-in of the app became a challenge. Most of the power theft was happening in suburban areas where the literacy rate was very low, which meant that our target market neither understood English nor had access to smartphones to download the android app and report electricity theft. Secondly, the partnering government department could not convince the utility company to use the app, which relied on law enforcement agencies to act upon the perpetrators. Having too many stakeholders involved resulted in the app not being officially deployed. This was a crucial learning for the Fellowship team. We realized that prior to developing a solution, understanding the end-to-end problem and existing systems is essential as is ensuring that the human and systemic pieces of the user journey are in place before developing any application. In addition, we also revised our problem statement selection criteria to ensure we assess and understand department dynamics and relationships prior to partnering with them to avoid situations which require the approvals of multiple stakeholders for the adoption and maintenance of projects;

- **Ensuring maintenance and sustainability**: One of the most significant steps in getting the solutions adopted is the transition period where the solutions are handed over to the parent government department. While many government departments promise to hire resources, the lengthy hiring process, lower pay scales, and unavailability of quality resources are major roadblocks in sustaining these solutions. One of the projects we developed for the department of Minorities in Khyber Pakhtunkhwa [69] was supposed to help the minorities in the province to get government grants. Despite the platform being used to disburse over PKR 10m while the Fellows were maintaining it, after the handover, the platform was never updated by the department and was abandoned. To confront this challenge we maintain a volunteer community of Fellowship alumni who work with us in relaunching such platforms, where the departments fail to sustain the solution due to lack of resources or technical knowhow. Moreover, we also stress early on that departments dedicate a resource person(s) to maintain the application post-deployment;

- **An uptick in user adoption through cheap smartphones & 3G/4G Launch**: Prior to 2018, smartphone penetration in Pakistan was considerably low, 3G/4G mobile internet wasn't widely accessible and consumer adoption was low. In 2018 the era of sub $100 smartphones revolutionized mobile internet in Pakistan. The number of 3G/4G subscribers crossed 61.61 million active users [181] by December of 2018. This had a direct impact on the civic tech solutions we were developing. A large consumer base came online and started using these digital solutions. An evident example of this was with the Raabta app[3], a consumer app developed for the traffic police department KP, the app gained tremendous popularity among the citizens of KP. With this learning we took a mobile first approach, where for every civic tech solution that creating a mobile app made sense for, we launched that first in order to gauge the usage trends among users;

- **Encouraging female participation**: In Pakistan, female representation in STEM fields is significantly lower than their male counterparts. This is also evident in the job market where fewer women pursue a career in IT and Engineering. In the Fellowship Program, we wanted to encourage female participation and bring more diversity to teams working on innovative civic tech solutions. We did this by introducing a 25% quota system for female applicants. This helped us include more female Fellows who in turn got the opportunity to work on innovative civic tech solutions. We are also planning to launch a female-centric Fellowship where the program will be focused primarily on onboarding female Fellows, with a focus towards enhancing their skills and introducing them to entrepreneurship;

- **Adapting to challenges**: Just before the launch of the sixth cycle of the program, Pakistan went into lockdown due to the global pandemic. The Fellowship team decided to execute the program remotely and facilitate front line government departments in relief work towards the containment of the pandemic [73]. Redesigning the entire program for remote execution and convincing government departments to agree to remote work format was a challenging task. This is especially so because many government departments continue to rely on manual in-person work. The Fellowship team convinced the departments by showing them the benefits of online work and ensuring that the deliverables and milestones are achieved on time. We are currently using phone calls and Google Meet for weekly meetings between Fellows and government officials with in-person meetings reserved for high priority discussions only. For these in-person meetings, instead of all team members being present, only a few Fellows go to government offices alongside Code for Pakistan's Government Liaison. This is so that we can avoid overcrowding, have important discussions in small groups and ensure that SOPs are followed. Now three months into the sixth cycle, the projects under development have reached the MVP stage and are nearing completion;

- **Spreading the word/marketing**: Pakistan is ranked as the 7th worst country in terms of access to sanitation with over 41 million Pakistanis without access to toilets [134]. In 2019, during the 5th cycle of the Fellowship Program, a Fellowship team created an app focused on promoting health and hygiene. The team developed the Public Toilet Finder[4], a digital solution to facilitate the public in finding, using, and promoting public toilets. The partnering department was the Water and Sanitation Services Department. Despite the buy-in from the department, the app could not bring about the transformational impact it initially intended to. The app has a low number of users and downloads, partly due to slow marketing campaigns from the department's end. The Fellowship team designed a comprehensive marketing plan, which included offline and online marketing, but lack of departmental resources and technical know-how around marketing resulted in poor outreach. To overcome a similar problem in the future, KPITB has started teaching marketing courses to graduates who can support government departments in marketing and outreach and a similar module has also been added to the Fellowship curriculum;

---

[3]https://play.google.com/store/apps/details?id=com.kptrafficpolice.trafficapp&hl=en

- **Government buy-in**: A digital solution can never work without the buy-in from the parent government department. One of the solutions our Fellows developed aimed to connect the public to doctors based on the reviews left by their previous patients [67]. Once developed and ready to be deployed, the parent department refused to share the doctors' public data despite committing to provide the data initially. As a result, now a written MoU is signed with the parent department before initiating the project and we have multiple meetings with departments to discuss expectations, deliverables and concerns early on so that we can work together to navigate challenges, build trust and ensure successful deployment. Also, through our Fellowship Government Liaison Officer we try to engage top tier government officials from secretary level to IT officials in each department, getting their buy-in along the way by sharing regular updates and the potential impact of the app so that the app is sustained after launch. Furthermore, we also work closely with government staff who are responsible for maintaining the application after deployment. We ensure that they test and use the app throughout the application cycle, receive any training required to maintain the app, and are comfortable with the solution by the time it is handed over to the department. This increases the chances of adoption and sustainability of the app post-deployment;

- **Extensive user research and continuous testing**: One of the biggest requirements to ensure successful adoption of the developed solutions is to perform extensive user research. One of the projects [70] that we developed for the Excise and Taxation department enabled the department officials to keep track of the seized and confiscated non registered and tempered vehicles. This system had multiple interfaces across multiple departments and required all the stakeholders' input to process a request. Given the complex workflows, the department and the Fellows were not able to interview all the stakeholders resulting in the collapse of the workflow post-deployment. This reinforced the importance of user research. User research is now a significant component of the Fellowship Program, and Fellows conduct surveys with potential end users of the system to ascertain the demand for a specific app or service and use their findings for adding desired features;

- **Approvals and policies**: Changing or introducing new policies to adopt a solution or getting approvals from the ministries' higher-ups can take a long time. Despite knowing that the state's existing laws did not allow crowdfunding [66], one of the departments requested a crowdfunding platform to connect startups with potential investors. Post-development, the department could not bring a change to the policy; hence, the platform was never utilized. This was a good reminder that civic innovation isn't just limited to solving the technology problem. Navigating legislative hurdles and campaigning for innovation in local policies is also crucial. Through the Fellowship Program's platform as well as Code for Pakistan's other initiatives and partnerships, we raised our voice on different platforms to convince the government to introduce policies that are business friendly where innovators can leverage the internet to scale their businesses and reach wider audiences.

---

[4] https://play.google.com/store/apps/details?id=com.watsoncell.publictoiletfinder&hl=en

### 3.2.3. Conclusion

How can we strengthen governments working in fragile political environments? How can cultural shifts be evoked and digitization be implemented in government departments when change is perceived as threatening and processes are mostly manual? How might we work with governments to develop policies that spark innovation? For the past five years, Code for Pakistan has been actively involved in setting up an inclusive civic tech ecosystem in Pakistan. Using technology as an enabler, Code for Pakistan is focused on building a stronger relationship between citizens and the government. Our experience of working with the KP government has enabled us to think along the lines of replicating the model across different provinces of Pakistan. We are also working on establishing a digital services unit in partnership with the KP government that will take up more ambitious digitization initiatives in the province. With our work on the Open Data Portal [72], we were able to help draft the province's first Open Data Policy that will be passed through the provincial assembly soon. Civic innovation does not happen in silos; to innovate in public services and to improve government technologies, establishing trust between key stakeholders is essential. Civic engagement cannot be solved merely through technology but by concerted efforts and lobbying at the policy level to help government departments build their infrastructure, increase their capacity, and be more responsive to openness and transparency. With the Fellowship Program, we have accumulated a great deal of knowledge on navigating these bureaucratic hurdles and innovating despite the structural and policy level issues that linger. Only through innovation and examples of digital civics practice can we influence a cultural shift in policy and framework.

### 3.2.4. Acknowledgments

On behalf of Code for Pakistan, we're thankful to the Khyber Pakhtunkhwa Information Technology Board and the World Bank for the incredible support and partnership, and to all our partnering government departments for the engagement and cooperation throughout these years. We would also like to extend our gratitude to the Fellows who have worked tirelessly to develop and deploy these digital solutions to facilitate the public.

# 3.3. Collaborative Commons: Civic Tech in Toronto

**Curtis McCord**
University of Toronto
Toronto, ON, Canada
curtis.mccord@mail.utoronto.ca

## 3.3.1. Introduction

In July 2020, Civic Tech Toronto (CTTO) celebrated its 250[th] weekly hacknight and its 5[th] year of existence. Since 2015, CTTO has carved out a space for residents and visitors to come together to learn and collaborate on shared matters of concern. In this time, CTTO has consistently offered a space for projects, for speakers, and for community to emerge. My research examines the way that 'civic' or 'public interest' technology practise occurs in Toronto, a large metropolitan city in Canada. CTTO has organized weekly hacknights where participants to work on civic technology projects, and brings civilians and public servants into direct and sometimes sustained contact. I offer to the workshop reflections and preliminary findings based on a year of action research as an organizer in CTTO, including interviews with participants, workshops, and participant observation.

The work of building democratic institutions can be adequately described as (but not reduced to) a process of sociotechnical systems design. Government institutions seeking to improve their policy-making processes now frequently turn to computer-supported practises in order to include citizens in a more participatory way. The technical and situational expertise of citizens suggests they can be extremely valuable collaborators in policy-making [63, 161], both in terms of their epistemic contributions, and to the democratic legitimacy that participatory governance can help to build. However, it is not enough to design a system to receive public contribution; the way that those interactions are structured will determine what interactions can take place [75], and the processes by which these contributions become decisions must be accepted as legitimate. Relationships of power and expertise can still inhere in 'participatory' or 'co-designed' processes, especially when corporate and governmental entities seek to maintain control of the agenda, which can be described as "pseudo-participation" [170]. For example, Sidewalk Labs and Waterfront Toronto ran a robust consultation process that spanned over years, but without really ceding any control over decision-making [144].

Civic technology is not merely about using (or designing) digital tools to interact with institutions (or to circumvent them), it is also about the production, discussion and development of technologies *as a form of civic engagement*. In other words, civic technology describes processes just as much as it does products, and for some, the prioritization of process over product is a defining feature [194]. Focusing on civic technology groups, such as CTTO, as well as civic technologies, can show us how community and cooperation are essential to realizing the democratic potentials of civic technology.

## 3.3.2. Methodology and Workshop Themes

CTTO hacknights are a construction of space and time where civilians offer their views or expertise on matters of public interest, and actively work to address these matters through voluntary commitment to ongoing projects and discussions. Reproducing this space is key to creating and maintaining relationships between civilians and the state [143]. In this respect, studying civic technology practise requires approaches to observe and analyze gatherings that constitute and maintain the group, and to understand the technologies that enable their cooperation. However, focusing on a single site for this research poses risks, especially if our singular case excludes other examples of contemporary practise, omits the way that the CTTO community relates to other similar communities, or where critical reflection is lacking.

To structure this engagement, I have adopted the metamethodology of Critical Systems Practise [101] (CSP), which has four nonlinear phases: Creativity, Choice, Implementation and Reflection (see Sidebar 13.2). I have conducted the research as two phases of CSP, the first intended to prime a theoretical, social, and geographic understanding of the state, or situation, of civic technology in Toronto, and a second phase focused on intervention and participation within CTTO. Below, I provide a bit more detail on the research process, relating it specifically to the workshop themes.

| Creativity | Researcher builds understanding of the situation alongside participants using modelling techniques, interviews, etc |
|---|---|
| Choice | Participants choose a framing and approach to an aspect of the situation (potentially including specific techniques from systems thinking) |
| Implementation | Participants agree on a course of action, and undertake to enact it |
| Reflection | Researchers reflect on the other phases, both throughout involvement, and at the conclusion of the action research |

Table 3.2: Phases of CSP}\cite{jackson2003systems}

### 3.3.3. Social Worlds and Situational Analysis – Locality and Infrastructure

CTTO is not the only group in Toronto that centres their discourse on 'technology', or that makes technology development or deployment core to their practise. Furthermore, in some cases CTTO's goal is less to intervene in a specific situation *per se*, but to create the conditions by which participants can address situations, often, but not always, with computational or design interventions. Using 'situational analysis' (SA) methods from Science and Technology Studies (STS) [41], I explore how CTTO is located within a larger social arena of contemporary and historical groups in Toronto using technology and design practises to address matters of shared concern. By describing the relationships amongst these groups, as well as with public institutions, I believe that researchers and practitioners can explore if and how public service organizations can build empowering and democratic ways of interaction.

SA has proven quite useful for explaining some of the particulars of my localized case. In particular, the concept of a *social world*, composed of participants operating and working together to address matters of shared concern, supported by "shared discursive spaces" [41, 211, 212] has allowed me to explore how CTTO bridges gaps between professionals, public servants, laypeople, and activists. CTTO has, at least at times, created a uniquely heterogeneous space for projects to form, to grow, and even to outgrow CTTO entirely. Part of my work has been to reveal a history of connections and collaborations. In particular, this has meant examining how the community, in terms of hacknights and projects, interact with government. In some cases, this has been based on formalized partnerships, the in other cases, has been based on the informal overlap between CTTO members and the government. There are many interesting examples from CTTO to compare and critique.

### 3.3.4. Action Research – Methods and Strategies

Cooperative work is central to CTTO, required to intervene in an identified situation through a process of 'making' where information systems or other technology products are created to serve some purpose, or to explore a specific topic. I also adopt and methodological framing from Soft Systems Methodology [33, 34] (SSM) which connects action research to the elicitation of emergent research themes, and is well suited to CTTO as a group. This environment is highly pluralist, consensual, based around personal choice, discussion amongst peers, and the need to continually maintain work on projects as well as to maintain the group as a whole. In that spirit, my research has situated me in CTTO as a participant, where my research agenda advances in light of conversations with participants, and my understanding of the groups organization is in part constituted by the experience of doing the work of organizing events, booking speakers, running projects, conducting outreach, etc.

In this capacity, I argue that CTTO can be productively considered as a sort of commons, where the reproduction of the community assures that the goal of producing more social relations is assured [89]. Distinct from public and private ways of governance, commons are governed by the communities that use them [20, 165]. In this case, CTTO produces community as a resources which can be contributed to, and drawn on, by any participant. Volunteers organizers work to reproduce the hacknights, and to do so in such a way that is consensual (because voluntary commitment is required), and transmissible, as organizing roles are in continual churn, with the average tenure between 6 months and a year. In particular, workshop participants may be interested in hearing about CTTO's hard pivot from in person to virtual hacknights in response to the COVID-19 pandemic, which has forced the group to systematically reconsider the hacknights in terms of purpose, accessibility, and community support. They may also be interested in hearing how online assets support the in-person hacknights, in some cases allowing them to be run with minimal effort, while at other times becoming a major support for transformation and community organizing.

### 3.3.5. Closing

*The Relational State* [44] neatly underscores a key problematic of democratic governance that the workshop draws attention to. Are states actually the kind of entities that can relate directly with citizens, or should they focus on supporting local improvements in diverse and pluralistic contexts through common institutional standards that allow citizens to pursue much of this work themselves? The thematic tension here is one of centralization and decentralization: should civic tech groups orbit around the state, addressing themselves to a central bureaucracy, or do they manage to achieve their goals and projects with a more autonomous relationship with regimes of policy-making and allocation of resources?

Either way, the need to develop and document practises, relationships, and activities for engagement, as well as the technological objects to support them, is key to realizing the democratic potential of civic tech. The character of these relationships matters, especially when we consider how these practises might constitute enduring features within our societies. I look forward to discussing how trust can be built between civic commons and public institutions, particularly insofar as trusting commoners means giving them power and responsibility to address civic issues. In general, I believe I would be a suitable participant in the workshops, and really look forward to meeting more scholars engaged in civic tech research, and to become acquainted with the CSCW community.

## 3.4. Designing Open Government Data Portals for the Universal Access and Effective Use of Ordinary Citizens

**Briane Paul V. Samson**
De La Salle University
Manila, Philippines
brianepauls@acm.org

As citizens around the world clamor for more transparency and accountability from their national governments, we can see a steady increase in published open data and freedom of information laws being passed. The Philippines is spearheading this movement by encouraging more agencies to publish open government data and more recently, responding to Freedom of Information (FOI) requests. However, it is still far from achieving universal participation because most of their implementations focus on use cases for data workers and scientists. This leaves the data they host not maximized to their potential and with no guarantee if available data can address people's information needs. Here, I argue that there is a need to improve the usability and inclusiveness of open government and freedom of information data portals that will allow for more universal access, without explicit need for data literacy and experience.

### 3.4.1. Introduction

The term "*open government*" was first used in the 1950s and was generally used to refer to making previously confidential government information public with the goal of having a government be more politically transparent [228]. Under this umbrella movement, governments have employed either one of two popular options to promote transparency and accountability: voluntarily publish them as open data and or provide citizens with a "right-to-know" process. Most of these open government initiatives started with the United States right after the World War [228]. But in 2011, countries from all over the world gathered together and formed the Open Government Partnership (OGP) with the goal of promoting "*accountable, responsive and inclusive governance*" [163]. Being a founding member of the OGP, the Philippines launched Open Data Philippines (ODPH) in 2014 as one of its key commitments [19]. Unlike the open data movement, lobbyists of freedom of information seeks for governments to establish constitutional guarantees for a "right-to-know" process. In the absence of publicly available government data, a passed FOI law ensures that data requester that they will get a response within a specified number of days. However, it is still the discretion of the government agency whether they will provide the data or not.

While we are still far from having truly open governments, we are already witnessing small successes of open government and FOI data being used for policy, research and civic innovations. For example, projects like Citygram[5] in the USA allows citizens to get updated with recent reports in their area while GoodGuide[6] in the EU lets users check how healthy grocery items are. Several technology companies like Sakay.ph[7] that provides commute routes in the Philippines and Rentlogic[8] that grades apartment buildings in New York got started with open government data as well. Aside from these, open government datasets are mainly accessed by data workers who use them for research and data analysis. In countries with more advanced open data catalogs, citizens are able to do statistical analyses [38] and build visual analytic tools to explore the given open data [117, 183]. However, Choi and Tausczik (2017) found that most data work are limited to shallow analyses such as descriptive and exploratory analysis even though there is potential for these data sets to be used for inferential or even predictive analysis. Governments have so far focused on the needs of an expert subset of the population, with access to advanced tools and grand ideas that can definitely drive economic development and promote societal changes. However, these datasets also contain valuable information for ordinary citizens that they cannot readily access and use because of a need for high technical expertise. That said, open government data have yet to achieve universal participation, especially among sub-national governments and ordinary citizens where they can have the most impact [28]. Thus, it is critical for the success of these initiatives that they expand their reach and explore other ways that citizens can make use of these rich data sets without any barrier to entry and high learning curve.

---

[5] https://www.citygram.org/
[6] https://www.goodguide.com
[7] https://sakay.ph/
[8] https://www.rentlogic.com/

Despite the extensive usage of civil society organizations and the private sector, open government data still suffers from low utilization which suggests a disconnect between the published data and the information needs of the citizens [78]. With some data publishers merely "*dumping*"[9] their data, they seem to be unaware of the potential and relevance of their data to those can potentially benefit from them. On the other hand, citizens are unaware of the published open data [21]. The Freedom of Information initiative could be a way for data publishers to understand demand through its requests for data access but the current disconnect with the open data initiative does not allow publishers from either sources to get hold of this information. Currently, datasets are spread across multiple sources with inconsistent to non-existent links between them. For example, there is a National Crime Statistics dataset for 2013 in the Open Data portal[10] and there are several successful requests for the same data in the FOI portal[11]. This makes it relatively challenging for users seeking the right answer to fulfill their information needs.

### 3.4.2. Philippine Open Data Systems

As one of the pioneering members of the Open Government Partnership in 2011, the Philippines has taken steps to disclose data to the public through the Full Disclosure Policy [29]. Since then, the Philippines has launched various portals containing OGD [29, 39, 223]. In this study, we focus on two specific portals: Open Data Philippines (ODPH) and the Electronic Freedom of Information (eFOI). These portals cover the two ways citizens can gain access to OGD: government-led publication of data and citizen-led request of information. ODPH[12] hosts data provided by different agencies and allows users to search, preview, and download these without any restrictions. In the dataset preview, users may also create exploratory graphs with the fields available in the file. On the other hand, eFOI[13] serves as an online tracker of citizen requests for government information that has not yet been published. eFOI requires citizens to register and to provide a valid proof of identity before making a request. Requests must be filed to the correct agency otherwise it gets rejected. An email would be sent notifying the requester of the status after submission. Other citizens may also browse through existing requests. Each request shows a conversational thread of the FOI officer and the corresponding agency.

### 3.4.3. Initial Findings

We conducted a needfinding survey of 119 respondents to understand the information needs and awareness of Filipinos regarding the open government data portals provided by the Philippine government. We also investigated the information seeking behavior of 13 lay citizens and 8 data workers to compare the behavior of both groups. In addition, we also conducted usability tests and semi-structured interviews with these 21 participants to identify gaps in the design of the two portals. In terms of the information seeking behavior of citizens, regardless of data experience and level of personal involvement, participants from both studies still relied heavily on Google to find needed data or information as seen in previous studies [115]. Although some participants believed that some of the available data from the portals are useful and relevant, most of the available data were mostly incomplete and outdated. This could be mainly attributed to the pace of digital transformation of the Philippines and the capability of the government to adapt to new technology to better manage and publish data. To address this issue, strict policies on data management and awareness campaigns are needed. We believe that design improvements could also be made to the current open data portals as majority of our participants shared common design and interactivity concerns with both portals. Both our results and previous literature [108] suggest that existing search techniques from web searches do not apply to data search. Although some of the participants were able to adapt to the search engine, we believe that an **improved search engine** would enable faster and more efficient searches. As spatial and temporal keywords were common in search queries, we recommend that the search engine looks through the metadata of uploaded data. However, this would require proper documentation of published data which may only be controlled by the data provider. Our results also show that regardless of data experience, citizens prefer a straightforward answer when seeking information. To support this, we emphasize the need for **descriptive visualizations** to communicate the content of the data. The majority of our lay participants expressed interest in visual aids to help them understand what the values in the data mean. Although storytelling and visualizations have been determined effective in communicating data [61], further research is needed to understand how Filipinos perceive data visualizations. Citizens need to be able to know how to read graphs and visuals presented. As previous literature suggests, there are various factors that affect the attitude and perception of individuals about visualizations [173].

---

[9] https://www.stateofopendata.od4d.net/chapters/regions/seasia.html
[10] https://data.gov.ph/dataset/national-crime-statistics
[11] https://www.foi.gov.ph/requests?status=&search=crime+statistics
[12] https://data.gov.ph/
[13] https://www.foi.gov.ph//

# 3.5. Digital Interventions in the Global South: A Case Study from Rural India

**Eric Gordon**
Emerson College
Boston, MA, 02116 USA
*eric_gordon@emerson.edu*

**John Harlow**
Emerson College
Boston, MA, 02116 USA
*john_harlow@emerson.edu*

*Water, sanitation, and hygiene (WASH) factors kill hundreds of thousands of children every year. This paper reports the results of a school-based toilet use and handwashing intervention in Tamil Nadu, India that enrolled over 2,500 students aged 7—11. Over 1,000 of those students received a robust, participatorily-designed, play-based intervention with a mobile game developed in direct partnership with the producer of Chhota Bheem, India's most popular children's character. Results indicate that provision of soap increased handwashing even in the control group, but that the intervention arms only improved advocacy outcomes with statistical significance. We interpret this to mean that play-based interventions can contribute to public health advocacy in the Global South.*

## 3.5.1. Introduction

In 2016, water, sanitation, and hygiene (WASH) factors caused over 300,000 preventable diarrheal deaths of children under 5 [182]. WASH factors include adequate access to clean drinking water, household sanitation that safely disposes of excreta, and soap and water for handwashing. In 2020, the United Nations Children's Fund (UNICEF) estimated that 3 billion people lacked soap and water at home and 900 million children lacked soap and water at their school [219]. Handwashing with soap at critical times (e.g. after toilet use, before eating and cooking, etc. [4]) can prevent about one third of diarrheal diseases [90].

World leaders addressed WASH issues in Sustainable Development Goal 6.2: "adequate and equitable sanitation and hygiene for all and end open defecation" [157]. People practice open defecation for many reasons, e.g. cultural norms, perceptions of cleanliness, and access to clean infrastructure. In rural India, over 1/3 of households practiced open defecation in 2017 [226], and in 2014, only 15% of people in India regularly washed their hands with soap after defecation [175].

In 2017, our team was awarded a grant by the UBS Optimus Foundation to develop a play-based learning program and mobile game to encourage hand washing and toilet use: Hygiene with Chhota Bheem (`http://elab.emerson.edu/hygiene`). This intervention focused on children because shifting the norms and perceptions of adults is more difficult. Although children cannot make decisions for their households or communities, they can be persuasive advocates, leading to the research question of this paper: how might locally tailored, play-based mobile interventions in the global south improve knowledge, attitudes and behaviors about handwashing and toilet use?

## 3.5.2. Methods

Early participatory design sessions with children and teachers revealed potential benefits of connecting the game to popular culture. This led to Green Gold Animation, the makers of Chhota Bheem, with "over 40 million viewers" [55], agreeing to provide access to their animation assets (e.g. characters and backgrounds). The Engagement Lab @ Emerson College created the game with Green Gold's support. Project partners also included the Indian Red Cross Society, a WASH working group of NGOs in Tamil Nadu, India, and the Mary Anne Charity Trust (MACT).

MACT's deep ties in Tamil Nadu attracted over 100 children and a dozen teachers to game design workshops in regional schools. A weeklong charette with teachers, government employees and health advocates followed. Participants were provided room and board if needed, and compensated for their time. Our Boston-based team traveled to Chennai three times in Year 1 to iterate the intervention design. During that process, we recognized the limitations of a mobile only intervention, and pivoted to include analog elements.

The result was a four-week play-based curriculum. In each week, children watched a short animated video (or read from a picture book) that introduced them to Chhota Bheem and his pals battling and defeating an evil germ wizard who hypnotized all the villagers in Dholakpur to "poop in the field" and not wash their hands. With the help of the Clean Wizard, the heroes learned "spells" to defeat the Germ Wizard, i.e. songs and dances of the program's WASH learning goals. Children also completed challenges (e.g. teaching others the spells, handwashing with soap before eating) to earn stickers and help defeat the Germ Wizard.

The intervention was evaluated in a randomized control trial of 30 schools in Tamil Nadu. Twelve schools in the Bheem arm received the full intervention. Twelve schools in the Analog arm received a simplified version without the digital game or videos. Six schools in the Control arm with no intervention. The focus herein will be on the full intervention Bheem arm.

The Institute of Public Health in Poonamallee hosted two lead teachers from each treatment school for "train the trainer" curriculum training. The US-based team helped facilitate, but the training was administered entirely in Tamil by the project field manager. Bheem arm teachers reviewed facilitation guides, videos of the lessons, Chhota Bheem story videos, and the digital game. They were then presented with certificates and the intervention materials: a bag of liquid soap, extra copies of the storybooks, facilitation guides for other teachers, and four Lava A44 mobile phones running Android operating system 7.0. The devices were kept in the school office, and signed out by participating class-rooms weekly. Control group teachers were also trained on handwashing protocols in their schools.

Surveys were administered to students prior to the intervention (baseline), following the intervention (post), and six weeks after the intervention (follow-up). The survey assessed WASH knowledge, attitudes, and behaviors using simple language (and pictures when possible) to make it as easy as possible for children to respond. Trained data collectors administered the surveys in Tamil, reading each question and its response options aloud. Students sat in groups and circled answers on sheets for entry into a tablet on site.

Table 3.3: The Hygiene with Chhota Bheem intervention addressed the psychosocial determinants of behavior described in the Integrated Behavioral Model for WASH [57].

| |
|---|
| Why toilet use is important |
| How to properly use a squatting toilet |
| Why handwashing with soap is important |
| When to wash hands with soap to prevent germs |
| How to wash hands with soap |
| Empowering children to talk to friends and family about toilets and washing hands with soap |

### 3.5.3. Results

The mobile game was downloaded over 23,500 times, with over 90% of those in India. During the study, the game was played 11,119 times in the intervention district, for 12,564 hours across 32 unique user IDs and 128 unique session IDs. Average user rating was 4.6/5 stars, with 88% of players rating it 5 stars.

These results are based primarily on the survey of n=2,614 students who completed both the pre- and post- intervention survey: Bheem (n=1,051), Analog (n=1,106), and Control (n=457). Analysis controlled for baseline characteristics that did not change between baseline and endline. Students who were absent at endline (n=247) were dropped from the analysis.

Positive attitudes toward handwashing with soap (0.57 SD) and toilet use (0.48 SD) increased with statistical significance across all study arms. Students also reported that they wash their hands with soap at the same times most days across all three study arms (0.09 SD, p<0.01). This is likely because soap was provided to all of the study schools. Students in the Bheem arm showed an average of 0.578 SD higher than the Control arm (p<0.01) for how to wash hands, and 0.596 SD higher (p<0.01) for how to use the toilet. If we combine the Bheem and Analog arms, there are a number of areas of statistical significance. We tested the assumption that the mobile game would reinforce learning outcomes, and that immersion in a familiar story would do the same. Those things alone did not seem to have much impact. However, play, which is common across the A and B arms, did.

### 3.5.4. Discussion

We began this project with the assumption that a mobile game would be an impactful intervention. Even though the game garnered a significant amount of attention among the study participants and beyond, it failed to create outcomes on its own. Rather, the play-based nature of both the Bheem and Analog arms suggest that grounding the intervention in play supplied children with confidence to advocate to their families and peers. Immersion in the play space empowered children to talk about the taboo topic of open defecation. These outcomes were especially apparent with young girls, for whom the topic can carry significant shame. While traditional measurements of knowledge, attitudes and behavior did not produce significant results, the correlation between play and advocacy is powerful.

### 3.5.5. Acknowledgements

We gratefully acknowledge the grant from UBS Optimus Foundation, Mary Beth Dawson, Wade Kimbrough, Jahnvi Singh, Divya Dhar, David Levine, all MACT staff, and the teachers, government and health professionals, and children that helped with the design.

# 3.6. Diseñando Asistentes Virtuales Inteligentes para el Gobierno Federal

**Saúl Abraham Esparza Rivera**
Civic Innovation Lab, UNAM

**Norma Elva Chávez**
Secretaría de Relaciones Exteriores, México

**Saiph Savage**
Civic Innovation Lab, UNAM

Inteligencia artificial o I.A. por sus siglas, es un constructo que se ha desarrollado desde hace décadas, con el avance de la tecnología y el desarrollo de componentes de hardware y software, las diferentes sociedades del planeta han estado involucradas en el desarrollo y crecimiento de esta tecnología. Hoy en día podemos encontrar Asistentes Virtuales o Bots encargados de llevar a cabo tareas con el fin de facilitar procesos y servicios que hasta hace unos años requerían de la intervención directa de un ser humano, lo que trae consigo muchas de las complicaciones de ejecutar tareas plenamente con mano de obra humana. Si bien se sigue tomando en cuenta al ser humano en la realización de ciertos servicios su participación se ha visto reducida gracias a los avances en el campo de la IA.

Alrededor del mundo, los gobiernos y empresas de diferentes países se preocupan por apoyar a la ciudadanía para que pueda tener acceso a información y servicios de manera más eficiente y sencilla, gracias a esto es que el auge de los asistentes virtuales en años pasados ha tomado tanta fuerza e influencia ya que la automatización de servicios y tareas ha logrado que cada vez más las personas estén en contacto con las instituciones que así deseen. En el caso de México, la Secretaría Relaciones Exteriores (SRE) ha tomado la decisión de implementar un asistente virtual con el cual pueda mejorar las tareas de trámite, renovación y consulta de citas de pasaportes, con el objetivo de que la ciudadanía tenga a su alcance la información y las herramientas necesaria para llevar a cabo dichas tareas y no sólo esto, la carga de trabajo se verá reducida debido a que el usuario no tendrá que interactuar con un agente humano o ir presencialmente a la cancillería para para realizar sus consultas, lo podrá hacer desde la comodidad de su casa y de así desearlo podrá pedir asistencia a un agente humano para resolver sus dudas.

## 3.6.1. Introducción

No es nuevo decir que en el mundo imperan La tecnología, desde teléfonos inteligentes hasta computadoras personales todo mundo tiene acceso a un trocito del mundo tecnológico. Aunado a esto tenemos la expansión digital de la cobertura del internet alrededor del mundo y los esfuerzos de los gobiernos porque dicha cobertura sea lo más grande posible [37, 87]. Con esto en mente decidimos abordar la problemática de llegar a la máxima cantidad de usuarios dentro de la República Mexicana haciendo uso de las tecnologías antes mencionadas. Así pues, decidimos utilizar una plataforma con la cual pudiéramos construir un asistente virtual de acuerdo con las necesidades que tienen los ciudadanos Mexicanos y que tuviera la capacidad y el alcance que un proyecto como este requiere.

La plataforma escogida para realizar la construcción y diseño del asistente virtual fue Botpress, una plataforma enfocada a programadores, con y sin mucha experiencia, con la cual pudimos construir, experimentar y realizar las pruebas necesarias para qué asistente virtual cumpliera con los objetivos que la Secretaria Relaciones Exteriores propuso.

Además de los trámites que la Secretaría de Relaciones Exteriores tenía en mente también incluye secciones de información dentro de cada apartado de acuerdo con el trámite incluyendo un apartado exclusivo para preguntas frecuentes y otro más dedicado a la comunicación del usuario con un agente humano. En cuanto a la distribución y medios la difusión del Asistente Virtual la plataforma Botpress nos brinda con las herramientas necesarias para colocar nuestro Bot o Asistente Virtual en una serie de plataformas ya sea en redes sociales como Facebook o WhatsApp o bien ponerlo como un pequeño widget o acceso directo en nuestra página web. La conversación con el Asistente Virtual resulta ser bastante ágil debido a que los mensajes que éste le mando el usuario son sencillos de comprender además de brindar información y guías entre preguntas haciendo que el flujo de la conversación se mantenga en un mismo sentido desde el principio hasta el final de esta procurando que los errores sean mínimos y el objetivo del usuario al ponerse en contacto con el Asistente Virtual sea cumplido.

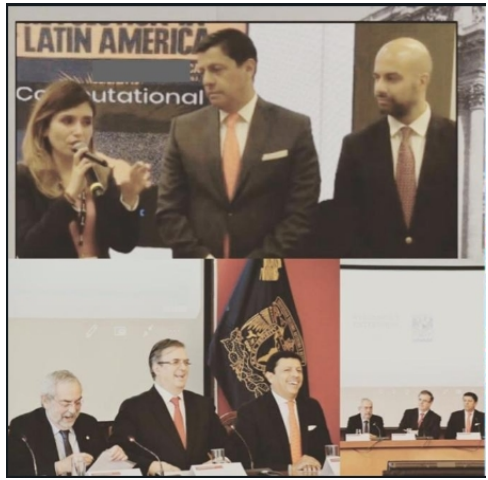

Figure 3.2: Establecimos una colaboración entre la Universidad Nacional Autónoma de México (UNAM) y la SRE para diseñar tecnología para el Gobierno Federal de México.

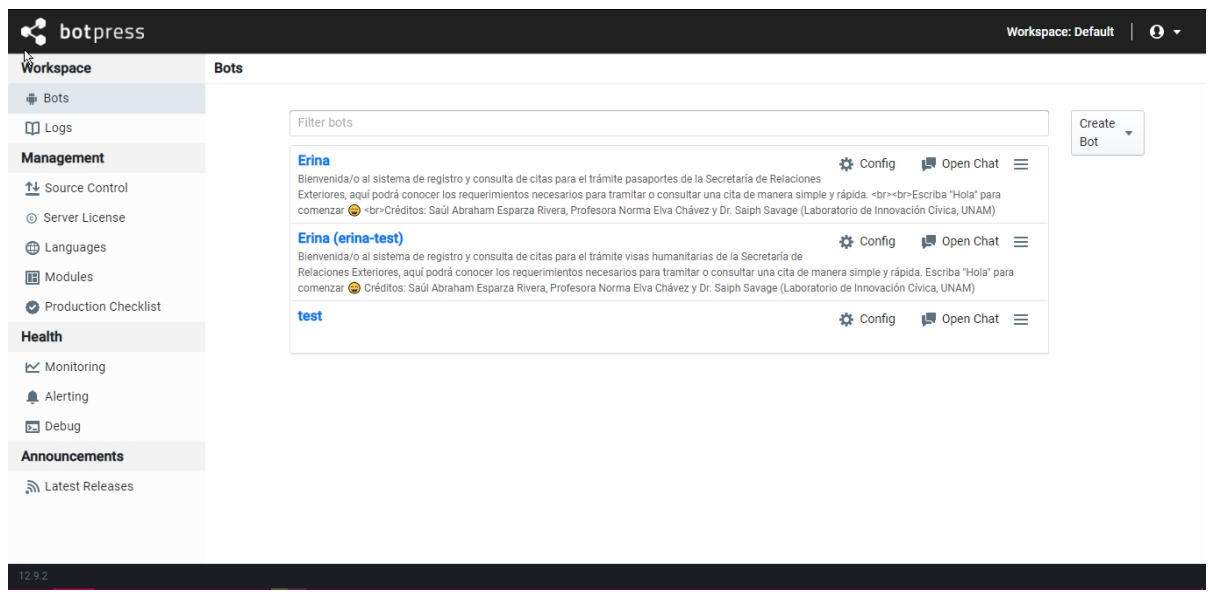

Figure 3.3: Panel principal de Botpress.

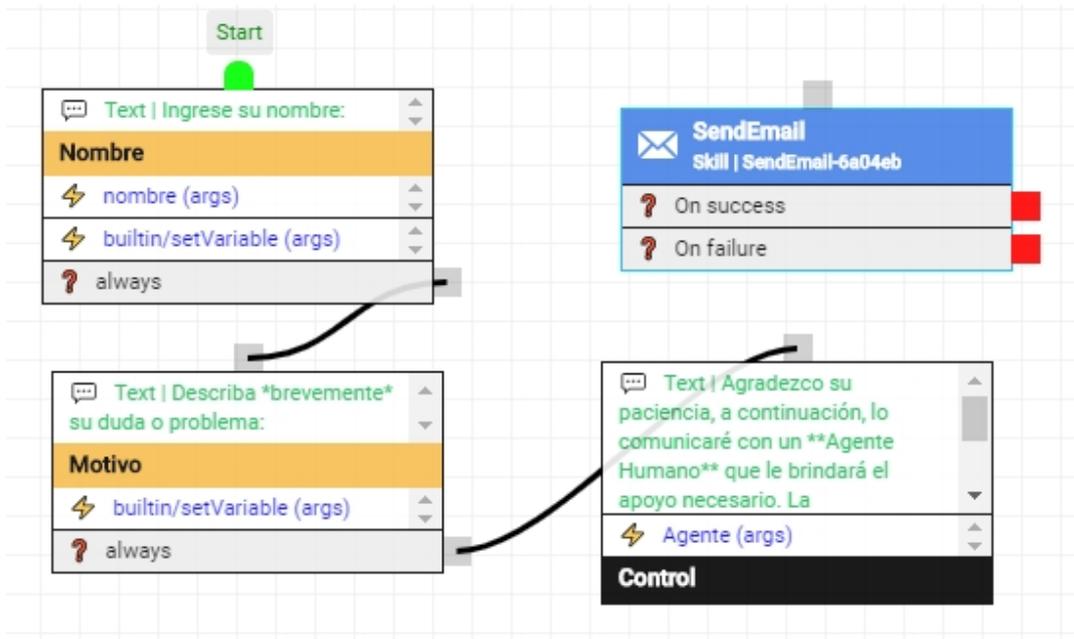

Figure 3.4: Todo flujo debe tener un nodo inicial, los nodos siguen reglas establecidas por el diseñador y que al cumplirse permiten la correcta comunicación entre los nodos involucrados y el funcionamiento deseado del flujo.

### 3.6.2. Botpress

Si bien ya hemos mencionado antes a la plataforma Botpress, resulta pertinente hablar un poco más de ella para comprender mejor su funcionamiento y capacidades. Botpress nos ayuda a generar un Bot por medio de una interfaz gráfica muy intuitiva (Figura 3.3) con la que podemos diseñar y construir un Bot o asistente virtual que cumpla con una tarea simple o con un conjunto de éstas para satisfacer alguna necesidad. Si lo que se busca es crear una conversación fluida y fácil de seguir Botpress es la herramienta correcta ya que nos proporciona de módulos básicos integrados por defecto en la plataforma desde mensajes de texto simple hasta la capacidad demandar elementos personalizados o más complejos que incluyen elementos visuales o conexiones con sitios y aplicaciones el usuario pueda disfrutar de una experiencia más llevadera. El entorno de desarrollo de Botpress es una gran red de nodos interconectados que se comunican entre sí (Figura 3.4) para poder realizar un flujo de conversación.

Uno de los principales retos al momento de crear un asistente virtual en la plataforma BotPress es la versatilidad de los requerimientos establecidos por ejemplo, el cuidado de la información de los usuarios y el manejo de esta (Figura 3.5). De modo que se implementó una base de datos segura en la que dicha información reside y es utilizada por el mismo asistente virtual para mostrarla en caso de que el usuario lo desee.

### 3.6.3. Composición

Desde el comienzo del proyecto se decidió por estudiar y posteriormente atacar los procesos y trámites más populares que son consultados en la SRE es por esto que los servicios que se iban a ofrecer vía el asistente virtual serían: renovación, trámite y consulta de citas para un pasaporte.

Renovación de pasaporte
Para el flujo renovación se tomaron en cuenta las medidas indicadas por los sitios oficiales de la SRE en los cuales se especifican los documentos e información a considerar para la renovación de un pasaporte. Con base en esto, el asistente virtual se encarga de recabar información por medio de preguntas al usuario y también mostrarle los documentos y la información necesaria a tomar en cuenta previo a la generación de una cita (Figura 3.6).

Una vez que el usuario decide proseguir con la generación de su cita el asistente virtual se encarga de guiarlo por todo el proceso facilitando el trámite y generando una experiencia totalmente única que resulta rápida y sencilla de comprender. Al finalizar la información del usuario se guarda en una base de datos. Esta información puede ser consultada en la sección "consulta de citas".

Figure 3.5: Módulo NLU: Usando expresiones regulares se pueden capturar datos que sigan un formato específico.

Figure 3.6: Renovación de pasaporte

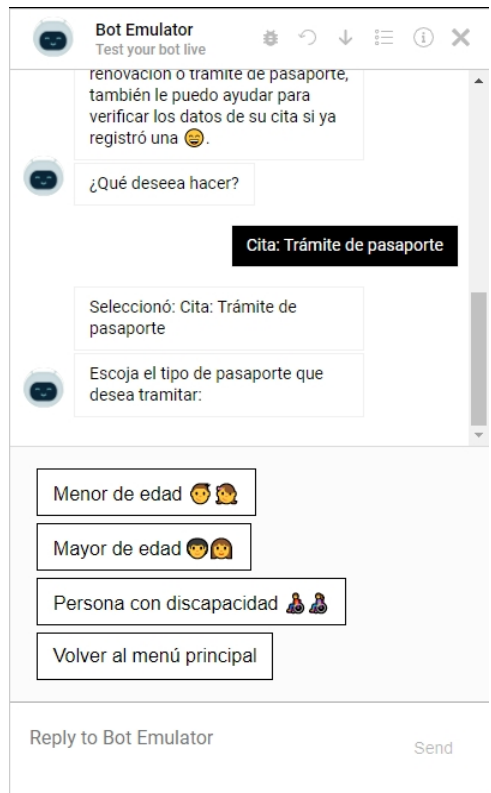

Figure 3.7: Trámite de pasaporte

Trámite de pasaporte
En la sección para un trámite de pasaporte nuevo oh por primera vez el usuario encuentra opciones definidas de acuerdo con los tipos de pasaportes que se pueden tramitar en la SRE sí ya sean para mayores de edad, menor de edad o personas con discapacidad (Figura 3.7). A cada caso se definieron secciones en las que pueden encontrar información necesaria de acuerdo con los distintos casos que contempla la SRE, así como requisitos y factores a considerar antes de generar su cita.

La opción para generar una cita le presenta al usuario con diferentes preguntas con las cuales se busca agendar una cita para llevar a cabo el trámite de pasaporte, el proceso resulta similar al de renovación en cuanto a la experiencia en general pero la información guardada de la cita del usuario está marcada según sea el caso.

Consulta de citas
Para consultar una cita el usuario deberá contar con el CURP que ingresó al momento de registrar su cita de trámite o renovación de pasaporte, una vez ingresada esta información el asistente virtual contestará con un mensaje de texto que incluye toda la información recabada al momento del trámite incluyendo datos cómo CURP, fecha de la cita, lugar de la cita y horario de la cita (Figura 3.8).

Secciones informativas
El asistente virtual tiene la capacidad de brindar atención informativa (Figura 3.9), ya sea usando las secciones dentro de cada trámite con las cuales se pueden conocer más detalles acerca del trámite en qué se está interesado o bien accediendo directamente a la sección preguntas frecuentes en la cual Obtendrá información pertinente a la pregunta que esté consultando.

Otra sección informativa de gran importancia es la sección contactar a un agente humano, en la cual como su nombre lo indica el usuario solicita ayuda a un agente humano directamente. En esta sección el usuario se comunica única y exclusivamente con el agente humano y la conversación con el asistente virtual se ve detenida y sólo podrá ser resumida una vez que el agente humano así lo decida (Figura 3.10).

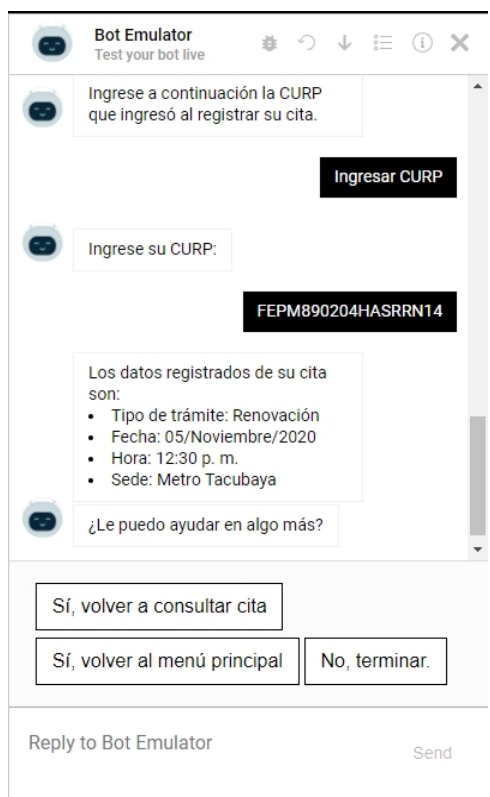

Figure 3.8: Se muestra la información al usuario.

### 3.6.4. Mejorando la experiencia

Desde el principio se tenía claro que el objetivo de este proyecto sería mejorar la experiencia de obtener información, generar y consultar citas tanto para los usuarios cómo para las personas que anteriormente se encargaban de manejar todas estas consultas.

El asistente virtual brinda ese apoyo y sirve como contacto entre la ciudadanía y la Secretaria de Relaciones Exteriores creando un nuevo paradigma en el cual las personas ya no tienen que salir de su casa para obtener esta información ni tampoco tienen que investigar por jornadas de tiempo extensas toda la información necesaria para realizar sus trámites o dispensar sus dudas ahora todo queda concentrado en un único asistente virtual encargado de brindar información y de solventar las dudas de la ciudadanía así como aligerar las cargas de trabajo y automatizar los procesos detrás de todos los trámites y consultas con los que la Secretaría trabaja día con día.

Gracias a la accesibilidad que tiene este asistente virtual se mejorará la experiencia en general de los usuarios y se impondrá un nuevo estándar dentro de la sociedad en el cual las personas ya no tienen qué gastar más tiempo del que desean investigando por información sobre los servicios que ofrecen las instituciones de sus gobiernos ahora podrán tener esa información cada vez más accesible para ellos.

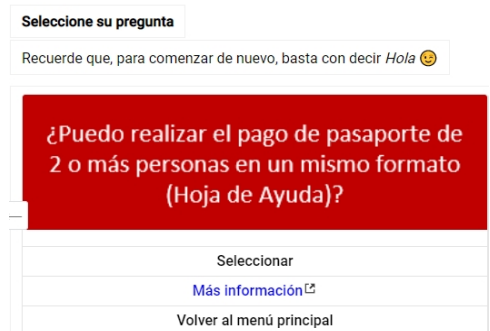

Figure 3.9: Las preguntas frecuentes son mostradas en forma de tarjetas, el asistente virtual contesta a dichas preguntas mediante un mensaje.

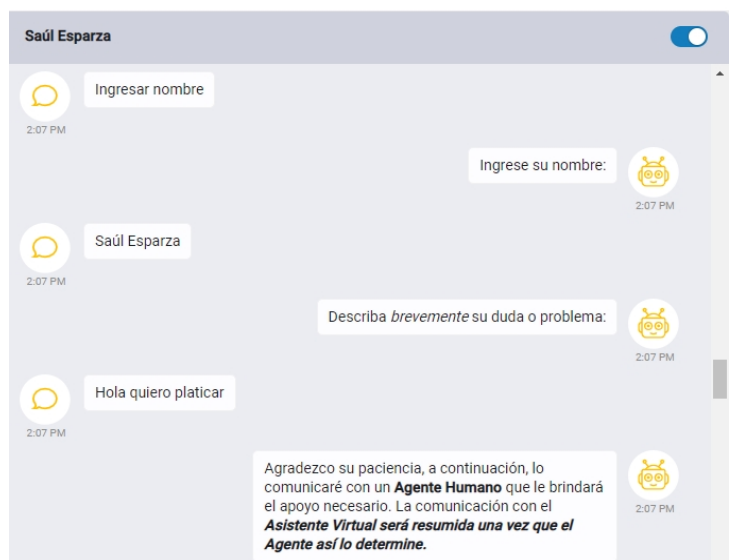

Figure 3.10: Se muestra la interfaz desde la perspectiva del agente humano, en ella la comunicación queda estrictamente entre el agente y el usuario.

# 3.7. Evidence Socialization in a Community based Dengue Prevention initiative

**Delsy Denis**
Universidad Nacional de Asunción
Asunción, Paraguay
*delsy.denis.21@fpuna.edu.py*

**Sofia Rivas**
Universidad Nacional de Asunción
Asunción, Paraguay
*sofiarivas1993@fpuna.edu.py*

**Julio Paciello**
Universidad Nacional de Asunción
Asunción, Paraguay
*julio.paciello@pol.una.py*

**Cristhian Parra**
United Nations Development Program
Asunción, Paraguay
*cristhian.parra@undp.org*

**Luca Cernuzzi**
Catholic University Nuestra Señora de la Asunción
Asunción, Paraguay
*lcernuzz@uc.edu.py*

*Dengue prevention efforts that involve the community in collecting and socializing evidence about Aedes Aegypti breeding sites have been proven effective to reduce the risks for dengue and other arboviruses. Key for their success is to make sure evidence (i.e., information about breeding sites and infestation levels in the community) is properly socialized with volunteers and the community at large. Here, we analyze one case of weekly messaging notifications as the means to achieve this goal.*

### 3.7.1. Introduction

Well designed community participation strategies have been proven effective in reducing the risks of dengue and other arboviruses, by facilitating the elimination of breeding sites for its vector, the *Aedes Aegypti* [7]. One of the key challenges of these programs is to ensure awareness on the evidence (i.e., information about breeding sites and infestation levels in the community) to influence collective action. This paper discusses our experiments, using weekly messaging notifications as the means to socialize evidence with volunteers of a community mobilization entomological surveillance program in Asunción, Paraguay. The TopaDengue project engaged local volunteers of a community in a citizen science participatory program to monitor the presence and evolution of the *Aedes Aegypti* in the community (i.e., community-based entomological surveillance), using the DengueChat platform to collect and visualize data.

Table 3.4: Dengue prevention or control ICTs.

| Project | Description |
| --- | --- |
| *GraviTreks [2]* | Quick scanning and instant report of adult mosquitoes surroundings. |
| *Spectra [1]* | Georeferenced information of breeding sites |
| *Mosquito Habitat Mapper [128]* | Part of *Nasa Globe Observer*, it facilitates sampling and counting of mosquito larvae (as DengueChat, but without the social component) |

### 3.7.2. Related Work

Interactive systems for better engagement between citizens, communities and governments, also known as Civic Technologies (CivicTech) [14] have oriented our work. Platforms like Ushahidi, SeeClickFix, FixMyStreet, and Urban Decor are closed examples to the type of participatory mapping that takes place in TopaDengue. More closely connected to our goals of using messaging and notifications, we build upon previous work from our team that has explored how these might encourage participation [13]. In the context of Dengue prevention, some of the most salient examples of interactive systems to both collect evidence on breeding sites and socialize it in some form are listed in table 3.4. Most of these systems are geared toward the report of mosquito breeding sites or areas where dengue cases exist to the pertaining authorities. However, these systems do not focus on promoting community action and collaboration.

---

[14] https://civictech.guide/

### 3.7.3. Context: The TopaDengue Project

The community mobilization as a prevention strategy adopted for the TopaDengue project takes inspiration from the methodology applied in *Green Way* [8] and it has been contextualized with a social volunteering program specifically designed for a vulnerable community in Asunción *Bañado Sur*, Paraguay.

   We worked with volunteers who are residents of this community, training them to host once a weekly house-by-house visits in search of potential breeding sites. They were asked to document the process with support of tablets and paper forms, to upload and visualize the data in *DengueChat (DC)*.

### 3.7.4. Methodology and Proposed Solution

The methodological aspects of our research and our engineering process can be framed as a Design Research [203] project.

**Preliminary Exploration**

We participated in the fieldwork with the volunteers [107], collecting qualitative notes about their experience with both fieldwork and the use of DC within this initiative. One of the design challenges that was identified, and over which the design and development efforts focused was that of evidence socialization:

- **Problem:** Not knowing the impact of the work done by the volunteers (that is, the evolution of the *Aedes Aegypti mosquito* infestation levels in the community, represented through the concept of *'casas verdes'*, are houses that for two months do not have active breeding grounds). This limits the possibilities of action of the volunteers, and can result in them losing the motivation to continue with their voluntary contribution, which in turn limits the number of *'casas verdes'* that the project can achieve.

- **Design Challenge:** How could we increase the amount of *'green houses'* in the community through risk socialization strategies in the community?

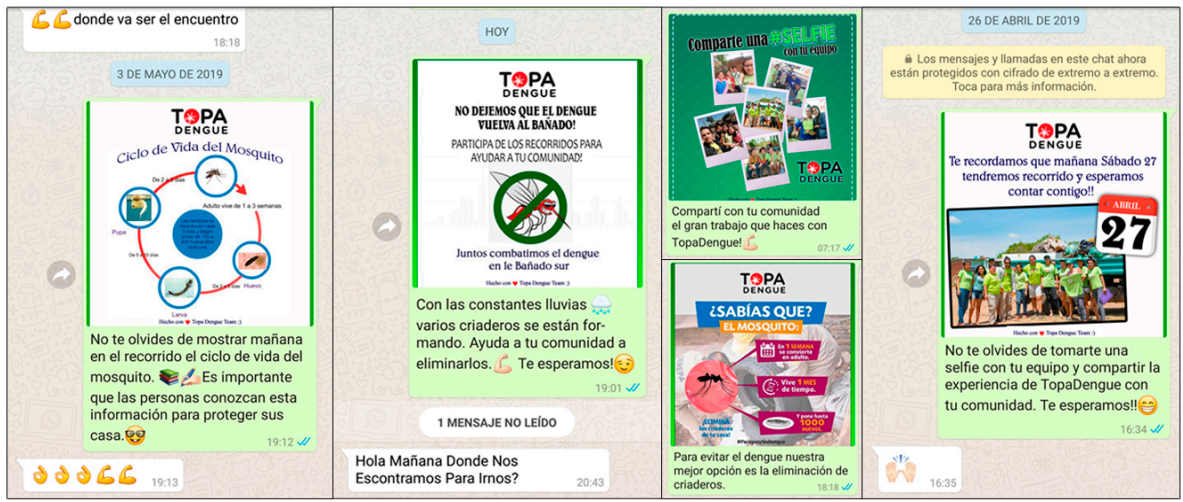

Figure 3.11: Messages sent by WhatsApp.

**Socialization via messaging strategy**

We implemented a Wizard of Oz (WOZ) prototype that consisted of a calendar to coordinate the delivery days of each message:

1. Information: about the project's progress.

2. Education: ways of preventing and eliminating mosquito breeders.

3. Motivation: for volunteers to participate of TopaDengue project activities.

The WOZ methodology simulated the auto-send functionality for messages, but in reality, the researchers manually sent them to the volunteers. Figure 3.11.

### 3.7.5. Results

Before launching the WOZ study, and in order to assess feasibility, we collected information about the communication channels used by volunteers and facilitators. A first survey was carried out in which 31 volunteers participated. Based on all these exploratory results, WhatsApp Business was chosen as our broadcast channel. A total of 48 messages were sent to 32 volunteers, from 2019-03-09 to 2019-08-02. To measure the impact of the messages sent via WhatsApp Business, a survey was delivered to 31 volunteers in total, in multiple times throughout the period of this case study. The most important results from these surveys are listed below:

- 71% of volunteers said that the messages were useful, with 35% (11) choosing educational messages, 35% (11) choosing active messages, and 32% (10) choosing informative messages.

- When asked whether they shared the message outside of the group of volunteers, 50% (19) did so using WhatsApp, and 15% (6) through Facebook. Also, 29% (11) talked about the project with someone upon receiving these messages.

- On a closer examination, using a grounded theory approach [32], we codified the responses to the open-ended question of how exactly the messages helped volunteers looking to decipher which of the message types were useful. We found that both informative (11) and educational (11) purposes were common, whereas active messages were not registered, contradicting the previous result.

- On a closer examination, we further codified the responses we got in the open-ended question to deepen our understanding of how exactly the messages helped volunteers. After manually classifying these answer, we found that both informative (11) and educational (11) purposes were the most common reasons why the messages were helpful. A minority of users reported that the messages helped in motivating them to participate (3).

- Most of the volunteers (66.7%) were better informed about the project in face-to-face meetings at the beginning and end of the fieldwork days, WhatsApp messages and social networks appear as a second option.

Considering the total number of visits to houses done by volunteers throughout the *TopaDengue* project, 45.2% were performed from April to July 2018. In the same period of time for 2019, this percentage rose to 54.8%. Interestingly, there were less volunteers available in the later period, and even despite this, more visits were performed.

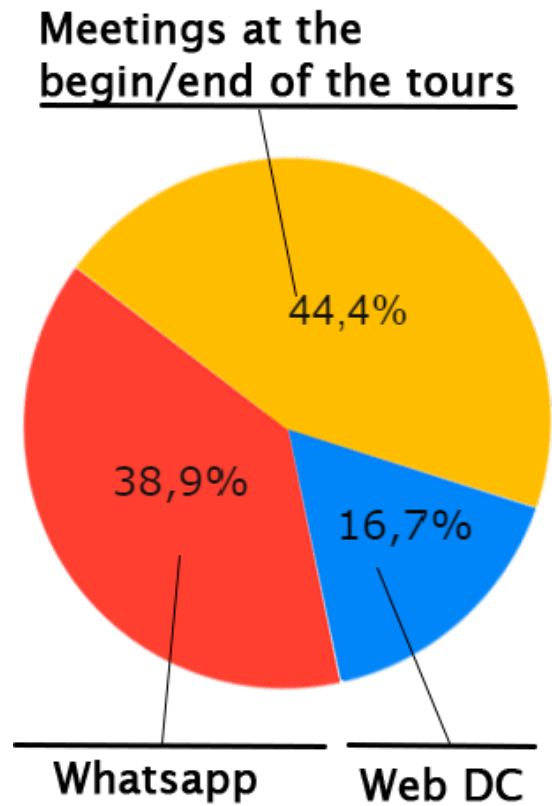

Figure 3.12: Information Media.

### 3.7.6. Conclusions and Future Work

Starting from an open-ended exploration, within the fieldwork activities of the *TopaDengue* project, we got valuable lessons. The messages strategy was welcomed by participants, generally considered useful in their activities with the project. Interestingly, while volunteers reported that educational and active messages were the most useful for them, the analysis revealed that the usefulness of informational messages were on par with the educational design, while the active messages were not registered. This is interesting because it validates in part one of our hypothesis (and the core value of the DC model): the most important role for ICTs is to ensure that the information (or evidence) is socialized, because this is what will mobilize the community to action. However, in order to fully validate this hypothesis, and particularly, the role of the notifications strategy in making it happen, a controlled experiment will be needed. Both results invite to further design explorations of both the content of messages that are sent to participants, and the strategy to deliver them.

Moreover, we asked users to select the best mechanism for them to be informed about the project. Figure 3.12 shows an interesting result which partially validates the results from the messages study. However, it opens up a new space of exploration: information sharing and socialization at the beginning or at the end of the field activities. The group meetings at the beginning and at the end of their activities are recognized as key moments of socialization. Finding ways in which ICTs can support this space, represents an opportunity for design and engineering. ICTs have a role in resolving these challenges, and the path we laid out in this paper is one in which a continuous feedback loop between research, design, development and community action is in place throughout the prevention initiative.

# 3.8. Flattening the Curve with Civic Technologies: A Case of Open Innovation during COVID-19

**Marta Poblet**
RMIT University
Melbourne, Australia
marta.pobletbalcell@rmit.edu.au

**Fatima-Ezzahra Denial**
Impact For Development
Tanger, Morocco
denif910@newschool.edu

**Pompeu Casanovas**
La Trobe University
Melbourne, Australia
p.casanovas@latrobe.edu.au

**Víctor Rodríguez Doncel**
Universidad Politécnica de Madrid
Madrid, Spain
vrodriguez@fi.upm.es

**Tarik Nesh-Nash**
Impact For Development
Tanger, Morocco
tarikn@gmail.com

This paper draws from previous research on crowd-civic systems to present the case of Mobadarat, an open innovation platform to crowdsource ideas and solutions to contribute to the COVID-19 response. Mobadarat leverages local capacities and digital infrastructure to channel ideation and solutions sourced from the local community. The paper also considers the issue of building trust between different stakeholders in this process and suggest further lines of research on civic technologies.

## 3.8.1. Introduction

COVID-19, the first pandemic of the digital age, has crudely revealed the limits of our emergency systems. In many countries, the response to the event has been hampered by unpreparedness and lack of both coordination and resources. Yet, the pandemic has also accelerated digitally enabled forms of collective intelligence that have been tested and deployed over the last decade: crowdsourced crisis mapping, citizen science, and peer-to-peer networks of open innovation. This paper briefly examines the role of civic technologies in the pandemic emergency by presenting the case of Mobadarat, an open innovation platform developed in Morocco to address COVID-19 related issues. The structure of the paper follows the questions that this workshop on civic technologies addresses: (i) role of local contexts and infrastructures (and case study); (ii) key elements in building trust, and (iii) methods and strategies for conducting research on civic technologies. We conclude by suggesting further lines of research in this area involving open data.

## 3.8.2. The Role of Local Context and Infrastructure in Designing, Implementing, Adopting, and Maintaining Civic Technology

For some years now, crowdsourcing and open data have evolved as a powerful method for civic engagement in crisis and emergency situations [177, 178]. In this context, crowd-civic systems—broadly defined as systems blending digital technology, design, and public data [146]—leverage crowdsourcing tools and techniques to enhance the discovery of relevant local information about issues of public concern. Previous research on crowd-civic systems [179] has proposed to consider them as part of broader "linked democracy" ecosystems with some specific properties: (i) contextually-bound (defined by an inner environment); (ii) open-ended (adaptive); (iii) blended (offline and online interactions); (iv) distributed (networked structures); (v) technologically agnostic (based on needs); (vi) modular (composable modules); (vii) scalable; (viii) knowledge-reusing (taping on multiple sources); (ix) knowledge-archiving (keeping knowledge accessible); (x) aligned (consequential). By leveraging digital technologies, therefore, crowd-civic systems can produce collective, reusable commons-based knowledge with consequential effects. In other words, when decisions are made or solutions are reached, the outcomes can be consequential and extend their reach to the outer context, aligning with and informing external processes of decision making.

Case study
Mobadarat.ma was born in March 2020, one week after the declaration of lockdown in Morocco. Launched by Impact For Development (IFD) in partnership with the School of Collective Intelligence of University in partnership with the School of Collective Intelligence at University Mohamed VI polytechnic and Alakhawayn University, the platform aimed to create a space where the community's collective intelligence can be utilized to address the effects of COVID-19 in Morocco. Mobadarat.ma1 leverages crowdsourcing as a method combined with both bottom up and top down strategies:

- **Ideation**: The use of crowdsourcing to enable participants to share their ideas and proposals regarding the emerging Covid-19 related challenges;

- **Initiatives Observatory**: a bottom-up approach that enables cross learning and knowledge exchange. It also provides open government stakeholders with a platform to collectively and proactively tackle the challenges at hand. In turn, the tool collects and publishes lessons learned;

- **Open challenges**: a top-down approach consisting of a space where decision-makers can share their challenges and ask the community to submit proposals and solutions.

Mobadarat has been adopted by the United Arab Emirates University in the UAE context. This expansion opens the door for international collaboration and learning in open innovation. In this regard, and as part of its effort to breaking international knowledge silos, IFD, in partnership with GovRight, has also launched OpenDev Library, a benchmarking platform combining initiatives, policies and approaches in various policy areas, co-created and undertaken by stakeholders as part of their efforts to foster further development.

### 3.8.3. Key elements of the configuration of trust among government, citizenry, and local organizations

The erosion of citizens' trust in democratic institutions is one of the most noted trends in recent years, together with the rise of populist solutions [97]. Mobadarat.ma aims at introducing a new relational paradigm based on an epistemic relationship between citizens and governments. This relationship builds on the capability of "a group of individuals to envision a future and reach it in a complex context" and the idea that "knowledge is openly shared, used and remixed" [179]. This paradigm requires further engagement from citizens to become a driving force in creating proposals and solution-driven approaches to challenges.

The strategy to leverage the capacities of citizens to collectively produce innovative solutions has long manifested itself in fragile states, such as the recent emergency in Lebanon shows. Yet, these citizen-driven solutions are often dismissed as mere reactions to the inability of those states to fulfill their needs and thus they can only exist in fragile contexts. In our view, citizen-driven initiatives may offer a wide array of lessons learned beyond its immediate local context. The goal of mobadarat.ma is reusing that knowledge and adapting it to the context of other developing countries.

### 3.8.4. Methods and Strategies for Conducting Research on Civic Technologies

Research on civic technologies is multidisciplinary and applies different methods and strategies, depending on the theoretical and empirical focus. In recent work [176], for example, we analyse mutual help as a digital commons and consider whether digitally-enabled mutual help aligns with Ostrom's design principles for the sustainable management of common-pool resources [165].

Artificial Intelligence has also been adopted in civic technology research and applications. In our view, this requires a middle-out approach rather than top-down or bottom-up ones. As stated in [169], forms of engagement require coordination mechanisms that are in-between, middle-out, as neither co-regulatory models nor self-regulation are adequate to comply with the ethical, legal and technological requirements needed in the interplay of civics and technology. Linked democracy designs should endorse the set of principles for Open Governance—e.g. traceability, transparency and accountability—and the recommendations for a responsible AI design—security, scalability, adaptability, modularity, interoperability. This is the common framework in which the tools and metrics of digital data analytics—Natural Language Processing, Machine Learning and Representation Languages—can be implemented and embedded to facilitate the emergence of sustainable civic technology ecosystems.

# 3.9. From Passive Viewing to Active Listening: Civic Technologies for Peace

**Anna De Liddo**
Knowledge Media Institute, The Open University
United Kingdom
*anna.deliddo@open.ac.uk*

**Philipp Grunewald**
King's College London
United Kingdom
*philipp.grunewald@kcl.ac.uk*

In this paper, we introduce Democratic Reflection, an audience feedback technology to promote active listening, deeper reflection and personal learning through interactive video replays. With the help of Aegis Trust, a British based Non-Government Organization working for the prevention of genocide and mass atrocities worldwide, we engaged 44 citizens in a study that tests how our audience interaction technology can support the enhancement of critical thinking and active listening capacities, as well as influence understanding of the contents and emotional engagement with audiovisual materials. By critically reflecting on the project we highlight some pragmatic challenges of using this digital tool with stakeholders in Rwanda.

## 3.9.1. Introduction

In 1994 Rwanda experienced large scale genocide. Since then the society has attempted, via various approaches, to work with this experience to prevent a repetition of such an event. Video archival materials are part of the collective memory of the country, and are used extensively by NGOs and piece building organisations as digital storytelling tools to promote empathy and to build peace. One of the main challenges these organisations encounter is to measure the impact of digital storytelling videos, and the changes they affect in the viewers perceptions of the genocide. It is also unclear to what extent the viewing experience improves people's critical thinking, active listening, and other capabilities for peace. Measuring changes in critical thinking and active listening skills is difficult at individual and collective levels. We therefore resorted to Democratic Refection, a new civic technology to gauge viewers' instant reactions to video replays.

## 3.9.2. Democratic Reflection: A Novel Audience Engagement Technology for Critical Thinking and Active Listening

Democratic Reflection (democraticreflection.org), is an innovative audience feedback tool that allows people watching a video, to instantly express their inner cognitive and emotional reactions to the viewing experience [49]. From this personal interaction, the tool creates a blueprint of both the personal and collective experience viewers are going through, which is then used to generate a series of analytics and visualisations to enhance both personal learning and reflection, and enable the collective assessment of the viewing experience. Democratic Reflection was tested with over 2000 UK citizen in the 2015 [86], 2017 and 2019 [110] political election debates, and was part of a suit of hypermedia technologies that were reported to improve self-reflection and learning, "change the way people felt about political leaders", and showed to "change personal assumptions" that people had before using the tool [50].

## 3.9.3. User Study in Rwanda with Aegis Trust

In 2018 with the support of the ISOOKO EU project (http://isooko.eu/) and Aegis Trust, Figure 3.13 we used Democratic Reflection to engage a small group (44 people) to actively listen and reflect on a testimonial video ("Ubumuntu") telling stories from the Genocide in Rwanda [168]. We had two main goals. Firstly, we aimed to capture the impact of the video material on the audience. This information is particularly valuable for Aegis Trust to select the most impactful videos to use in their peace building programmes. Secondly, we aimed to test to what extend Democratic Reflection enabled users to more proactively engage with the video material, to develop empathy, and better understand the genocide.

## 3.9.4. Testing DR with Video Testimonies from Rescuers

Two groups of Aegis Trust stakeholders were invited to participate in the trials. All of them had existing relationships with the organisation (a potential bias). These two groups represented a stakeholder group called 'youth' that Aegis Trust regularly engages with and another called 'parents' that Aegis Trust regularly engages with. All participants had previously been to events run by Aegis Trust (a significant number had received training from Aegis Trust). These two groups were subdivided and split across the control and trial groups. The control group watched the Ubumuntu documentary on their laptops (provided by Aegis Trust at the Kigali Genocide Memorial) in the browser on YouTube. This is the conventional way of accessing the content. The trial group watched the video in the same way but used the Democratic Reflections interface for consumption and engagement with the video. The outcomes were then measured via a range of research methods and afterwards trial and control group results were compared to allow for evaluation findings with regards to the Democratic Reflections tool.

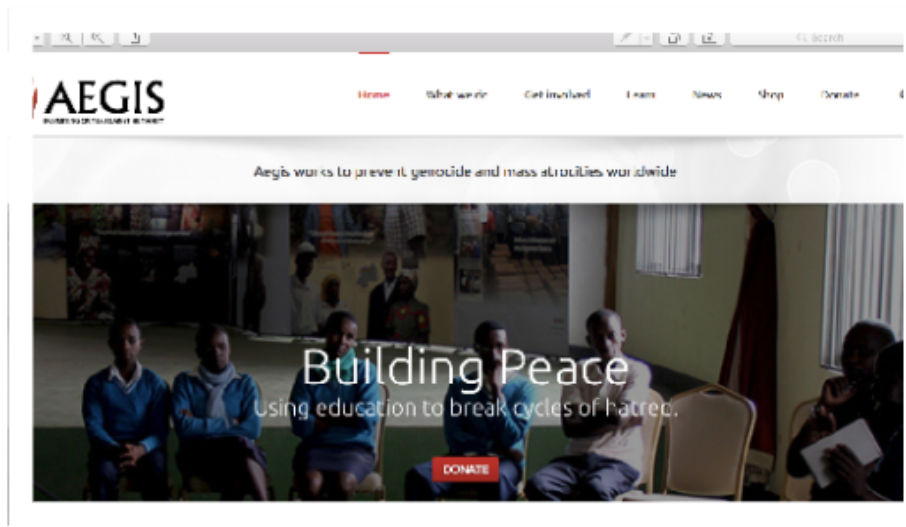

Figure 3.13: Aegis Trust works in the UK, USA, Rwanda, Central African Republic, Kenya and South Sudan. Aegis Trust has been present in Rwanda since 2001. It worked with the Government of Rwanda to establish the Kigali Genocide Memorial (KGM www.kgm/rw ) in commemoration of the 1994 genocide against the Tutsi in Rwanda. )

### 3.9.5. After the Test: Visual Analytics and Time Analysis of Audience Feedback

The test was extremely useful to design and refine an awareness rubric and to assess the extent to which the video promoted awareness and empathy. In relation to our aim to use Democratic Reflection to capture the impact of the video replay on the audience, we were able to develop detailed insights on the interaction. For instance, we were able to quantify that people empathic feedback to the video were quite high, with 51 percent of audience reactions showing from medium to very high personal connection to the video. Whereas in terms of understanding of the genocide, the video did not seem to be very effective in promoting personal commitment to act.□The data gathered by Democratic Reflection also allowed comparison of the different users' reactions by demographic. For instance, it enabled the analysis of the difference in empathic and cognitive engagement between man/women or younger/older people. Additionally, the tool also provided timeline visualisations of the audience reactions, to analyse reaction moment-by-moment and identify picks of reactions, which may highlight particularly engaging passages within the video (See Figure 3.15). This demonstrate that the tool was particularly useful as video assessment technology and has therefore the potential to be used by NGOs for the selection and design of video material to promote piece building education. The statistical analysis from the Rwanda's test and the quantitative analysis of the multimedia experience facilitated by the tool is object of a wider publication. In this short paper we focus on challenges and opportunities of testing digital civic technologies, like Democratic Reflection, in such a dividing context as a post genocide community in Africa. Our reflection suggests the following lessons learned.

### 3.9.6. Discussion

Building capacity for peace after a genocide is an incredibly difficult challenge, and doing so with technology is even harder. The Rwanda's community we worked with has very low digital literacy. Whereas people use mobile devices to support basic life functions, they have very limited access to desktop com-puters and internet connections. Therefore, we encountered some very pragmatical challenges in running our test: 1) narrow and costly engagement, 2) very limited interest in the technological interaction. 3) cultural framing which may bias the results. Engagement is Costly. AEGIS Trust, as many NGOs, works mostly on a voluntary base, and needs to mediate between a longstanding commitment to the community, and their temporary role of information brokers between designers, researchers and local people. This means that practitioners on the field have very limited time and very limited resources to support the technology testing and the research. Also, the selection of participants is done on a very opportunistic basis, favouring people that have more time or resources, which per se represents a bias in the sample of people that can be really engaged on the field. Technological Scepticism. We registered people's lack of interest in the technological interaction. In fact, to access the AEGIS Trust computers, and minimise effort for the practitioners in the field, we created a "fake" interaction context, and asked people to sit in front of a computer and watch a video while being all in the same room. This denaturalised the usage of the technology, which was initially designed for distributed settings, or live staged events interactions. It does not surprise then that users found it a "cold" way to interact and would have rather "talked to each other". Nonetheless, the tool was highly usable, and during the interaction people clicked the feedback cards a lot and with steady engagement. We recorded almost 3000 clicks from 22 people in 20 min. This is basically more or less 137 feedbacks per person, at almost a rate of 7 cards per minute (which is more than 1 card every 10 seconds). The engagement was steady from beginning to end of the video replay, which means that people did not get "bored" of interacting with the tool. This shows that despite people's lack of initial interest in the technological modality of interaction, this was still an experience which engaged them in a proactive way. Considering that one of the objectives of our research was to change the viewing experience from passive viewing to more active listening experience, this was per se a very positive result. Cultural Framing. Practitioners reports that, when local people turn up to the testing workshops and interact with international researchers, which are in charge of the data gathering, they always take that very seriously. They really try to do a good job and there is a strong sense of respect toward both the host (AEGIS Trust) and the international partners. This is part of the local culture and means that when we (the researchers) ask local people to use a technology, they will really try to use it, perhaps even just to "please" the people that invited them to the event. This typology of participant is indeed a very different one from the one we previously tested the technology with (such as for instance UK citizens replaying televised election debates). It is an open question the extent to which this may have biased the results and their interpretation. And it is an even harder question what technology or testing design solutions can be found which allow to adapt the interaction to different cultural framing.

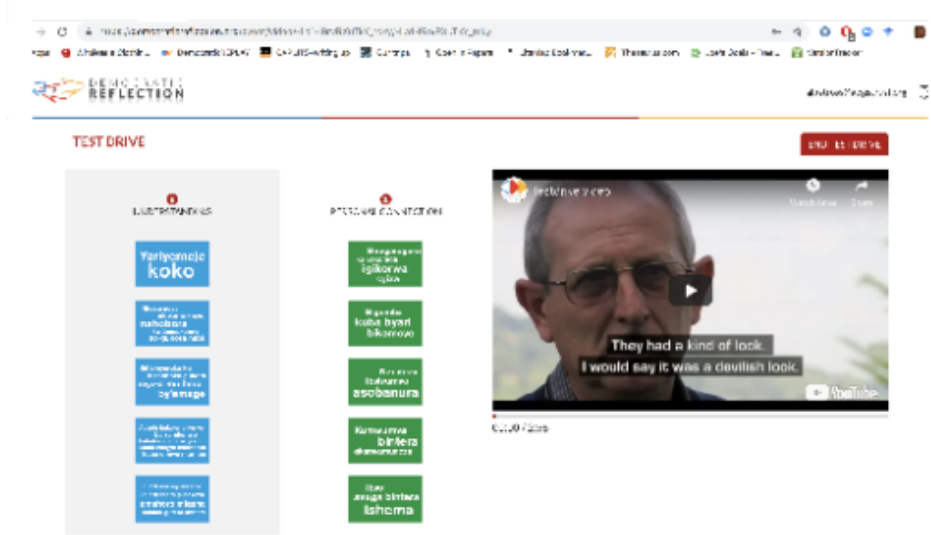

Figure 3.14: Democratic Reflection Interface, showing on the left the 10 reflective statements cards in the local Language and on the right the Video Testimony. )

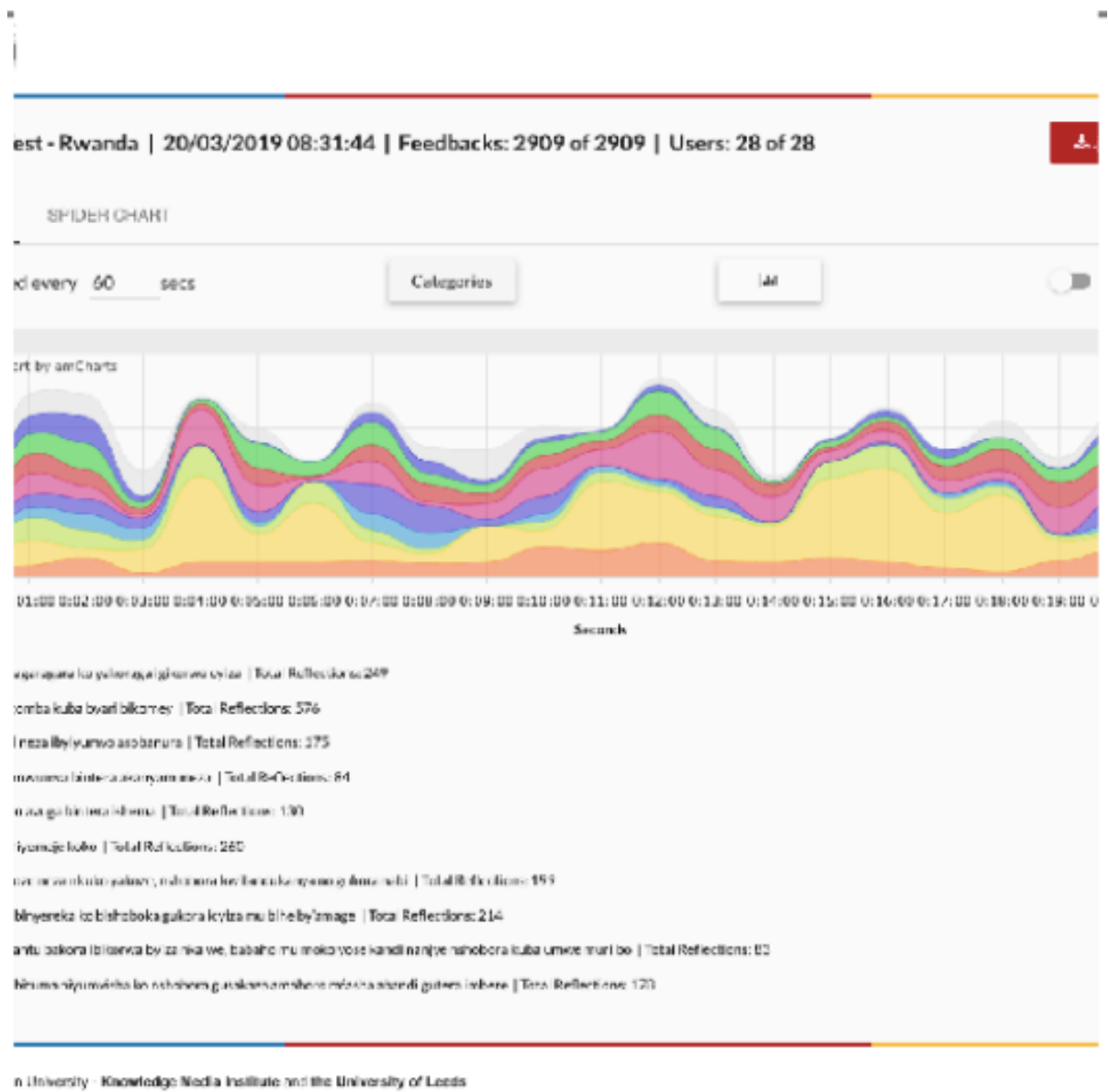

Figure 3.15: Feedbacks Timeline by reaction type, showing how many cards were clicked in a given moment, and picks and waves of reactions. y. )

# 3.10. Governing the Commons of Platform Labor Data Assets

Samantha Dalal
University of Colorado Boulder
Boulder, CO 80027, USA
samatha.dalal@colorado.edu

Brian C. Keegan
Department of Information Science
University of Colorado Boulder
brian.keegan@colorado.edu

As the platform-mediated digital economy expands, so does the invisible and distributed labor force behind it. But the workers in this economy have few legal or economic protections to advocate for their rights. Part of the difficulty in creating labor protections in the platform-mediated digital economy is that information pertinent to building court cases or reliable employment statistics is inaccessible by design. These data assets should be made available to workers so they have equal power as the platform operators in deciding their working conditions. While tensions between labor and capital are not new, I argue that these data assets are an example of a commons that should be jointly and democratically governed.

## 3.10.1. Introduction

The platform-mediated digital economy facilitate dual value production, resulting in the creation of data assets that are extremely valuable to platforms and workers. The platform-mediated digital economy's value and growth are continually increasing: the World Bank predicts that by 2020 it will be a $25 billion-a-year industry [84]. While tensions between labor and capital are not new, these data assets have properties that lend themselves to being treated as common, rather than private, goods. Commons do not always end in tragedies and can be sustained through governance strategies prioritizing accountability and participation.

## 3.10.2. Dual Value Production and Data Assets

Companies in the platform-mediated digital economy, like Uber, Lyft, Doordash, and Amazon Mechanical Turk, extract rent from the transactions between customers and workers. However these platforms also extract a far more valuable asset from these transactions: data about the preferences, contexts, and behaviors of both customers and workers. This production of data assets can be characterized as "dual value production: the monetary value produced by the service provided is augmented by the use and speculative value of that data produced before, during, and after service provision" [220].

Data Asset Generation and Use
Workers in the platform-mediated digital economy create "dual value production", generating an immense volume of data assets about their work patterns. However, this data is not shared by platforms with workers or customers for a variety of reasons like privacy, computational complexity, and competitive advantage [220]. These data assets include rich information about worker behaviors, such as if and when they log on or accept an request, the locations they operate in, the speed with which they complete tasks, and images of their environment [58]. These data assets assist platforms in creating dynamic maps of the workflows of their highly distributed infrastructure, giving platform designers an omniscient view [111] In essence these platforms "enact their programmability to decentralize data production and recentralize data collection", echoing the techniques used by industrialists following the Taylorist practices of decentralization, delegation, and centralized oversight [91, 111].

The Value of Data Assets for Firms
For platforms, the value of this data is twofold: they are able to feed worker-generated data into their software infrastructures to improve their AI systems and they are able to reduce the value of workers' expertise through algorithmic approximation and communication. Data assets generated by laborers are integral to the software infrastructures of platforms, as Veena Dubal puts it: "central to the infrastructure of AI is the labor of dispersed and atomized workers in global supply chains who create, gather, pick, clean, label, and/or otherwise process the data that informs and shapes AI systems." [58]. This reliance plays into how platform companies portray themselves as technology companies, thereby increasing their valuation as investors expect these companies to be able to leverage their data assets to move towards a full automated infrastructure by making human labor fungible and cutting these costs, essentially "convert[ing] data into money" [190]. From a managerial perspective, granular data generated by workers through the platform economy is key to creation of a centralized management system that allows for an omniscient view of a widely-distributed labor infrastructure and easily-implemented experimentation with new workflows [100]. In essence, workers produce value both in the form of the tasks completed but also in the form of metadata encapsulating the process of that task completion that is fed back to platforms to improve both their software infrastructure and their algorithmic management systems.

The Value of Data Assets for Workers
Workers are not privy to most of the data assets that they generate, and what assets they can access are typically formatted to make it infeasible to derive meaningful insights [220]. This is an example of a common information asymmetry between corporations and laborers: one of the key assets in negotiations is data regarding worker activity [58, 111, 220]. Given that the management techniques of platform-mediated digital economy corporations closely mirror those of the Taylorist scientific management techniques employed by manufacturers during the 1940s-1960s, Vera Khovanskaya et. al have argued that union negotiation techniques that were successful in advocating for workers' rights in the past may be useful for the current situation. Through data transparency, wage contestation, and strategic participation unions during that time period were able to successfully negotiate stronger laborer protections to counteract the scientific management techniques at play [111].

### 3.10.3. Governing the Commons of Labor Data Assets

Defining common-pool resources
Collective action problems are among the most pervasive challenges in social behavior but can be overcome with appropriate incentives for cooperation and coordination [116]. Economists differentiate goods into four classes based on two dimensions: excludability and rivalrousness. Excludablility is the ability to prevent consumption of a good or service while rivalrousness is the extent to which the consumption of a good or service reduces the amount of consumption for others [166]. Common-pool resources like forests and fisheries exist within the non-excludable but rivalrous quadrant where it is difficult to prevent people from using resources but overuse can lead to depletion of the resource, what is commonly known as the "tragedy of the commons" [88]. However, commons do not necessarily end in tragedy and Elinor Ostrom's work emphasizes how accountability, participation, and conflict resolution can create sustainable common-pool resources [165].

Data Assets as Enclosed Common-Pool Resources
Conventional wisdom treats many digital goods and services as public goods like radio or air that are both non-excludable and non-rivalrous. However, this framing obscures the very real materialities and complexities for developing and sustaining information infrastructures: development, maintenance, delivery, and management are examples of marginal costs. Accounting for these rivalrous features shifts many types of digital and information goods away from public and into common-pool goods requiring different kinds of governance and oversight.
    But the data assets generated by customers and workers within the platform-mediated digital economy are currently treated by firms as private goods. This is really an example of *enclosure* where common-pool resources are converted into private goods through the erection of exclusionary property rights [25]. As described above, both firms and workers can benefit from the value of the data assets generated in platform-mediated digital economy but the logics of enclosing and excluding workers from accessing and using them enables the accumulation of value for entrepreneurs and investors rather than workers. What precedents from social and economic history exist for converting enclosed goods back to common-pool resources? Or developing common-pool counterpowers that can out-compete enclosed goods?

The Role of Civic Technology
A key difference between labor organizing during the 1940s-1960s and today is that platform-mediated gig economy workers are not protected by the same labor laws and allowed to unionize due to their worker classification. This exclusion of workers from legal protections and privileges in addition to the atomized nature of this work force makes communication, information sharing, and collective action extremely difficult. However, marginal success has been achieved in gathering data from platforms through subject access requests (SARs) which have been used to support legal arguments for fair wages and workers rights in lawsuits both in the EU and in New York City [93, 220].

The emergence of civic technology platforms such as the Worker Information Exchange (WIEx) utilizes a technique similar to the wage surveys used by unions in the 1940s-1960s. By encouraging workers to submit SARs for their data and then giving workers access to this data in an aggregated form. Platforms like WIEx facilitate data transparency by creating a data commons, but there is still work to be done in regards to the type of data gathered, the ownership and management of this data, and the governance of this data. As workers in the platform economy are not legally allowed to unionize, a data commons would provide a digital platform for both restoring agency over data assets to workers and facilitating collective action to address the power asymmetry between workers and platforms.

### 3.10.4. Conclusion

Civic technology can play a role in addressing the trust and privacy concerns of platforms who are reluctant to disclose data assets, citing privacy concerns and proprietary information. In addition civic technology can provide the infrastructure for a data commons where aggregation, analysis, summarization, and distribution can be implemented to provide information pertinent to addressing power asymmetries in the platform-mediated digital economy. I urge civic technologists to incorporate Ostrom's eight guiding principles of governing the commons into their development and construction of civic technology to create and govern this data commons [165]. Ostrom's framework outlines the requirements for governance of a complex system, incorporating the importance of context dependent infrastructure, nesting of governance structures, institutional variety of stakeholders, amongst others relevant to the successful maintenance of a data commons in a distributed global context [165].

# 3.11. Improving logistic ICTs for community-based dengue prevention

**Delsy Denis**
Universidad Nacional de Asunción
Asunción, Paraguay
*delsy.denis.21@fpuna.edu.py*

**Sofia Rivas**
Universidad Nacional de Asunción
Asunción, Paraguay
*sofiarivas1993@fpuna.edu.py*

**Julio Paciello**
Universidad Nacional de Asunción
Asunción, Paraguay
*julio.paciello@pol.una.py*

**Cristhian Parra**
Universidad Católica Nuestra Señora de la Asunción
Asunción, Paraguay
*cristhian.parra@uc.edu.py*

**Luca Cernuzzi**
Universidad Católica Nuestra Señora de la Asunción
Asunción, Paraguay
*lcernuzz@uc.edu.py*

*Organizing a community to participate in weekly entomological surveillance activities that reduce the risk of Dengue and other arbovirosis was the focus of the TopaDengue project, a cuasi-experimental impact evaluation of a community-based prevention program, supported by ICTs. This paper discusses our learnings while addressing a key challenge of this project, which is likely a common problem in participatory mapping initiatives supported by civic technologies.*

## 3.11.1. Introduction

Once a week, and equipped with paper and digital tools, volunteers of TopaDengue in Asunción, Paraguay, visited houses in their communities in order monitor the presence and evolution of mosquito breeding sites. This type of prevention program, which relies on the active participation of residents in the entomological monitoring of their communities (i.e., systematic collection and socialization of evidence about the presence of *Aedes Aegypti*), is a promising approach for sustainable dengue prevention, proven effective in previous studies [7]. Based on the *Green Way* [8] and supported by DengueChat[15] to map, process, and socialize data, TopaDengue was a research project that designed, evaluated and found evidence of positive impact of a community mobilisation program for dengue control [171]. The program was tailored to the characteristics of a flood-prone neighborhood in Asunción, and used a cuasi-experimental design to evaluate its impact by comparing infestation levels in an intervention area versus a similar control territory. Drawing on design research within TopaDengue, we identified that one of the key challenges driving the efficient implementation of the program was the lack of good support for coordinating the activities (i.e., selecting city blocks to visit, assigning volunteers, enabling offline data collection and synchronization, etc.), a need that is likely common to other civic technology participatory mapping initiatives. In this paper, we discuss how we came to focus on this challenge and the lessons we learned while designing and testing solutions to address it.

| GraviTreks | Quick scanning and instant report of adult mosquitoes surroundings [2]. |
|---|---|
| Spectra | Georeferenced information of breeding sites [1]. |
| Mosquito Habitat Mapper | Part of *Nasa Globe Observer*, it facilitates sampling and counting of mosquito larvae (as DengueChat, but without the social component). |

Table 3.5: Dengue prevention or control ICTs.

---

[15]www.denguechat.org

### 3.11.2. Related Work

DengueChat is similar to *civic technologies* in the engagement and civic data spaces[16] like Ushahidi, FixMyStreet, Urban Decor, and others used in participatory mapping of issues or resources for collective action. In the Dengue domain, previous studies have designed and evaluated similar ICTs to support entomological [155][16] or epidemiological [131] [132] surveillance through citizen engagement. However, none of them addressed the challenge of improving logistics of the participatory program in which they were integrated.

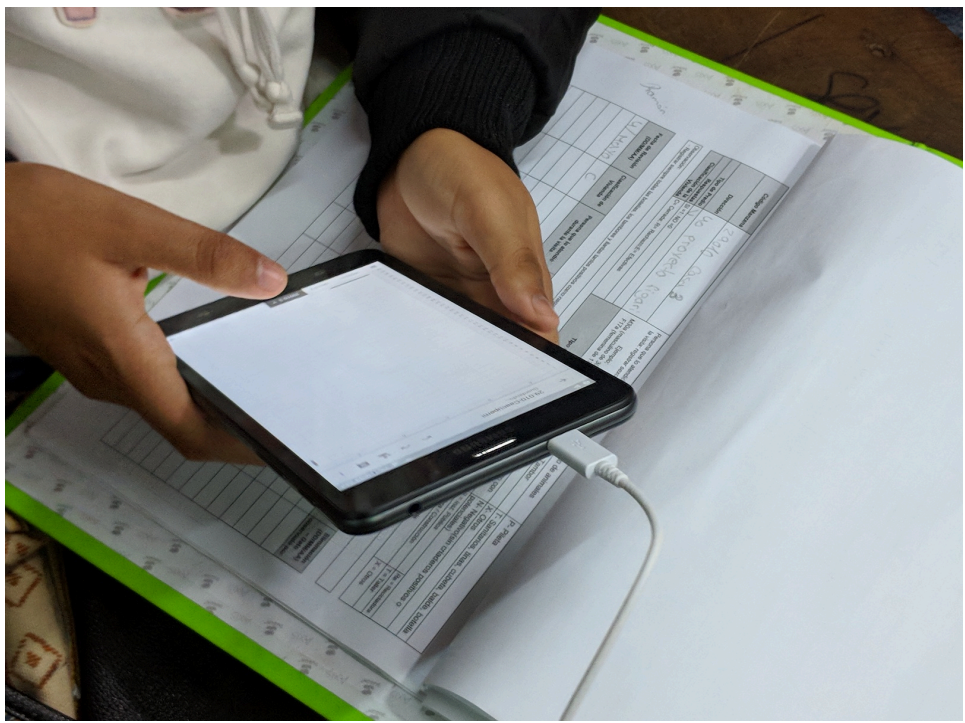

Figure 3.16: Evolution of paper forms

### 3.11.3. Identifying the Challenge

By accompanying volunteers and facilitators in their fieldwork, we observed the experience in detail. The first important observation was the need for a closer look on the usability of the mobile app for DengueChat, used to collect breeding sites data at house level. Through heuristic evaluation[159], we identified major usability problems. Development work required to address these issues led us to consider other tools for data collection. Once we identified alternatives, we conducted user testing through four training sessions and one field test day where we shadowed volunteers to document their experience using the tools in the context of real house-by-house visits. We followed up with semi-structured interviews to further deepen our understanding of the experience. Insights from these activities led to the selection of the Open Data Kit (ODK)[17] toolset for data collection, coupled with an automated synchronization system that moved data from ODK to DengueChat. Concurrent to this exploration, we also observed different issues on the use of the paper version of the data collection forms, used for backup and places less safe to walk around with tablets (see Figure 3.16). Combining ODK with better paper forms gave us a robust and lightweight data collection workflow that worked offline, was easy to use for post-activity uploads and digitization, and required no development on our end, shifting coding efforts entirely to the DengueChat server components, particularly, to a single synchronization module.

Mobile app development and maintenance plus paper form issues of use and integration into a digital workflow were the first hints that fieldwork coordination and seamless orchestration and integration of

---

digital and analog tools were key to the success of the program. This would become salient when we realized that up to one third of every fieldwork day was used to select city blocks to visit and decide assignments for volunteers, which was done manually, at the beginning of the day, by looking at physical binders with paper forms documenting previous visits and printed maps. Our focus then became how to improve logistics and increase the time spent visiting houses rather than planning to visit them.

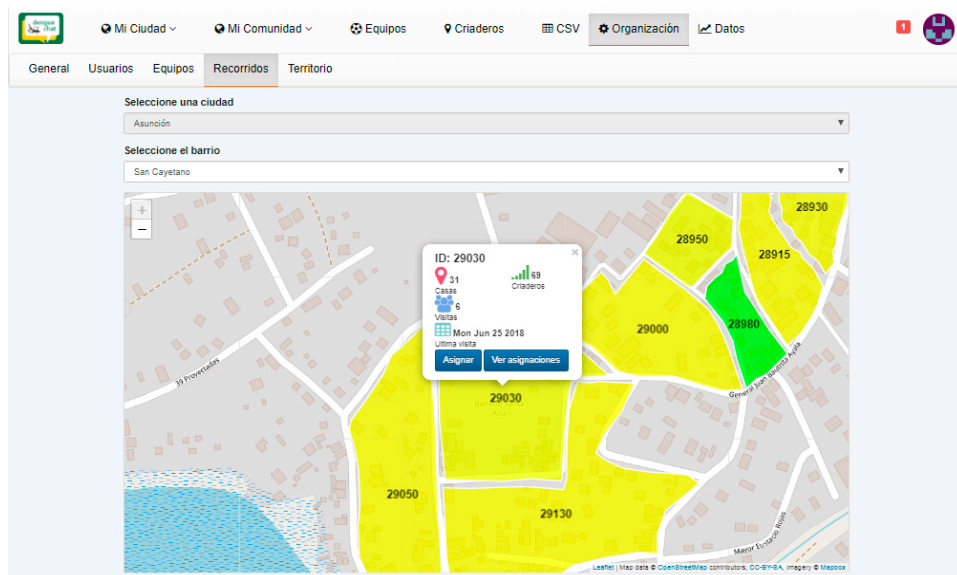

Figure 3.17: Route Assignment Module.

| **Compatibility:** compatibility issues on many low-end smartphones prevented the app from installing, or resulted in partial functioning (e.g., camera not working). |
| --- |
| **Storage:** storage needs for installing were not met in many low-end smartphones. This caused some of the volunteers to uninstall the app in order to accommodate other data, like personal photos. |
| **Usability:** some problems were found with respect to all of the design principles of Nielsen, requiring a development team to re-design the app, if it were to address these problems. |

Table 3.6: Problems in the original mobile app

### 3.11.4. Addressing the challenge of logistics

To improve logistics, we designed a map to schedule fieldwork days, explore city blocks status and previous visits statistics, and assign them groups of volunteers. Each city block polygon displayed its number of houses, its current infestation level, its number of *'green houses'* (i.e., houses free of breeding sites for over a month), the date of the last visit, and the number of previous visits (See Figure 3.17. We also developed UX improvements on several pages of DengueChat and a new points distribution algorithm to reward the collective effort of volunteers, a feature only feasible now based on the logistics data the new module allowed to store and manage. The new gamification process rewards all the volunteers assigned to a city block when new green houses appear on that block. In the old mobile app, only the user who uploaded the data received the rewards.

| |
|---|
| **Problem:** Fieldwork by volunteers to implement the program entomological surveillance activities were characterized by a large amount of time devoted to logistics (i.e., organizing how carrying on the fieldwork, selecting city blocks to visit, assigning tasks, etc.). |
| **Design Challenge:** How might we minimize the amount of time that was devoted to the logistics of the fieldwork to implement the program's entomological surveillance activities?. |

Table 3.7: The logistics challenge

### 3.11.5. Evaluation and key results

We organized 12 think-aloud sessions (4 facilitators, 8 volunteers) in which they evaluated these improvements.

Half the participants tested the initial version of DengueChat, and the other half tested the new version. Half of each group completed the tasks on a laptop, and half on a tablet. We asked them to try several tasks, but mainly, the route assignment process was observed. Three key results emerged from these tests: first, the new logistic module was praised by both facilitators and volunteers as good way to organize the work, second, the module made it possible to follow explicit city block assignment strategies (i.e., assign the least visited city block, assign the most infected ones, etc.), and third, implicit knowledge of the territory and previous visits was no longer paramount to organize the activity. Facilitators and volunteers who used the old version depended on their implicit knowledge of the territory and previous weeks of work: facilitators knew where volunteers went in previous weeks and used that to organize the routes, while volunteers knew only their assignments and struggled to come up with assignments. In the new version, volunteers who did not have prior experience with organizing this work, showed their own strategies to organize the routes based on the existing information on the screen and had now difficulties doing so, which means that volunteers themselves could potentially self-organize with these improvements.

### 3.11.6. Conclusions and Future Work

Our work within the TopaDengue project has resulted in an interesting design research loop between community, researchers and volunteers of a community-based dengue prevention programs, built upon civic technology in the engagement and civic data design spaces. Actively involving facilitators and volunteers early on the research and design activities ICTs, we uncovered key challenges driving the efficiency of the program's implementation, making it possible to respond and adapt quickly, introducing new and better data collection tools and synchronization algorithms, and finally introducing new features to the main platform that led to interesting new practices of self-organization and gamification strategies that reward collective effort of volunteers working together. Because our new designs were introduced at the end of the program's implementation, it remains for the future to replicate these results beyond the laboratory think-aloud sessions. However, we see values in these insights for the development of civic technology and initiatives for participatory mapping and collective action. They also point to the importance of a continuous feedback loop between research, design, development and community action when deploying civic technologies and community participation processes in a local context.

# 3.12. Inclusivity in Town Halls: Challenges, Paradigm Shift, and Opportunities

**Mahmood Jasim**
University of Massachusetts Amherst
Amherst, MA, USA
*mjasim@cs.umass.edu*

**Amy X. Zhang**
University of Washington
Seattle, WA, USA
*axz@cs.uw.edu*

**Ali Sarvghad**
University of Massachusetts Amherst
Amherst, MA, USA
*asarv@cs.umass.edu*

**Narges Mahyar**
University of Massachusetts Amherst
Amherst, MA, USA
*nmahyar@cs.umass.edu*

Traditional face-to-face town halls remain a popular choice for community consultation for government officials at local and national levels due to their ability to foster discourse and help the officials understand community members' viewpoints and aspirations. However, current town halls suffer from a lack of inclusive participation where vocal attendees often dominate the conversation, and silent attendees' opinions remain unspoken. In this article, we draw from our experiences of designing civic technologies to address this problem and articulate existing approaches and their shortcomings. We also highlight the paradigm shift in town halls during the COVID-19 pandemic and identify new opportunities for designing civic technologies to facilitate inclusive participation in future town halls.

## 3.12.1. Background

Government officials' preference towards face-to-face synchronous town-hall meetings to gather public opinion towards civic issues and agendas can be attributed to their objective of creating a two-way communication channel between the public and themselves to gauge the dynamics of public perception [60, 98, 136]. However, fixed time slots and physical locations for co-located meetings alongside lack of resources such as time, budget, and human resources restrict government officials to fully utilize the town hall meetings [99, 122, 137]. Furthermore, for the community members who attend the town halls, social dynamics such as shyness and tendency to avoid confrontation with dominant personalities might restrain their abilities to share opinions [26, 218]. This lack of inclusivity often induces disinterest in the general public to participate in town hall meetings [46, 77]. Investigations by Bryan [27] and Gastil [79] corroborate with this scenario as they highlight a steady decline in civic participation in town halls due to a growing disconnect between the local government and community members.

## 3.12.2. Existing Approaches and Recurring Challenges

In recent years, several civic technologies have been proposed to increase engagement in town halls. For instance, some researchers have experimented with audience response systems (ARS) [22, 129]. Bergstrom et al. used a single button device for attendees to anonymously vote agree or disagree on civic issues. They demonstrated how back-channel voting can help underrepresented users to get involved in the meeting discussion [22]. The America*Speaks*' public engagement platform, 21 Century Town Meeting®, also uses audience response systems to poll the attendees during town halls or share their opinion by voting on proposed agendas [129]. However, in these works, the ARS are used for binary voting or polling [22, 129], or to get responses on Likert scale-like questions [123] prompted by organizers at specific times in the meeting. As such, they do not enable attendees to provide real-time feedback.

Other researchers have proposed various solutions to increase participation in face-to-face meetings such as design charrettes, or general group-meetings, by using interactive tabletop and large screen surfaces to engage attendees [96, 135, 198]. For example, UD Co-Spaces [135] used a tabletop-centered multi-display environment for engaging the public in complex urban design processes. Memtable [96] used a tabletop display to integrate attendees' annotations on top of multimedia artifacts. IdeaWall used real-time visualizations of thematically grouped discussion contents on a large screen to encourage meeting attendees to discuss such topics [198]. However, large displays and real-time visualizations might distract attendees from concentrating on meeting discussions [9], and lead to fewer contribution [83]. Furthermore, using expensive tabletops and large interactive displays might be financially and practically infeasible in a majority of cities.

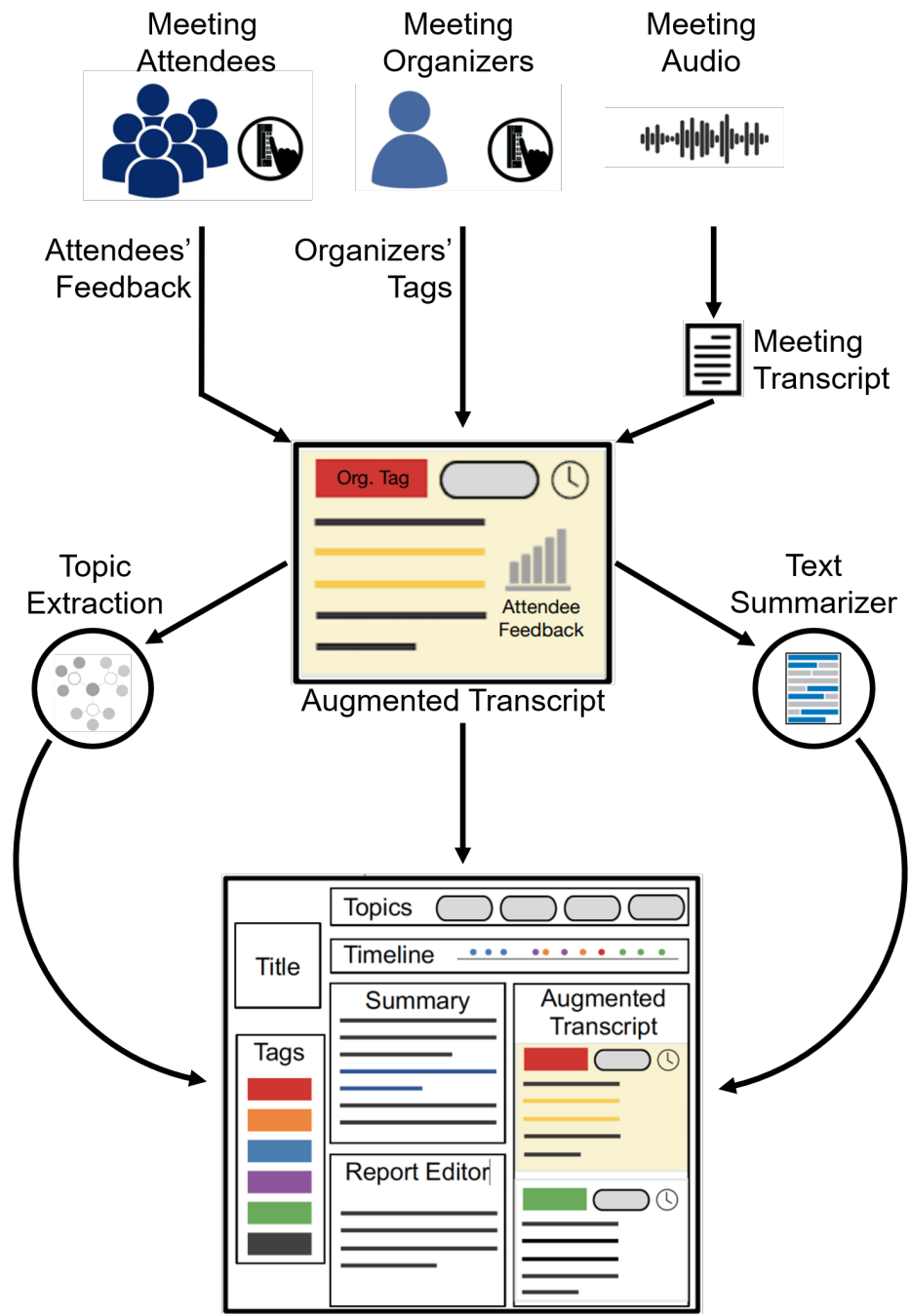

Figure 3.18: A snapshot of CommunityClick's workflow. During the meeting, attendees and organizers can use iClickers to share feedback and tag the meeting. The meeting is also audio-recorded. The recordings are transcribed automatically and then augmented with the organizer's tags and attendees' feedback. Furthermore, we generated the feedback-weighted discussion summary and extracted the most relevant topics. The interactive interface enables the exploration and utilization of augmented meeting discussions, which is available online for organizers to examine and author meeting reports.

In our recent work [103], we partnered with local government officials to design civic technologies to help them collect more inclusive public data from town halls. Inspired by the success of audience response in the education domain [153, 225], we decided to use iClickers but performed additional modifications to allow reticent participants to provide real-time feedback on ongoing discussions silently by clicking five customizable options on the iClickers. We integrated the attendees' feedback with the automatically generated meeting transcript and applied text analysis methods including topic modeling and a novel feedback-weighted text summarization for a more inclusive analysis process. We deployed this tool, CommunityClick, in a town hall and interviewed eight organizers to evaluate our approach. The results of the evaluation demonstrate that our approach has the potential to create an equitable platform by enabling silent attendees to share their opinions anonymously in real-time during town halls and allowing organizers to account for silent attendees' feedback in their reports. Figure 3.18 shows the workflow of our approach in capturing attendees' feedback and enabling analysis of such feedback. Despite the initial success of our approach, there were concerns around the use of iClickers as the mechanism for capturing silent attendees' feedback as it can create distractions and provide an unfair advantage to younger and tech-savvy attendees who might be more receptive of novel civic technologies in town halls. Furthermore, there are logistical overheads associated with the procurement and maintenance of such physical devices. In addition, while CommunityClick allows attendees' to share their opinion in real-time, they are still limited to using five options and cannot provide entirely open-ended textual comments.

### 3.12.3. COVID-19 Conundrum and Future Opportunities

During the early 2020, the COVID-19 pandemic originated from the SARS-CoV-2 or more colloquially termed Coronavirus forced the overwhelming majority of world's nations into lockdown, which still continues to several regions in the world to date. During this time, the overwhelming majority of the face-to-face public consultation methods including town halls came to a stand still due to the danger associated with public congregation and the risk of spreading the virus. As a result, in hopes of maintaining a functional participation of community members in local governance, the majority of the states in the United states and governments around the world sought after alternative solutions to hold town halls meetings. Many of them moved towards synchronous online virtual town hall meetings [3]. However, this major paradigm shift has rendered previous face-to-face in-situ civic technology requiring physical interactions, such as large interactive surfaces, wall interfaces, and even audience response systems such as clicker devices ineffective, since the discussions have moved online. As the world struggles to recover and find a new normal to adapt to this shift in daily life, new possibilities have opened up for innovation and development of civic technologies as research shows that virtual town halls and online engagement in general can attract and accommodate a larger audience previously disenfranchised by the challenges associated with face-to-face town halls [104, 216].

Despite being online, the virtual town halls often do not allow the virtual attendees to provide real-time feedback silently or anonymously as the organizers are often restricted in different ways by the tools used to arrange the virtual town halls. To learn more about these new challenges, we interviewed 3 organizers (O) who evaluated our CommunityClick tool [103]. One organizer (O1) mentioned, *"We use Zoom for our virtual town halls and the organization setup for these meetings have the chat options disabled. The only way for attendees to participate is to identify themselves and then speak up."* Another (O2) revealed even more constrained practices, saying *"During the discussions, all attendees are muted. They can raise their hands and an organizer will moderate the discussion by allowing people to contribute by letting them speak at specific times. However, these meetings take one hour or two hours and many don't get the chance to say what they want."* In essence, these virtual town halls have changed the mode of communication, but the challenges towards inclusivity including the inability to provide real-time and anonymous feedback still prevail as articulated by O3, *"Like regular town halls, we keep listening to a handful of people while the rest stay silent or just leave the meeting."*

To translate CommunityClick's success in regular town-halls to virtual town-halls and transcend the requirement of having physical iClickers, CommunityClick could be augmented with a virtual clicker developed as a web-application, which can be accessed online. Figure 3.19 shows a prototype that enables virtual attendees to share their opinion using different options similar to physical iClickers while attending the virtual town hall using the tool used by the organizers (Figure 3.19B). In addition to providing options to provide feedback on the ongoing virtual discussion, the attendees can provide further explanations regarding their choices by providing open-ended text comments (Figure 3.19C).

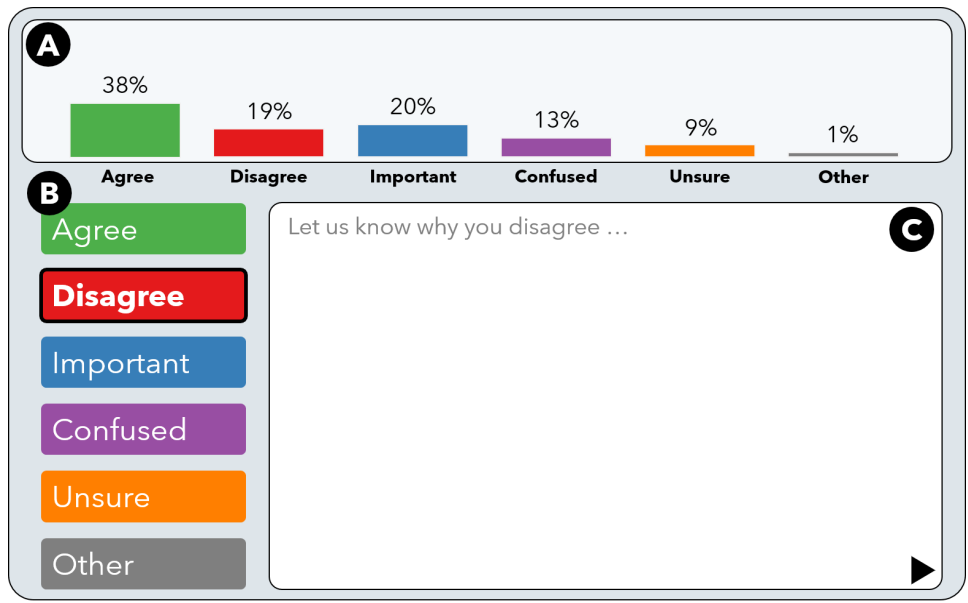

Figure 3.19: A prototype for CommunityClick's virtual clicker. A) The bar chart provides live statistics based on the feedback of all attendees during the meeting. B) A set of customizable options for attendees to provide feedback on the ongoing virtual town hall discussion. C) A textbox where attendees can provide open-ended text feedback according to the feedback option they selected. In this example, the attendee has chosen to disagree with the ongoing discussion and can provide justification if they choose to.

Furthermore, the prototype could also provide overall live statistics of how all the attendees have been responding to the meeting discussion which could lead to further engagement while keeping the attendees updated to the flow of the discussion (Figure 3.19A). In future, we intend to engage and further our partnership with organizers who are grappling with the new challenges around virtual town halls to collectively design, develop, and evaluate novel civic technologies that can address these challenges. Future civic technology researchers and designers could also explore ways to provide support for facilitating online synchronous group discussion by designing automatic or semi-automatic chatbots. Previous works have seen success in deploying chatbots to help facilitate goal-oriented synchronous online discussions [120], increase engagement [229] and promote attendees to contribute more [112].

As the world gradually recovers and returns to normalcy, the pandemic lockdown and rethinking of town halls could motivate the local government to open up to online virtual town halls as a complementary method of community consultation, given the recent success of engaging more attendees without the physical space restrictions. It is critical for future researchers to adapt to this new possible paradigm shift and rise to the challenge of designing and developing new civic technologies to facilitate more inclusive engagement and opinion sharing in town halls, virtual or otherwise.

# 3.13. Integrating Culture Theory and HCD to Design Global Government Interfaces

**Anqi Cao**
HCDE, University of Washington
Seattle, Washington, USA
*cc0051@mix.wvu.edu*

**Antonio Aranda-Eggermont**
HCDE, University of Washington
Seattle, Washington, USA

**Mojin Yu**
HCDE, University of Washington
Seattle, Washington, USA

**Patriya Wiesmann**
HCDE, University of Washington
Seattle, Washington, USA
*cc0051@mix.wvu.edu*

**Stephanie Blucker**
Civic Innovation Lab, UNAM
Mexico City, Mexico
*cc0051@mix.wvu.edu*

**Shazia Ansari**
Civic Innovation Lab, UNAM
Mexico City, Mexico

**Diego Flores**
Mexico's Ministry of Foreign Affairs
Mexico City, Mexico

**Fabian Medina**
Mexico's Ministry of Foreign Affairs
Mexico City, Mexico

**Saiph Savage**
Civic Innovation Lab, UNAM
Mexico City, Mexico
*saiph.savage@mail.wvu.edu*

*Governments are continually seeking new ways to connect with citizens, and a number of platforms have emerged. However, most of these platforms do not follow a human-centered design (HCD) approach. As a consequence, these applications are not always user-friendly, and might not be empathetic with users.*

*To address this problem, we propose a framework that integrates culture theory and human-centered design into the design of government applications. Culture theory helps to provide the necessary context and initial lens to study citizens living across different countries, in an inexpensive way. Human-centered design helps to understand them in greater depth, but combined with culture theory it can become much more inexpensive to integrate. We study our proposed framework within the context of designing a virtual assistant for Mexico's Ministry of Foreign Affairs. The goal of the virtual assistant is to be usable and appropriate for both local Mexican citizens and Mexican citizens living abroad. With our framework, we propose an approach to design more culture- and user-friendly interfaces for governments across the globe.*

## 3.13.1. Introduction

A number of government applications have emerged to deliver services to citizens. However, most of these interfaces do not follow a human-centered design approach. Many times, it is because governments lack the infrastructure and resources to thoroughly conduct mixed-method user experience researches such as usability testing and interviews to make informed design decisions from a human-centered perspective; they might also have personnel that are not familiar with human-centered design, and hence not able to implement it. Furthermore, when thinking about human-centered design, a critical part that we need to take into account is how the design adapts to the needs and context of users. For government platforms it is particularly important to design interfaces that are adapted to the needs of the citizens. We argue that it is crucial to integrate culture theory into the research process, so that the governments can provide their citizens with adequate services.

In this work we present how to integrate human-centered design and culture theory into the creation of government interfaces. We present our framework within the context of designing a virtual assistant for the Federal Government of Mexico, specifically Mexico's Ministry of Foreign Affairs. Our framework provides governments across the globe with an effective way in which they can design new usable technological solutions for their citizens that are adapted to their culture, local context, and needs.

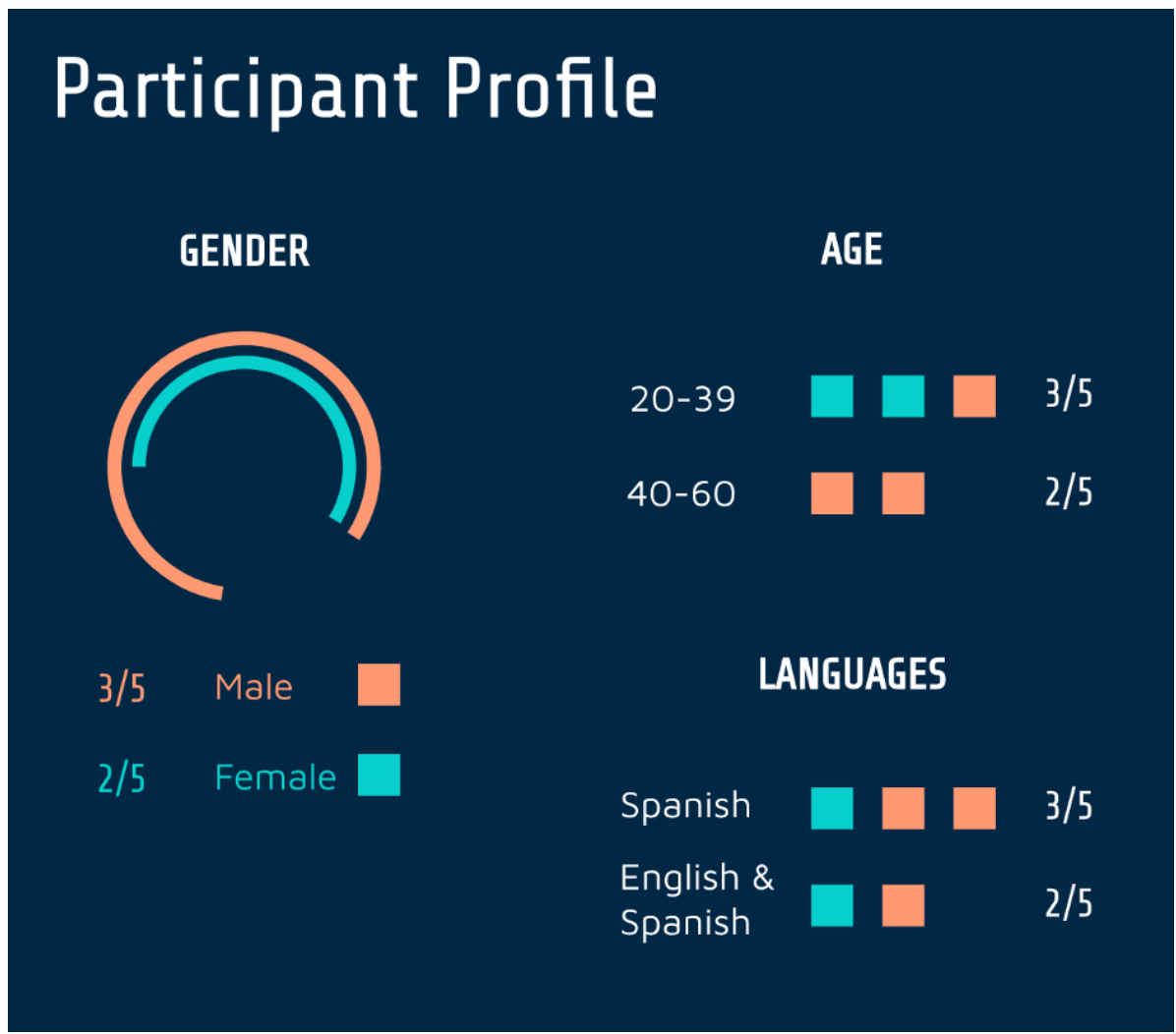

Figure 3.20: Participant Profiles

### 3.13.2. Framework: Combining Culture Theory and HCD

We argue that to design effective government interfaces we need to combine Human-Centered Design and Culture Theory. For this purpose we present our framework that combines the two to guide the creation of usable government interfaces.

Using Culture Theory to Drive the Initial Design
To effectively design a government product, understanding the citizens and their experience is critical. For this purpose, we use the culture theory of Geert Hofstede who has characterized countries and their citizens based on different metrics.

**Culture Theory | Hofstede's Cultural Theory**   Hofstede's cultural theory offered some valuable insights for our team regarding what design affordances might make sense for this experience.
First, Mexico's high power distance and low long-term orientation scores indicate an established hierarchy and system of rules that is not often questioned. This could mean, for example, that many people trust the government with their information or personal data, and respect governmental requirements.

Second, according to Hofstede's model, Mexico has a low score on individualism, meaning the culture prizes relationships especially those of family or extended family groups. Because of this, pursuing a supportive, personal, and conversational experience like the one the government had envisioned seems like a good way to improve the relationship between the government and its citizens. Third, Mexico scores high on masculinity, meaning that it's a work-driven society that values decisiveness and assertiveness. Combined with Mexico's high uncertainty avoidance score, it indicates that citizens might be frustrated with a process that lacks clarity and an unsure outcome.

There are two other important cultural theories that impact design: high vs. low context communication and polychronic vs. monochronic. Mexico is both high context and polychronic, meaning that personalization is important and that people may prefer to multitask as opposed to focus on one thing at a time.

Based on the Hofstede theory we argue that an initial interface to explore is a conversational and web experience that blends into one interaction that feels guided, informative, and friendly.

Using HCD to Improve and Iterate the Interface

Once an initial design based on culture theory is created, we argue that we can use human-centered design to drive the details of that interface.

Human-centered design is an approach to develop products, services, and interactive systems that are focused on user needs. It involves deeply getting to know the people who will use the design: how they think, how they'll feel when they're using it, and what they're trying to achieve.

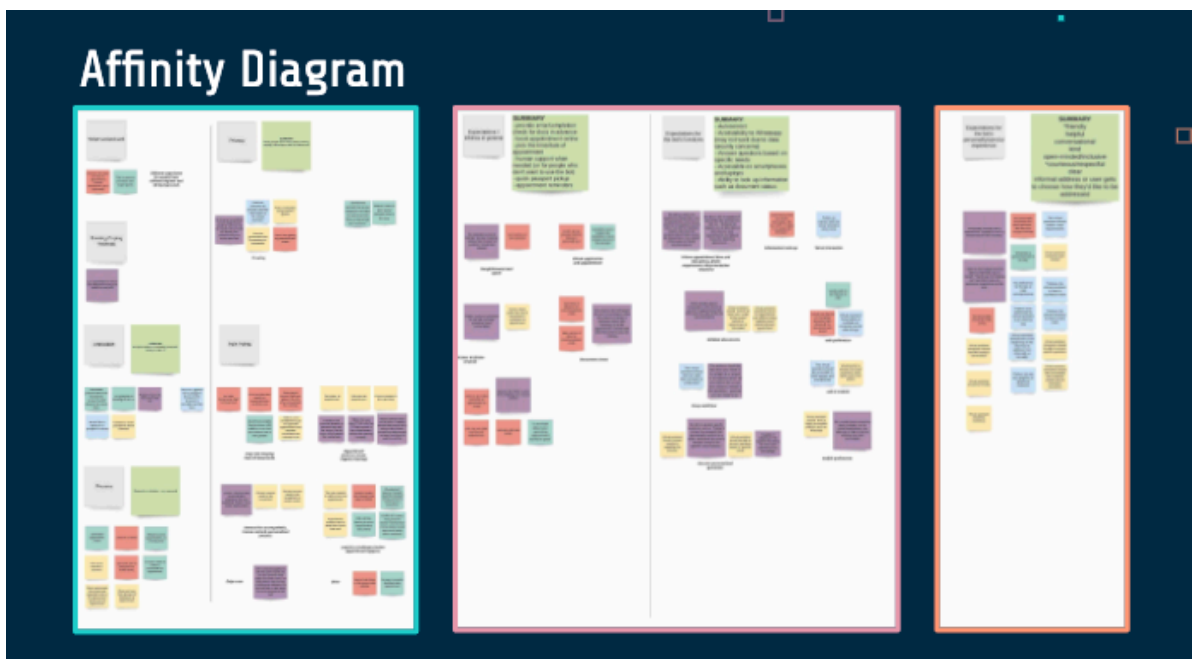

Figure 3.21: Affinity Diagram showing different themes

For this purpose, it is important to conduct interviews. In the case of designing an interface for the government of Mexico, we gathered insights from real people through qualitative research. We chose a semi-structured interview method as it allows for more open-ended responses while retaining some planning, which allowed us to gather more in-depth information on the participants' experiences with passport renewal in Mexico.

The interviews were conducted remotely via Zoom. A moderator led each session in collaboration with a translator and a note taker. The participants were first invited to share background information about themselves, and then guided through an open discussion on topics such as their last passport renewal experience, their evaluations of the process, and their thoughts on virtual assistants and chatbots.

**Human Centered Design | Participant Profile** A total of five interview participants were recruited through snowball sampling. Our interviews included Mexican citizens who were living abroad and

Mexican citizens living locally in Mexico. We aimed to recruit these broad range of Mexican citizens as we wanted to understand the perspectives of Mexicans who would interact with the virtual assistant (both locals and those living abroad.)

**Human Centered Design: Data analysis**   We collected qualitative interview data in two languages (Spanish and English) from our five participants. To familiarize ourselves with the data, Antonio (native Spanish speaker on the team) translated the Spanish transcript into English so that the English speakers on the team could interpret the interview data. After a thorough review of the text and audio recordings, we extracted a series of thoughts, suggestions, and feelings from the participants' responses. These snippets of data were then group by themes and visualized into an affinity diagram.

### 3.13.3. Findings

**Trust in authority:**  Overall, 4/5 of the participants said they trusted the government with their personal information, and that they would not have any issues with supplying the bot with any personal information it requested.  This is supported by culture theory which suggest that Mexicans trust and follow the government that they are part of a hierarchical society.

**What's difficult about the current experience:** People were overwhelmed by documentation and wanted help keeping track of what documents were needed.  The top pain point we identified was around errors in documentation. Participants said these errors were often identified over the course of several in-person visits, which made the passport renewal process frustrating, stressful, and inefficient. They also wanted empathetic individualized service, since different documents may be required based on specific circumstances.

**Expectations of the passport renewal experience:**  The high uncertainty avoidance in Mexican society manifests in the unwillingness of Mexican citizens to continue experiencing anxiety and ambiguity in the current passport application experience.  People would rather avoid all uncertainty and be very clear as to what documentation is required or any prior procedures before going for a passport appointment.  This means that passport application rules need to be set at the beginning of the process to avoid deviation and confusion.  Because of the frustration around documentation, people wanted a way to check their documents beforehand to make sure they are complete and error-free. They also wanted to be able to book an appointment online, where they could pick the time and date of their appointment followed by an appointment reminder.  Moreover, people preferred to have access to human support when needed.

**Expectations of the bot's functionalities:** People wanted to be able to access the bot on both web and mobile devices. Some participants specifically mentioned WhatsApp. People also wanted the bot to be able to answer questions based on their specific needs/situation, and to be able to look up their info, if possible.

**Bot personality:**  Participants expressed that they would like to interact with a bot that is friendly, courteous, helpful, conversational, inclusive, and clear.

### 3.13.4. Future Work and Conclusion

We argue that our framework might be especially beneficial for governments in developing countries where they might not have the budget to conduct extensive human-centered design studies (using culture theory to guide the design might be helpful and impactful).  This framework could also help governments who want to better understand their citizens who are abroad, but they might not have the budget to conduct excessive interviews with participants. Using the lens of culture theory could be an initial step to drive their design.

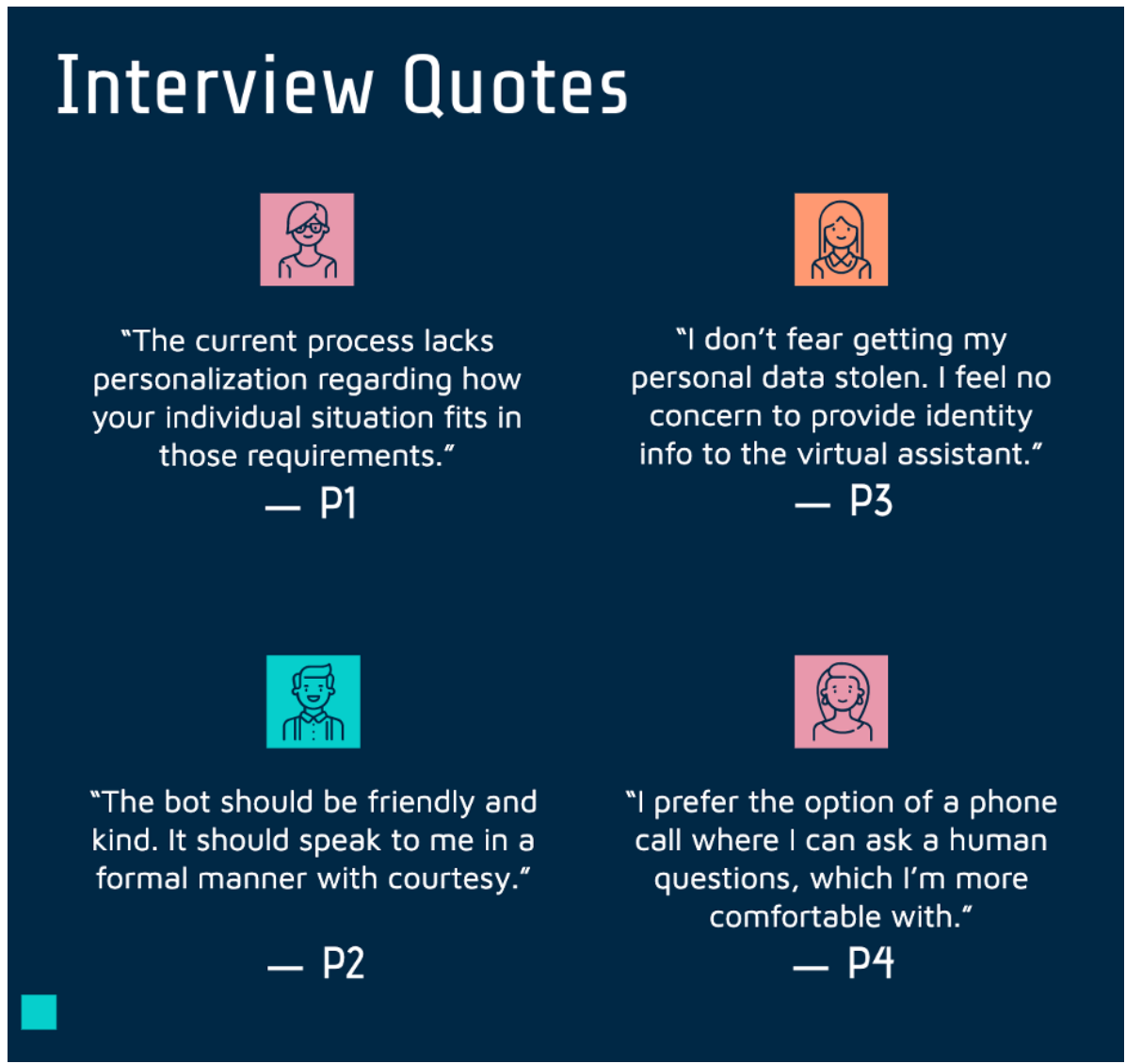

Figure 3.22: Salient Quotes

### 3.13.5. Limitations

One limitation of the Hofstede model is that it relies largely on generalizations. It is likely that the model we referenced does not account for indigenous populations in Mexico. Given more time, we would have liked to interview more people due to the complex nature of passport application and renewal. Additionally, we did not have the opportunity to interview the participants in person - valuable contextual cues might be missing and could have limited our understanding of their experience.

## 3.14. Local Solutions with Global Reach — Can Civic Tech Benefit from Open Source Software Ecosystem Practises?

**Antti Knutas**
LUT University
Lappeenranta, Finland
*antti.knutas@lut.fi*

**Victoria Palacin**
LUT University
Lappeenranta, Finland
*victoria.palacin@lut.fi*

**Annika Wolff**
LUT University
Lappeenranta, Finland
*annika.wolff@lut.fi*

**Sami Hyrynsalmi**
LUT University
Lahti, Finland
*sami.hyrynsalmi@lut.fi*

This position paper identifies benefits of using open source ecosystem practices within civic tech projects, the barriers against it, and offers some technical solutions that could address some of these barriers. We also lay the foundation for looking into less tangible aspects such as mutual benefits between the communities and cross community learning.

### 3.14.1. Introduction

Civic technology refers to the diverse ways in which people are using technology to influence change in society [24, 113, 208]. There are a variety of creators of civic tech, ranging from commercial actors, governments, non-profits, volunteer organizations, and loosely organized communities [200]. These creators vary in purpose [192] and in how they identify themselves as practitioners [47] [18]. What is common in all of these projects, is that they address a societal need identified by the public, or together with the public. For example, the need to identify air pollution, increased transparency, or participatory governance.

Civic tech projects are often partly or fully driven by volunteers, they might lack involved technologists, and the people invested in solving social issues are not always well resourced. Due to this, it is important that software supporting civic technology would be easily available and shared by successful civic tech projects. This can be a challenge due to the fact that civic tech solutions are created together with or bespoke for the community, on the other hand empowering that specific community, but at the same time making them more difficult to share and customize. Furthermore, making software more adaptable requires more effort and resources, and that is rarely the main goal.

Several civic tech groups already consider Free/Libre and open source software important, seeing its values to be consistent with the goals of equity, justice, transparency, and sharing instead of competition [47, 206]. Despite this, poorly resourced projects might turn to commercial or closed software, which is easy to take into use, but at the same time lacks community accountability and transparency, and might ultimately work at cross-purposes with the positive social change the community seeks.

Currently civic tech groups share certain types of resources and knowledge, such as best practises and processes for better co-design or equability. We propose that examining civic technology projects through the lens of open source software ecosystems, could bring additional value in the form of thinking of new ways and processes to efficiently share software solutions, without losing the values that are central to civic tech.

In this position paper, we first briefly review the concept of open source software ecosystems, present different artefacts and methods that those ecosystems share, and then discuss how those methods could be harnessed by civic tech projects in future research.

### 3.14.2. Open Source Software Ecosystems

Open source software emerges from a loosely coordinated, unsupervised community of volunteering developers and other contributors to address a specific need [74]. If an open source software community grows, an ecosystem may grow around it.

Open source software ecosystems (OSSECO) have two fundamental factors: network of organizations or actors and a common interest in a central software technology [150] or a shared market for software and services [102]. OSSECO in turn can be defined as a *"a software ecosystem placed in a heteroge-*

---

[18]Terms that the groups identify as include free software, digital literacy, community technology, and inclusive design.

*nous environment, whose boundary is a set of niche players and whose keystone player is an open
source software community around a set of projects in an open-common platform"* [74, p.24]. A review
by Franco et al. [74] lists several characteristics unique to OSSECO, which include software distribu-
tion paradigms including source code and repositories, license schema facilitating the relationship of
keystone players (OSS community) and niche players (partners, providers, adopters), and the OSS
community dominating the development rather than an individual organization. Example OSSECOs
include for instance the Debian Linux operating system[19], or the Jitsi Meet call platform[20].

What is similar in OSSECOs and civic tech groups is that both have a community as the key player.
What is dissimilar is that OSSECOs centre the software and are formed of a decentralised network that
form an online community, whereas civic tech centres the problem (rather than the solution mechanism,
which may or may not entirely be tech related) and the community is at least more likely to be primarily
physical.

Some examples already exist at the intersection of OESSECO and civic tech. For example, it could be
argued that for example Luftdaten[21] is both civic tech movement for cleaner air and an open source
software ecosystem around a citizen sensing platform developing open source software for both the
platform and diverse measurement devices. Luftdaten also provides other open resources, such as an
open data platform.

### 3.14.3. Discussing Opportunities at the Intersection of Civic Tech and Open Source Software Ecosystems

In this section, we present four discussion points on OSSECO practises and relate them to civic tech.

- **Generalizing and sharing common components or services.** Current software engineering
  practises allow modular architecture and sharing software components through technologies such
  as micro-architecture design and containerization. This would allow sharing underlying software
  components without compromising co-designed functionality.

- **Providing community-controlled deployment options through niche players.** In OSSECOs,
  niche players can enhance resources by providing a better user experience. For example, various
  Linux distributions provide a graphical way to install open source software from repositories. The
  Alphabet company provides an automated way to install open and secure networking software
  to a server of user's choosing. Better deployment and configuration features would reduce the
  need for technical expertise, while allowing the community to retain control.

- **Supporting capacity building and resilient solutions.** Currently most resources and attention
  to go new tools, despite there being a need for resilient solutions [47]. If communities center-
  ing on maintenance, upgrades and support were supported better, it would help making more
  sustainable solutions.

- **Managing and cultivating the ecosystem.** Many successful civic tech projects acknowledge
  the community as a central actor and center its needs. Similarly, volunteers and community actors
  in software development require support. In software ecosystems, OSSECOs try to monitor the
  ecosystem health and support the ecosystems through diverse methods.

Lastly, we differentiate between the technical aspects that facilitate sharing, such as the use of open
source practises within civic tech communities, and other less tangible methods of sharing. These other
aspects, such as how open source projects could learn from civic tech's equitable design practises or
addressing technology biases, are important but out of scope in this particular paper. We propose that
these topics should be addressed in future research.

---

[19]https://www.debian.org
[20]https://jitsi.org/jitsi-meet
[21]https://luftdaten.info/

# 3.15. Perceptions of News Sharing and Fake News in Singapore

**Gionnieve Lim**
Singapore University of Technology and Design
Singapore
*gionnievelim@gmail.com*

**Simon T. Perrault**
Singapore University of Technology and Design
Singapore
*perrault.simon@gmail.com*

Fake news is a prevalent problem that can undermine citizen engagement and become an obstacle to the goals of civic tech. To understand consumers' reactions and actions towards fake news, and their trust in various news media, we conducted a survey in Singapore. We found that fake news stem largely from instant messaging apps and social media, and that the problem of fake news was attributed more to its sharing than to its creation. Verification of news was done mainly by using a search engine to check and cross-reference the news. Amongst the top three sources to obtain news, there was low trust reported in social media, high trust in local news channels, and highest trust in government communication platforms. The strong trust in government communication platforms suggests that top-down civic tech initiatives may have great potential to effectively manage fake news and promote citizen engagement in Singapore.

## 3.15.1. Introduction

Fake news has gained increasing global attention since the 2016 US presidential election as the circulation of fake news on online platforms, such as social media, saw an explosive increase [5]. This phenomenon is largely motivated by political and financial gains and has imposed social costs as consumers' understanding and interpretation of real events are confounded [5, 59]. Our study adopts the definition of fake news as "either wholly false or containing deliberately misleading elements incorporated within its content or context" [17]. We aim to understand the perceptions of news sharing and fake news in Singapore that, with its unique socio-political scene, serves as a case study that provides interesting insights into the fake news phenomenon.

## 3.15.2. Background and Related Work

**Fake News and Civic Tech.** Fake news and citizen engagement have an adversarial relationship where "informational uses of the Internet are positively related to … social capital, whereas social-recreational uses are negatively related" [196]. The consequence of fake news in digital media is that of social costs, where consumers may make poorly justified and unwarranted choices based on inaccurate knowledge [5, 59]. The proliferation of fake news hinders civic efforts as it compromises the basic unit of exchange involved in deliberation - that is, information [147]. This puts it at odds with civic tech, which are digital technologies that promote civic efforts. It is this framing of fake news and civic tech that motivates our study.

**Fake News in Singapore.** Singapore is an attractive and vulnerable target to the deliberate spread of online falsehoods as an open and globally connected country, and a multi-racial and religiously diverse society[22]. On 8 May 2019, the Protection from Online Falsehoods and Manipulation Act (POFMA) was passed to address fake news. The rising attention of fake news in the local scene has motivated various research on the topic. Related studies look into the perceptions and motivations of fake news sharing [35] and responses to fake news [214]. While these studies draw parallels to ours, we note that they focus on fake news found in social media, whereas our study investigates fake news in general media. Furthermore, we investigate the broad perceptions of both news sharing and fake news.

## 3.15.3. Survey Method

We created an online survey on Google Forms to collect data on the perceptions of news sharing and fake news in Singapore. In the survey, we asked respondents to share about their sources of news, their levels of trust in these sources and their familiarity with fake news. The media items considered include television, radio, word-of-mouth, local news channels, global news channels, streaming networks, social media, instant messaging apps, email, government communication platforms, and work communication platforms. The survey included 24 question items of which three were demographic questions and three were branching questions. The branching questions directed respondents to the next relevant section based on their answer. While this may result in skipped section items, each answered section is assured to be based on the respondent's experience and not their estimation, which enables us to obtain more accurate self-reported data.

---

[22]https://www.nas.gov.sg/archivesonline/government_records/record-details/
6797717d-f25b-11e7-bafc-001a4a5ba61b

| Category   | Number |
|------------|--------|
| Male       | 47     |
| Female     | 28     |
| 18-24 y/o  | 48     |
| 25-34 y/o  | 17     |
| 35-44 y/o  | 7      |
| >45 y/o    | 3      |
| Students   | 58     |
| Employed   | 13     |
| Others     | 4      |

Table 3.8: Summary of the participant demographics.

Procedure
The survey was disseminated to a local Singapore university through a mailing list that included undergraduate and graduate students. It was also disseminated to students, alumni, and faculty that were part of the university's informal Telegram supergroup, and further shared to personal contacts of the researchers through instant messaging and social media. The survey was made available to anyone with the hyperlink and was fully voluntary. In total, 104 responses were received.

Participants
Of the 104 responses, only 75 responses had respondents who were currently based in Singapore and were thus included in the data analysis. The details on the participant demographics are provided in Table 3.8.

### 3.15.4. Findings
**News Sharing Behavior.**   Out of 75 respondents, 59 reported having shared news online, and 57 reported having come across fake news. Out of the latter 57 respondents, 15 reported having shared fake news online before knowing they were fake.

**Trust in News Sources.**   Based on the 75 responses, the top three sources in which respondents obtained news were: social media (N=54, 72%), local news channels (N=44, 58.7%) and government communication platforms (N=43, 57.3%). Respondents then reported their level of trust for the 11 media items (see Figure 3.23) on a 1-5 Likert scale (1: strongly distrust, 5: strongly trust). We found a significant main effect of media on level of trust $\chi^2(10) = 338, p < .0001$. Interestingly, despite the popularity of social media as a news source, respondents tended to have a low level of trust in them ($M = 2.73$), while we observed a higher trust in local news channels ($M = 3.77$) and government communication platforms ($M = 4.17$). Instant messaging apps were considered the least trustworthy ($M = 2.15$).

**Motivations for Sharing News.**   Based on the branch of 59 responses, respondents reported the top three motivations on news sharing as: the news is relevant to the receiver (N=53, 89.8%), the news is important (N=49, 83.1%), and the news is interesting (N=49, 83.1%). News was shared largely to friends (N=55, 93.2%) and family (N=39, 66.1%) and the most popular mode of sharing was through instant messaging apps, including WhatsApp (N=47, 79.7%) and Telegram (N=37, 62.7%). Most respondents practiced good news sharing habits by either always (N=15, 25.4%) or frequently (N=28, 47.5%) verifying news before sharing. They do so mainly by using a search engine (N=53, 89.8%) and by checking official government channels (N=52, 88.1%).

**Fake News.** Based on the branch of 57 responses, the top two sources of fake news were reported to be instant messaging apps (N=44, 77.2%) and social media (N=40, 70.2%). Most respondents realised that the news was fake either immediately (N=19, 33.3%) or within the day (N=24, 42.1%) and the most popular method they employed to ascertain the falsity of the news was to use a search engine (N=49, 86.0%). The main action they took upon knowing that the news was fake was to send a correction message to those whom they have shared the news with (N=29, 50.9%). In other cases, they also either sent a warning message to inform others of the presence of the fake news (N=23, 40.4%) or simply did not do anything (N=25, 43.9%).

### 3.15.5. Discussion

In agreement with other studies, both instant messaging and social media are considered the main sources of fake news. That most respondents encounter fake news in instant messaging apps and that they rate the sharing of news as a greater problem suggest that communication with personal contacts, such as widely-circulated chain messages, rather than with the public media, such as social media feeds, is the larger issue in Singapore. There is opportunity here for digital literacy education to educate Singapore residents on identifying fake news and the ways to handle them. There is also opportunity to design interventions in instant messaging apps to flag and reduce mass sharing of fake news. While such methods are being considered or have already been implemented in Singapore, such as fake news awareness campaigns, there remains more to be explored in targeting vulnerable demographic groups and at a larger scale.

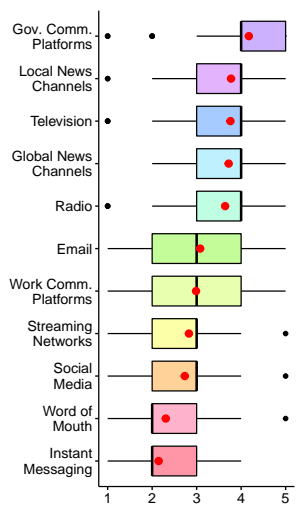

Figure 3.23: Reported levels of trust for different media items on a 1-5 Likert scale (1: strongly distrust, 5: strongly trust). Red dots indicate the average score.

The most interesting finding is that of the high level of trust and reliance on government communication platforms, e.g. government websites and hotlines, to provide truthful news and to debunk fake news. This signifies the relatively strong faith that Singapore residents have in the Singapore Government, and may be attributed to the successful ongoing efforts in making transparent government decisions and the readiness of the government in addressing public queries. This suggests a compelling opportunity for top-down civic tech initiatives that involve the government to drive civic efforts and citizen engagement amidst the sea of fake news.

### 3.15.6. Conclusion

Despite the growing awareness of fake news in Singapore, it remains an extensive problem. While the study found that most respondents are digitally literate and exercise good news sharing practices such as verifying news before sharing or sending a correction message upon knowing that they have shared a fake news, the mitigation mechanisms to identify fake news or to reduce the spread of them are underdeveloped. We believe that given the problem fake news poses to citizen engagement, efforts in addressing fake news, such as with civic tech, will be worthwhile.

# 3.16. Renová el Parque Caballero: A multi-modal civic engagement platform to revitalize an iconic public park of Asuncion

**Jorge Saldivar**
Barcelona Supercomputing Center (BSC)
Barcelona, Spain
*jorge.saldivar@bsc.es*

**Cristhian Parra**
United Nations Development Program
Asunción, Paraguay
*cristhian.parra@undp.org*

**Luca Cernuzzi**
Catholic University Nuestra Señora de la Asunción
Asunción, Paraguay
*lcernuzz@uc.edu.py*

**Luis Godoy**
Oficina Comunitaria de Arquitectura
Asunción, Paraguay
*luis@oca.com.py*

*This article reports on a multi-modal participatory process to propose, discuss and select restoration projects for a public park in Asuncion, Paraguay, and how it promoted the formation of a community of activists that remains active to this day*

## 3.16.1. Introduction

Democracy in Paraguay has developed over the past 30 years within a socio-cultural context of low participation and engagement of its citizens [62, 180]. Indifference and distrust in public institutions influence the perceived apathy of citizens to get involved in public affairs [140].

However, the pervasiveness of today's technology, especially smartphones and social media, is changing this by equipping Paraguayans with a variety of tools that facilitate engagement, from allowing access to public information to coordinating mobilizations [191]. Still, the use of technology for participatory democratic deliberation and decision making has been scarce.

In this paper, we discuss lessons learned from one experience designing and implementing a participatory process that combined digital platforms and offline deliberation spaces to revitalize an iconic public park in the capital of Paraguay, Asuncion.

## 3.16.2. The Caballero Park

Built-in the 1920's close to the city center of Asuncion, the park General Bernardino Caballero has been for decades a place of social recreation, sports, and cultural events for residents of Asuncion. In its splendor, this park of almost 20 hectares served as the epicenter of school class trips, family gatherings, civic and sports encounters.

The former president of Paraguay, Juan Natalicio González, described the park facilities in the decade of 30s with these words *"The Caballero Park offers landscapes of rare attractiveness. The fresh breeze from the river runs through its murmuring avenues. A coffee plantation bloodies of fruits in the fall. Yerba mate trees form artificial mounds. The strong lapachos with pink and yellow flowers are born along the routes. The water sleeps in the ponds and sings in the waterfalls, while some hieratical and long-legged bird meditates on its banks"* [85].

For generations of *Asuncenos* (citizens of Asuncion), the Caballero Park represents more than just a park is the iconic and historical place where they spent unforgettable moments of infancy and adolescence in the company of their lovely ones. It was at the beginning of the 90s when the situation of the park started to decline. The negligence and indolence of the several municipal administrations of different political signs destroyed all of the cultural, historical, and social legacies that the park symbolizes for the Asuncion's population. Figure 3.24 contrasts the current situation of the park with its appearance in its golden times.

Inspired and motivated by this place's historical value, we set up a participatory process where we invited the citizenry of Asuncion to discuss ideas and present proposals on how the Caballero Park can be revitalized to become a landmark in the Asuncion's social life again.

## 3.16.3. The Multi-Modal Participatory Platform

To achieve this goal, we set up a multi-modal civic engagement platform that combines online and onsite participatory channels. For the four months of experience (from April to August 2018), the process was structured in sequential activities that facilitate the participation: ideation, deliberation, refinement, and voting.

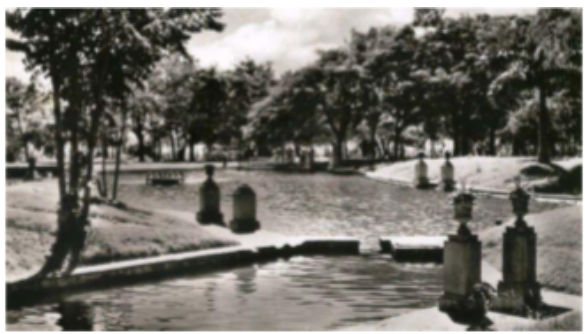

(a) The Caballero Park in its apogee

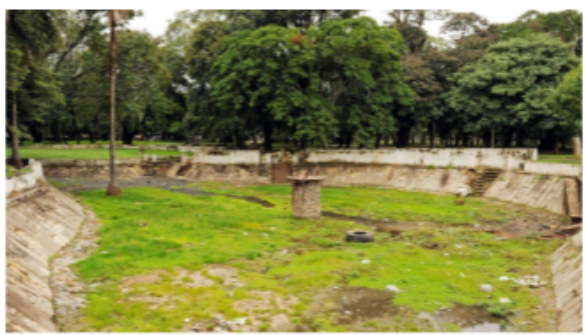

(b) The Caballero Park today

Figure 3.24: Appearance of the Caballero Park today and in the past. Photo: ABC Color

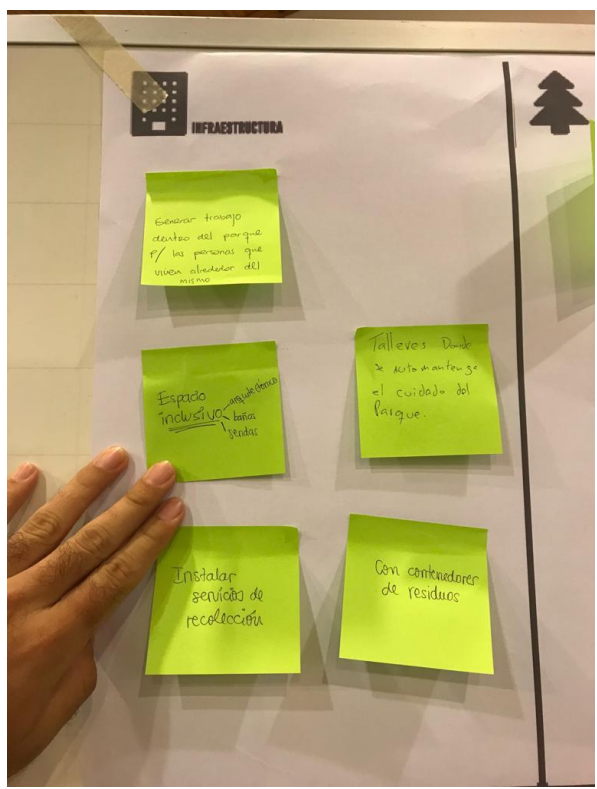

Figure 3.25: On-site ideation workshop

First of all, we asked the participants to focus their contributions on proposals to recover the infrastructure, landscape, nature, safety, sustainability, sport, culture and history, integration with the city, and education of the park.

In the ideation stage, citizens were invited to post ideas and opinions about revitalizing the Caballero Park. Ideas could be proposed through a Facebook page and via the participatory platform AppCivist. A face-to-face event was also organized in the park's vicinity to collect ideas and proposals from the neighbors on how the park can be recovered (see Figure 3.25).

At the deliberation stage, a deliberative workshop was organized. During the workshop, a set of pre-selected ideas were presented, and the participants had the opportunity to give feedback about the strengths and limitations of proposals.

In the refinement phase, the deliberated ideas were presented to the public on the AppCivist platform (see Figure 3.26). Authors of the ideas were invited to use the AppCivist's collaborative editing tool to expand ideas, turning them into proposals. They were instructed to consider, whenever possible, the feedback collected on the online platform and in the deliberative workshop.

The process finished with the voting phase. Here, the participants were invited to express their support or disapproval of the refined proposals. Voting was conducted both online, through the AppCivist platform, and in-person. We adopted a 1 to 10 scale for voting, and the participants were asked to evaluate the need, benefits, and feasibility of the proposals. During this phase, neighbors of the park were invited to a voting session celebrated in one of the park's dependencies. They were instructed to indicate with post-its their favorite ideas (see Figure 3.28). The top-five most voted proposals were presented to the municipality of Asuncion to be included in a recovery plan for the park.

### 3.16.4. Results

Almost 80% of the ideas during the ideation phased were shared through Facebook. A Facebook page was used to campaign the initiative. The participants primarily used the comment subsection of the promotion posts to left their opinions and ideas. In total, 346 people participated using both Facebook

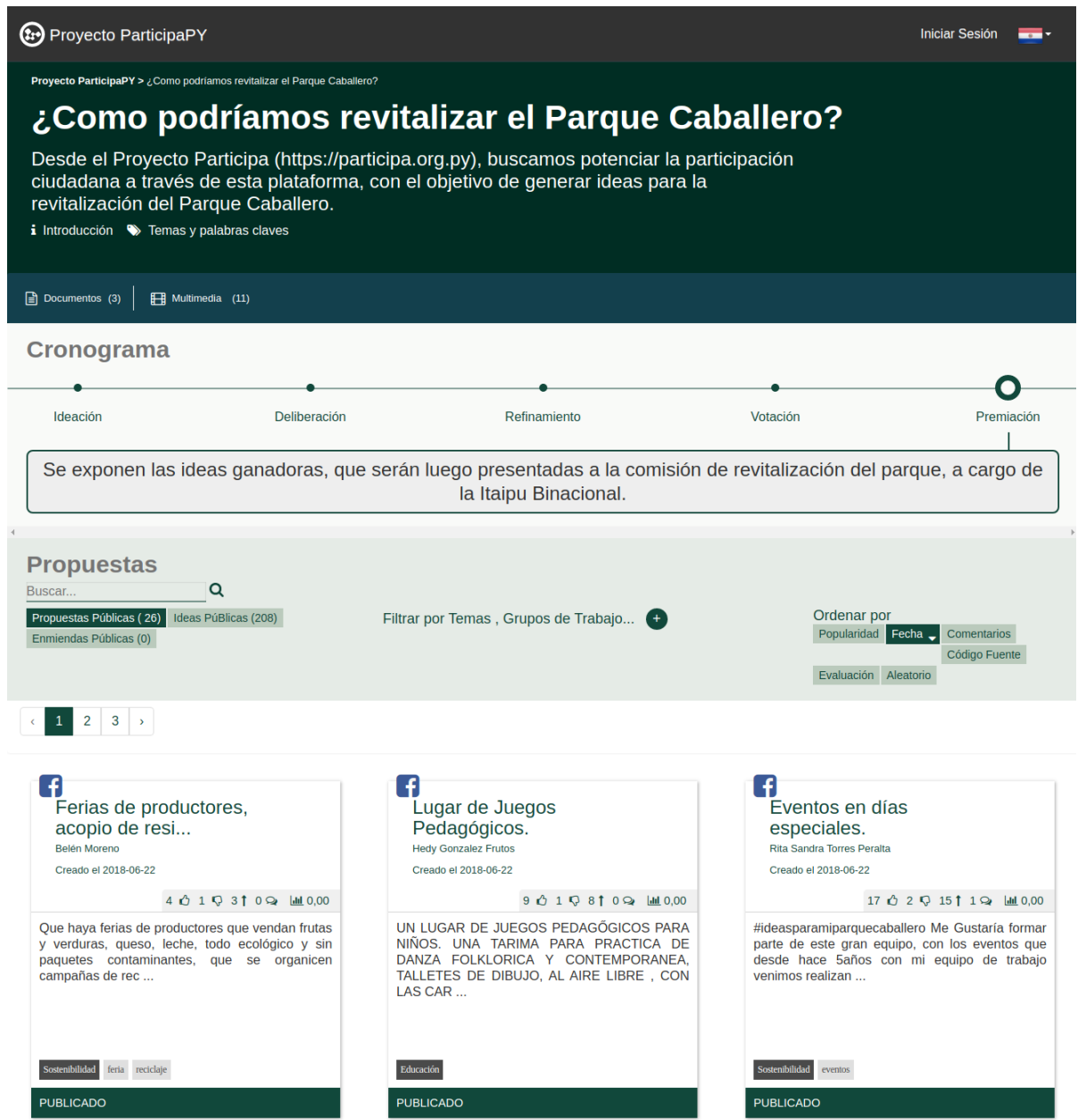

Figure 3.26: The AppCivist platform employed for ideation, proposal refinement, and voting. The Facebook icon indicates that the idea was originated on Facebook and replicated on AppCivist

and AppCivist. They generated 212 ideas, 159 comments, and 404 votes.

Almost half of the participants submitted ideas (44%, 154 out of 346) on renovating the Caballero Park. Less than 15% of the 346 contributors focused their participation only on voting for the ideas and opinions, while 7% posted opinions on other participants' ideas and comments.

Security was the topic most commented on by the participants, 35% (75 out of 212) of the ideas were related to this issue, as shown in Figure 3.29. The participants' interest in the park's safety is not surprising since the Caballero Park ceased to be a space for recreation and citizen gathering mainly because of severe security problems; from some time now, robberies and assaults were registered more and more frequently in the place. Notably, the most voted idea was not related to the most discussed topic but to promote sports activities in the park. Along this line, the idea that concentrated most of the comments was associated with recovering the park's nature by reforesting native trees.

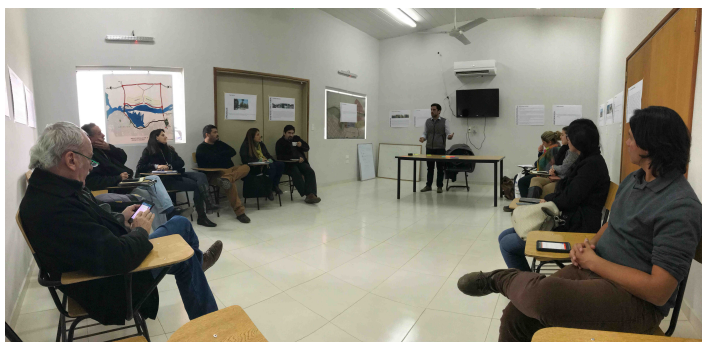

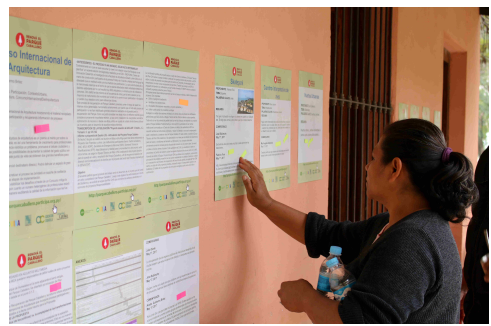

Figure 3.28: Face-to-face voting event celebrated in the dependencies of the park

Figure 3.27: Deliberative workshop

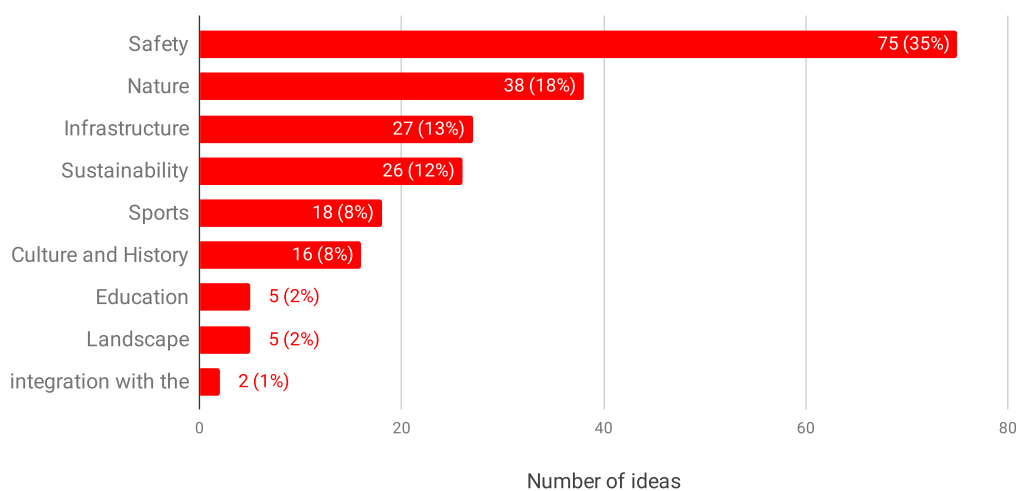

Figure 3.29: Distribution of ideas by topic.

**Profile of the participants.**   About 10% of the 346 participants filled out a post-experience survey, which inquired about their demographic characteristics and motivations to participate. The group of participants was divided between men and women in a ratio of 70-30, respectively. The average age was 43 years. Almost half of the survey respondents stated that they were attracted by the possibility of improving the park's situation. One-third of the respondents manifested that they were motivated by civic duty and the opportunity to contribute to a collective effort. In line with the patterns of participation, almost 80% of respondents manifested that they used Facebook to contribute to the initiative. Among the reasons they argued are: ease of use, familiarity, daily use, and little effort to keep abreast of process updates.

For the deliberation phase, the 203 ideas generated on Facebook were uploaded to AppCivist. Four ideas already existed in AppCivist, so the site was loaded with 207 ideas collected during the ideation phase. For a week, the ideas remained in AppCivist available to receive comments or votes. At the end of this period, the deliberative workshop was held to discuss a set of ideas, which were selected based on a combination of the following criteria: i) the number of comments received on AppCivist by the participants, ii) the engagement of the ideas' authors during this part of the process; iii) the level of detail contained in the idea, in relation to other ideas of the same topic. We also made sure that all topics were included in the selection. As a result, 26 out of the 207 ideas were selected.

The workshop was attended by 11 people, including four public servants, three authors of ideas, and four citizens of Asuncion who got to know about the event and decided to attend it. The authors of this paper facilitated the workshop. Figure 3.27 shows the participants of the workshop during the reflective moments. The participants were asked to contribute with opinions about the strength and limitations of the ideas. They were also required to provide feedback on how ideas can be improved. The goal was to use the workshop's inputs to guide the refinement phase later down the road.

After the refinement phase, the five most voted proposals were: i) organize in the park playful events on weekends; ii) build facilities for skating in the park; iii) improve the entrance area of the park; iv) organize public events on special days (e.g., holidays, national day); v) recover the pools and pond.

### 3.16.5. Discussion

*Renová el Parque Caballero* was one of the first participatory processes in Asuncion that used a mix of digital and physical participation. The resulting multi-modal posed challenges for organizers. That is to say, combining and analyzing multiple input sources while also enabling a highly engaging process that outlived its initial purpose through a new community of local activists who remained committed to the cause restating this park to its past glory.

Although not at the scale of other participatory processes, for the context of this community, the theme, and the relatively low amount of effort we invested in communication, the initiative's reach was surprising. Key to this reach was the use of Facebook as one of the means to collect ideas and comments, and the role of a communicator who actively engaged with the participants. A broad user base and good support for smartphones are two factors that make Facebook an interesting platform for scaling the audience of a participative process.

The unstructured nature of Facebook posts and comments, however, introduced the need for data extraction, transformation, load, and analysis before we could engage in productive deliberation and proposal development. The need to develop data processing and analysis capacities (both in human resources and tools) is a feature of these processes that integrate social networks.

Besides, we realized that Facebook is not the most appropriate means to deliberate and develop proposals in depth. Offline workshops and assemblies remain a better way to engage in deep and rich conversations. The offline events were also an excellent community formation strategy. People from the community came together, met, and later, by a WhatsApp Group, remained connected and engaged. The group *Amig@s del Parque Caballero* (Friends of the Caballero Park) remains active to this day, and it is even in the process of formalizing as an association to promote the restoration of the park. This is an unexpected lesson of this process: messaging platforms and groups are useful mechanisms to use when community formation and engagement is one of your goals.

Finally, but not least, one of the failures of our process represents also a key insight. Throughout the process, we were able to engage decision-maker actors in some moments (e.g., Municipality Mayor) but could not secure their commitment to take up projects to include them in the master plan for the park. The political will to establish binding participatory processes remains a challenge in Paraguay, making this type of participatory processes mostly not viable. The formation of the community and the active engagement of some of its members is probably the only reason why this process did not end in frustration, as most others do when participation does not influence decisions significantly.

## Acknowledgment

This research was sponsored by CONACYT, Paraguay through the program PROCIENCIA and resources of the Fund for the Excellence in Education and Research (FEEI by its Spanish acronym), through the grant 14-INV-102 "Participa: Fomentando la participación ciudadana en la innovación del sector público". European Union's Horizon 2020 research and innovation programme under the Marie Skłodowska-Curie grant agreement H2020-MSCA-COFUND-2016-754433.

# 3.17. To Pseudo-Participate or Not to Participate?

Victoria Palacin
University of Helsinki
LUT University
Helsinki, Finland

*The concept of citizenship is being redefined through the use of online tools. These digital tools and services embedded developers' biases and technological clientelist approaches, where participation is conceived as a use-case rather than a democratic human right. The lack of agency in decision making and agenda-setting is a growing phenomena in the design of digital public services. This position statement introduces the concept of pseudo-participation and opens up a conversation regarding how pseudo-participation is embedded in democratic socio-technical systems in current use.*

## 3.17.1. Digital Democracy

Active, substantive and meaningful participation is highlighted as a critical element to implement the 2030 sustainable agenda by the United Nations [151]. This has been the result of a significant evolution in the relationship between governments and the people in the past 60 years, from consciousness-raising in the 1960s to the incorporation of local perspectives in the 1970s, the recognition of local knowledge in the 1980s, the participation as a norm in the sustainable development agenda of the 1990s, and the e-participation concept in the 2000s [184]. Yet, despite the international commitment to advance the sustainable development goals (SDGs), *"public participation remains an overlooked key element of the SDGs"*, with no coordinated actions to build resilient participatory infrastructures [23, pg.1].

In the current era of participatory cultures [105], digital tools have been proposed as a way to mediate participation, in the design, development of solutions to common issues, with examples ranging from workplace information systems and city planning to environment and social policy issues [154, 209]. As a result, online participation has spiked in popularity across the world [51] as democratic societies continue to strive to embed participation in the direction and operation of political systems in different ways. But, participation has always been a complex concept in practice. Famous examples like the Ladder of Citizen Participation [12], the separation of informative, consultative and discursive interaction [31], different knowledge levels (e.g., [109]), and other typologies of participation and democracy (for example: [77, 130, 158, 186]), show the versatility of participation. Digital democracies are being implemented differently in a variety of contexts (for example municipal vs. national or international). A participatory initiative may have several aims, not all of which focus on engaging with people. Rather, sometimes the aim is to start a process to only give an impression of real engagement. This is known as technocratic clientelism (*"state-led regime with clientelistic mediation between the state and society"* [92, p.17]) and it is characterized by an appearance of political effectiveness by the creation of participatory processes, where both popular control and people's agency are virtually non-existent [92].

## 3.17.2. Pseudo-Participation

Online platforms aimed at public participation often embed a lack of agency and infuse pre-set agendas into features that resemble non-participatory degrees of participation such as manipulation, one-way information, placation [12]. My colleagues and I have conceptualized this contemporary phenomena as digital pseudo-participation by and in design - the configuration of digital artifacts and/or processes that can provide an illusion of participation but lacks supportive processes and affordances to allow meaningful participation to happen [170].

Pseudo-Participation by Design
*"Pseudo-participation by design emerges through the interaction with a configured artifact (i.e. digital service) that creates an impression of affecting change through digital interaction. In reality, these artifact's affordances have been pre-set by an agenda and do not offer any meaningful power to the people. They configure the role of the user (e.g. information consumer) and limit the ways they can interact with the tools. This is enabling digital participation without giving any real agency. This augments an existing lack of transparency in institutionalized participatory processes. In them the main focus is to collect as many opinions as possible as opposed to opening-up the mechanisms in play*

*during all decision making stages. Pseudo-participation by design hurts the willingness to participate, reduces trust in government, and diminishes the ability to create social capital. Through the design and implementation of an artifact (e.g. a website), city officials can embed assumptions about what are the expected roles of the inhabitants of the city [125]"* [170, pg.41].

Pseudo-Participation in Design
*"Participatory processes can result in limited involvement by those who will be affected by design decisions. Most typically, this manifests as introducing people's participation purely as an instrument - for example, incentivizing residents to report potholes in the city through a city website - and constraining other forms of participation. This phenomenon is what we call pseudo-participation in design. The power of making decisions about technologies for public use manifests also in the processes of design. Although many claim to be using participatory approaches such as participatory design or co-creation when designing digital services, but in reality, often those affected by the design decisions are marginalized and not involved in the design decisions loop [92]. The quality of popular control in pseudo participatory processes is very low [92]."* [170, pg.41].

### 3.17.3. What means to be a person in a digital democracy?

It is important to highlight that given the limitations of pseudo-participatory digital services to address common issues, people have developed crucial abilities to create, use and share digital resources that allow them to create their own livable cities (see, for example, the Environmental Data Governance Initiative). These practices of digital curation "in-the-wild" [48], are pervasive and evolve rapidly, especially in emergency situations such as the Chile protests in 2019 and the COVID-19 emergency. In these environments, a diverse set of actors have turned into curators of digital artifacts, data representations and interactions.

Through the concept of pseudo-participation, we want to highlight how important exploring participation not only through tools but in a wider context is. Scholars have different views on how the introduction of an online platform for participation will affect the citizens, politically [31, 158]. However, more work is needed to educate technologists regarding the power that hold to reconfigure people' roles in digital services. A first step towards that, is perhaps, to document how and where pseudo-participation takes place and the political affordances of socio-technical components and systems. Public technologies should not only enable people's participation but should benefit society and government in an equitable way.

## 3.18. Toward Understanding Civic Data Bias in 311 Systems: An Information Deserts Perspective

**Myeong Lee**
George Mason University
Fairfax, VA 22030, USA
*mlee89@gmu.edu*

**Jieshu Wang**
**Erik Johnston**
Arizona State University
Tempe, AZ 85281, USA
*jwang490@asu.edu*
*erik.johnston@asu.edu*

**John Harlow**
**Eric Gordon**
Emerson College
Boston, MA 02116, USA
*john_harlow@emerson.edu*
*eric_gordon@emerson.edu*

**Shawn Janzen**
**Susan Winter**
University of Maryland
College Park, MD 20742, USA
*sjanzen@umd.edu*
*sjwinter@umd.edu*

*While civic technologies for public issues and services such as 311 systems are widely adopted in many U.S. cities, the impact of the emerging civic technologies and their data-level dynamics are unclear. Because the provision patterns of civic issues to technological systems are different across neighborhoods and populations, it is difficult for city officials to understand whether the provided data itself reflects civic issues. Also, the disparities in the information provided to civic technologies in different neighborhoods*

*may exacerbate the existing inequality. To understand how civic data is created and how people's use of civic technologies plays a role as an intermediary process in shaping community performances, we take an information deserts perspective in studying 311 systems. The concept of information deserts is informed by a material understanding of local information landscapes, making it possible to distinguish local information's structural features from its social-construction process. Based on this theoretical lens, we suggest new opportunities for civic technology and data research.*

### 3.18.1. Introduction

Beginning in Baltimore in 1996, cities began using the telephone number "311" to enable residents to report non-emergency (i.e., not 911) issues, such as potholes, streetlights, and graffiti, among other urban problems [195]. Subsequently, cities have augmented telephone-based 311 services with email-, website-, and application-based reporting. Benefits from these 311 systems have included lower costs and better service provision [40], as well as more efficient resource allocation, such as moving calls from 911 to 311 [142]. Over time, 311 systems have become an important communication tool between local governments and publics, with 311 mediating public services provision.

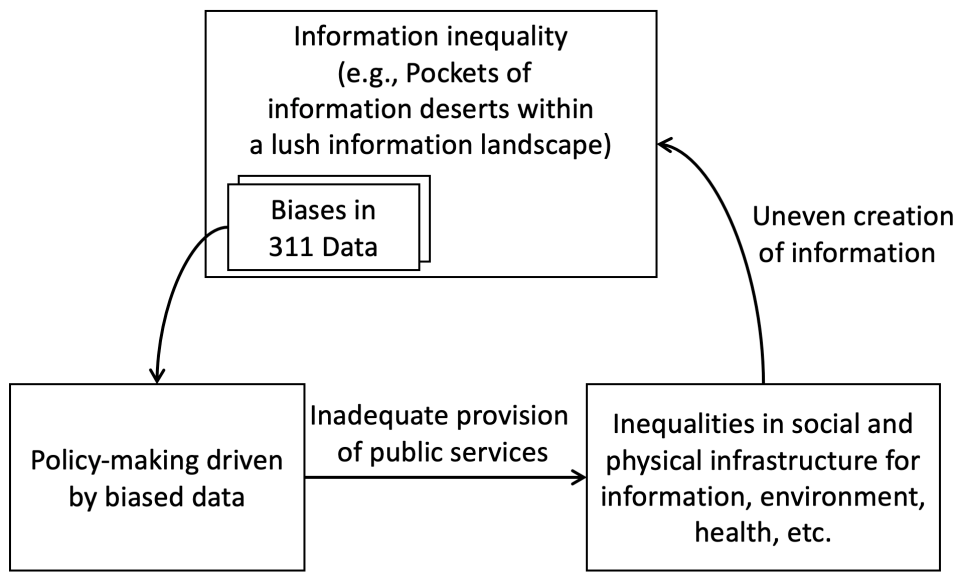

Figure 3.30: The relationship between data biases, policy-making, and inequalities.

While 311 systems have greatly expanded the number of people who contribute to urban data sets, researchers found that contributors' motivation is mostly not a sense of civic duty, but rather territoriality, or a desire to preserve and protect private space [167]. As a result, socio-economic divisions tend to be reinforced, not disrupted, through 311 systems [121]. Because a large portion of civic data is historically, demographically, or geospatially biased, computational social scientists often use algorithmic and data modeling techniques to understand and adjust for biases for predicting other community characteristics [222]. However, these approaches only address biases after the creation of data, rather than before or during its creation. In addition, the prediction models that rely on the biased features in the data could be less generalizable if civic technologies are revised or enhanced, which may alter the dynamics of human-system relationships in an ad-hoc manner. If it were possible to address biases as civic data is created, it could help researchers and governments improve the equity of municipal service provision as well as the robustness of civic data quality. Failing to understand and ameliorate biases in civic data can exacerbate the inequalities of the past and institutionalize them in the cities of the future (Figure 3.30) [127].

To understand the dynamics of smart cities of the future and minimize the negative effects of technical interventions on inequalities, this position paper proposes that it is imperative to:

- Identify the biases of the past in the datasets of the present

- Understand how those biases proliferate into hybrid computer-human systems

- Develop computational models that can operationalize the information deserts of civic issues to illustrate those biases

- Improve municipal decision-making and public service provision by mapping information deserts as a map-based visualization tool

- Produce design guidelines for future civic technologies to identify and address provision-level biases

Local Information Landscapes

To develop new approaches for modeling and visualizing the inequities in the pipeline of civic data creation, processing, and use, this position paper draws on a theory of local information landscape (LIL theory) [119]. LIL theory is a meta-theoretical framework that explains the community-level, material structures of local information (e.g. flyers, websites, block parties) and their relationships to other community characteristics. Mapping LILs can help cities better manage information provision processes for civic repair. Because 311 data is a result of ad-hoc design of civic technologies, studying 311 systems and their data from an LIL perspective may make it possible to examine some structural features of the civic issue landscape. Some particularly important aspects of that information provision include:

- The types of civic issues reported (e.g. potholes, broken sidewalks, abandoned vehicles)

- The volume of reporting across issues

- Individual reporting frequency (Boston median one report per year)

- Territoriality in reporting (range of report locations for an individual)

- Geographical coverage of individuals in reporting issues

Information Deserts

By studying the aspects of information provision as variables, we can uncover the information deserts of civic issues in cities [119]. Information deserts are conceptual and physical spaces where local information is poorly embedded in diverse infrastructures and/or less available than other areas of the city: the material pre-conditions of local information that can give rise to information inequality.

The reasons why people do or do not report issues to the 311 systems can vary significantly, and information about how and why people report issues to 311 is an important building block for understanding information deserts in cities. Only after understanding people's motivations to report civic issues and their information practices in their daily lives, does it become possible to understand information inequality and leverage 311 data in an efficient and equitable manner to further refine the civic technology. To thoroughly investigate the information deserts in cities requires (1) partnerships to obtain, interpret, and analyze 311 data, (2) individual-level data granularity (open data ideally, after eliminating privacy concerns), (3) a combination of social scientific and computational methods (to address data quality and missing variables), and (4) ongoing iterative interactions between city employees, researchers, and publics to define and visualize insights from this research that can improve municipal decision-making and public service provision. Our current partnership with the City of Boston makes the pilot of this approach possible as the individual-level dataset has been made it available to our research team.

## 3.18.2. Approach

To determine where and how information deserts are located, census and geospatial data complement 311 system datasets. By making use of computational models that describe individuals' information provision behavior and mobility, it is possible to identify a typology of Boston's information deserts based on community features that affect or are affected by information deserts. Then, it will be possible to assess relationships between information deserts and major demographic and geospatial features of data biases, as well as how those biases might proliferate into municipal decision-making. Building on the multi-dimensional LIL model, the interactions between components of LIL can be quantified by making use of computational models such as a flow network model for quantifying the degree of information fragmentation, an institutional network for measuring the embeddedness of information in diverse sources, or a comparative advantage model for measuring the relative impact of each information source. From an analytic perspective, a core part of this approach is that a series of these studies examines community characteristics, information provision behavior of individuals, information deserts of civic issues, and their outcomes separately as building blocks of the information inequality embedded in civic technologies.

### 3.18.3. Future Work

An early prototype visualization tool for 311 data, produced through this grant and a partnership with Supernormal, is available at `https://betablocks.city/discover`. As the work progresses, this tool will become the object of iteration in response to participatory design sessions with city officials to link its affordances with the needs of its potential users. Future work will include (1) refining and developing computational models of information provision behavior, (2) accessing 311 datasets from other cities through expanding our partnerships, (3) providing a typology of information deserts of civic issues through data analytics and interviews and (4) building data visualizations of information deserts of civic issues to help reduce bias and inequity in public service provision. This position paper is a call for civic technology researchers' attentions to the material, structural features of civic data that exists in diverse local infrastructures and civic technologies.

### 3.18.4. Acknowledgements

We thank all the municipal employees who participated in this work. We also gratefully acknowledge the grant from NSF (#1816763).

# 3.19. Virtual Assistant for Latinas Experiencing Domestic Violence

**Liliana Savage**
Civic Innovation Lab, UNAM
Mexico City, Mexico

**Luis F. Cervantes**
Acceleration Lab, United Nations
Mexico City, Mexico

**XueYan Chen**
Human Centered Design and Engineering
University of Washington
Washington, USA

**Ricardo Granados**
Civic Innovation Lab, UNAM
Mexico City, Mexico

**Saiph Savage**
Civic Innovation Lab, UNAM
Mexico City, Mexico

*Women from Mexico City that suffer domestic violence search help and counsel at the LUNAS centers, under the control of the Secretary of Women. Most of the service provided by these centers is in person, or by telephone, making it complex and slow for women to obtain help as fast and easy as possible. To help in that situation, we propose a new tool, a chat-bot to increase speed and easiness to provide the information needed by women in need and to reduce the workload for the people working at the centers. Women who attend to these centers need to fill the "Unique Registration Card", a registration that is fundamental to direct the women to the correct dependency, to allow work between different governmental dependencies, and to continue the follow-up of the cases. This chat-bot searches to increase the speed and the simplicity to collect the information needed for the registration. Furthermore, it intends to have a complete updated database, that could simplify the follow-up of cases and the successful entry of new cases on different government dependencies. This AI will be centered in an intersectional feminist perspective taking women as the center of the design.*

### 3.19.1. Introduction

Emerging evidence has revealed that domestic violence against women and girls has accelerated due to the recent global pandemic of COVID-19. In Mexico, 43.9% of women aged 15 years and over (19.1 million) reported being victims of intimate partner abuse over the last decades [76]. Both social and economic pressure, such as loss of income and extended domestic stays extend surged the incidents of domestic violence. With social isolation and restricted movement during quarantine, battered women are forced to live in confined spaces with their abusers and found no way to escape. Many of them were afraid to seek help or call police, knowing their aggressors might eavesdrop on their phone calls and might exacerbate the situation. The United Nations Development Program (UNDP) Acceleration Laboratory in Mexico is partnering up with the LUNA Centers of the Government of Mexico City (LUNAS) to advocate immediate action to end all forms of violence against women and girls.

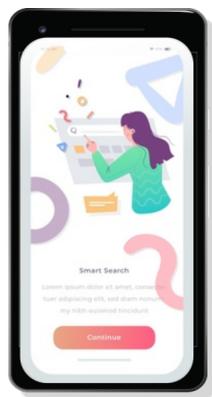

Figure 3.31: Interactive Feminist Interface

The first step for a woman to receive support, is to fill a "Unique Registration Card" (URC), which contains all the information needed by the institutions to provide the correct support and follow-up of the case. This form is a key tool for the Information Network of Violence Against Women (Red de Información de Violencia Contra las Mujeres - RIVCM). The file allows that other government dependencies that were not part of the filling of the URC access the information and proceed with the case. This registration card and network are the correct mechanisms to tackle the problem but there is still a lack of maturity in the technologies and methods.

Independently of the interdependency ability to work together, the filling of the URC poses also limitations. The URC is filled in person in the LUNA centers or by telephone. This filling is a tedious step, for the user and the worker who helps to fill it.

There are several layers of obstacles to tackle that are mixed in the same objective, to provide a tool useful for users, workers, and government dependencies to help in the current situation of domestic violence. We propose an Interactive Feminist Interface that will provide a framework where we can develop technology focused and centered in women. For this we take a feminist HCI approach [18], making participants all the women that will be possible users of the technology in a participatory design. We want that this technology adapts to the women using it instead of them adapting to use the technology, what is know as a Human-Centered Design [202].

With this we try not only to solve the direct problem of workload and collaboration in between dependencies, but provide a complete framework based in equity and intersectionality which could be used for other types of virtual assistants or machine learning in general [215].

### 3.19.2. Interactive Feminist Interface

As mentioned before the Interactive Feminist Interface has several layers of objectives. The first one is to provide a tool, a virtual assistant (chat-bot) that will help users to fill the information needed for the system to process their case. The second, to reduce the workload of the workers at the LUNA centers providing them with a tool that could help in orientation to fill the URC and to make it easier to access data and follow cases. And the last one, to provide a robust and reliable database handle for interdependency team work, allowing different government dependencies to access and follow the cases.

To propose the design of this virtual assistant, we have consulted users and workers, as well as people in the secretaries that are linked directly to the LUNAS centers, to be able to gather the sufficient knowledge to provide the correct tool.

### 3.19.3. Design Plan

Different laboratories, institutions and people are involved in the development of this virtual assistant. The United Nations Development Program (UNDP) Acceleration Laboratory in Mexico is partnering with the Civic Innovation Lab at UNAM, with governmental institutions, such as the Ministry of Women in Mexico City, and their centers to help women (LUNAS).

Plan of development

The development plan is carefully designed to allow feedback and participation of the main users of the virtual assistant. The virtual assistant development plan follows the following steps:

- Interviews with users and workers at LUNA centers

- Starter platform prototype

- Test phase

- Prototype real interaction tests

- Start platform architecture in production

- Training of LUNAS workers and people involved

- Technology tests in production

For each of these steps different "actors" are involved, For the first step and all the test phases the workers at LUNA will be involved. The Ministry of Women will help us to connect with workers from the LUNAS, and the women who utilize the services. We use a human-centered design approach to understand in detail the needs of the different stakeholders to design appropriate technology for them. Nevertheless, our scope is to involve all of these actors in the design, prototype testing and corrections, to be sure that the tool is not only focused to one group but is useful for each of the people involved and that fulfills and excels the needs.

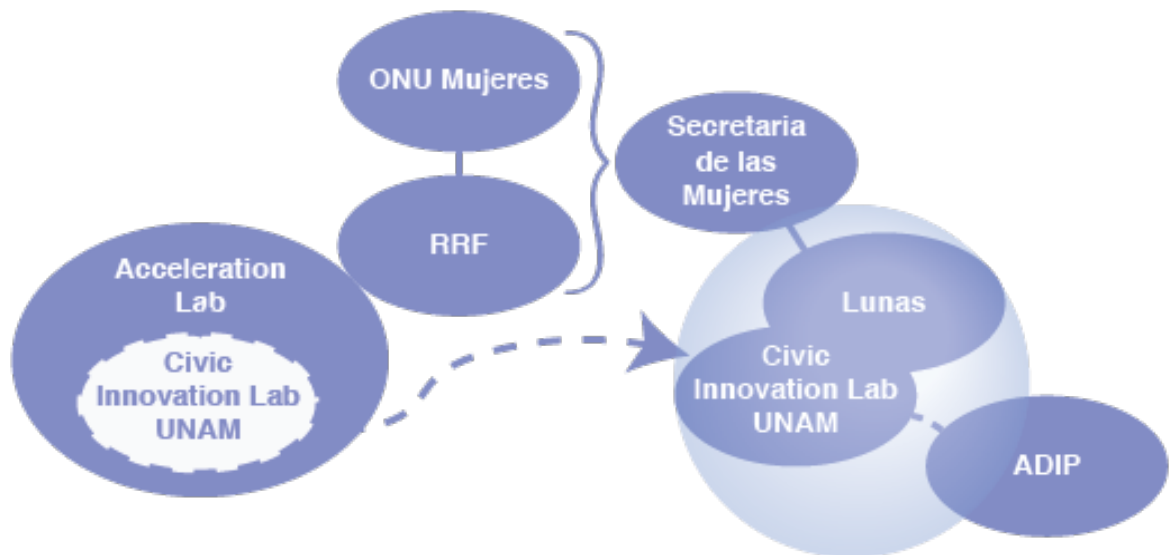

Figure 3.32: There are several institutes and people involved in this virtual assistant

Initial Needs Discovery.
We conected interviews to women in Mexico City, as well as to government workers at the LUNAS, and/or who worked in Mexico City's Ministry of Women. Through our interviews we identified some of the values and needs of these different stakeholders:

- **Women Users in Pandemic.** Some of the women who utilize the LUNAS centers during the pandemic expressed that they primarily called the centers to receive psychological and legal advice. The psycological advice is to help them cope with the violence they are exposed to. The legal advice is for them to manage economic situations where their children are removed; or they are having legal battles with their spouse where they do not want to pay alimony etc. The pandemic brought new circumstances where they cannot as easily express their situation over the phone as their partner is in the house with them.)

- **Government Employees.** The Government employees expressed that some of their biggest challenges involved not having the same protocols for registering calls, visits to the LUNAS centers (registering information about the women who came to ask for help), as well as having a concrete plan for the type of followup that would be given to women who called for help (followup involves what happens after the women visits or calls the LUNAS and expresses her situation.)

Based on this information we are now in the process of designing interfaces that address the needs of these different stakeholders. In particular, we are focused on designing dashboards that could help to have a homogenized way of registering the information of the women who visit the centers, and provide Mexico City's Ministry of Women with ways to visualize the type of followup that is being given to the different women. Through this we hope to improve the delivery of services and help that is provided to battered women. We will also aim to explore interfaces through which women can report their situation without having to call or visit a LUNAS center (based on the needs that we are identifying in our interviews.) In the following we present our plan for deisgning and deploying our tool.

Calendar of prospective prototype

- Interface design (3 weeks): Interface co-design with users and key actors.

- Fast prototyping (5 weeks): Tests with users with the interface, to gather information and iterate the design. In this stage the interface will be developed to production level.

- Training and workshops (4 weeks): Training of people involved in the use of the tool to provide services to users.

- Production: Distribution and installation of the interface. Testing on big scale.

# 3.20. A Tu Servicio Bogotá (ATS) – Marea Digital

**Estebán Pelaez**
Modelo de Involucramiento Ciudadano Participa+
(Movilizatorio & Fundación Corona)
Bogotá, Colombia
epelaez@fcorona.org

**Victoria Giraldo**
Modelo de Involucramiento Ciudadano Participa+
(Movilizatorio & Fundación Corona)
Bogotá, Colombia
victoria.giraldo@movilizatorio.org

*According to the goal of the workshop, we are glad to present hereunder an approach about how the local context and infrastructure affect the design, implementation and deployment of civic technology, based on two experiences; A tu Servicio Bogotá, and Marea Digital.*

## 3.20.1. Main assumptions

The hypothesis in which both civic tech experiences are based are the following: *"Civic tech is a mechanism for civic engagement which has the main objective of promoting citizen participation and advocacy in public decisions. The opportunity offered by civic technologies is the generation of data from the community for local development (community-driven data and development), seeking to deepen citizen engagement in the management of local problems".*

Civic technologies are meant to be scalable and replicable in different contexts, inspiring a cost-effective approach to costly technological developments. Nonetheless, the scale and replication of civic tech good practices have to consider an effective territorialization approach to prototyping and development, taking into account the following factors: problem identification, solution mechanism, local capacities, local barriers, alliance for implementation.

## 3.20.2. Description of the Civic Tech initiatives

A tu Servicio Bogotá is an civic and public innovation platform with three main purposes for the Bogotá health sector:

1. development of tools that promote informed citizen participation through community-driven-data;

2. technology that improves the interaction between citizens and the district government in relation with the quality of health services;

3. innovations that make the public policy in the health sector more efficient in the provision of different services.

Marea Digital is a civic tech platform which allows citizens to identify and report local issues that affect their communities and their quality of life, but also, to find and report local initiatives and actions to tend towards social welfare in the city of Buenaventura, Colombia.

Additionally, the platforms allow District Governments in Bogotá and Buenaventura, as well as other key stakeholders to recognize relevant information for design processes and decision making based on evidence.

## 3.20.3. Local territorialisation experience

Fundación Corona and Movilizatorio, building on the conceptual framework of the Citizen Engagement Model Participa+, have developed two civic and public innovation platforms that harness the realities of challenging territorialization experiences in Colombia.

A Tu Servicio Bogotá (ATS)

ATS is based on the following change hypothesis: if a lasting public-private partnership is harnessed under the framework of the project, and a civic innovation appropriations strategy is implemented through the main actors of the ecosystem, including the Health Secretariat of Bogotá, citizens will engage incrementally with the quality of the health services through the platform, overcoming institutional and trust barriers that diminish the effectiveness of interaction between citizens and the government in important issues that affect their quality of life like the health services.

The replication of A tu Servicio (`atuservicio.uy`) in Bogotá implements the best practices of the Uruguayan platform, taking advantage of the public data available in Bogotá and opens it through georeferenced visualizations. The 1.0 version of the platform is a product built upon the particularities of the Bogota and Colombian health sector, as well as the needs of the public, private and civil society stakeholders. The result was a platform built from scratch, but inspired by the ideas and good practices of the original Uruguayan platform.

The main challenge: how to implement a citizen driven data platform that includes thousands of health service providers (More than 16,000) with information that citizens can understand, and that's effective for public decision-making in the health sector. The prototyping process involved more than 15 multi stakeholder collaborative spaces to identify the main challenges relevant to the city´s context, the capacities of the local health providers, as well as the local authorities, to deliver and respond to citizen´s reports.

Marea Digital

In order to recognize the context and the infrastructure in which Marea Digital civic platform will be implemented, it was developed based on a diagnostic study on three fundamental components:

1. Citizen participation: analysis with the general aspects of the context which can facilitate or obstruct citizen participation in Buenaventura;

2. Technology: Access and use of tech in Buenaventura;

3. Technologies for citizen participation: Outcomes and learnings from experiences of citizen participation through at the local and national level.

Additionally, a diagnosis process based on challenges and opportunities related to communication and diffusion, training and effectiveness were carried out. This analysis had a crucial role in the design of the strategy for the civic tech platform.

## 3.20.4. Outcomes and learnings

The development of both platforms (ATS and Marea Digital) involved a complex process of territorialization of civic tech good practices. ATS was originally based on the Uruguayan platform, and Marea Digital on the Agentinian platform Caminos de la Villa. In both cases, the main outcome of the territorialization process were platforms made and prototyped from scratch, inspired by civic tech trends but taking special consideration on the priorities and particularities of local contexts.

For ATS, the territorialization process resulted in the following strategies that were particular needs identified for the Bogotá Health sector:

• Multi Stakeholder alliance to improve the quality of the city's Health Services (Public and private decision-makers, insurers, civil society organizations, health observatories).

• Engagement with health providers and insurers guarantees institutional responsiveness in specific areas that are tangible and close to citizens.

• Collection of detailed and specific information regarding identified health service categories are the key for institutional responsiveness and citizen´s trust in new participation mechanisms.

• The need to integrate - up to a certain level - a citizen driven data civic innovation for the public sector´s institutional decision-making mechanism.

• Local legislation is needed for continued-long term institutional support and responsiveness - sustainability of the territorization effort regardless of political leaderships.

• Engaging citizens through technological innovations need to combine digital and on-sight appropriation strategies. ATS has engaged more than 40,000 health users.

For Marea Digital, the territorialization process resulted in a civic tech - community-driven-data platform that adapted Camino de la Villa´s best practices in urban mapping for the particular needs of Buenaventura, with the following characteristics:

- Community-driven-data agendas prioritized (Humanitarian assistance, education, health, gender, infrastructure) based on two main variables: the possible medium term outcome of a rude accountability mechanism (tangible institutional responsiveness for specific community needs) in Buenaventura; and relevant policy areas for the territory´s sustainable development, according to the city's public policy cycle (New local development plan 2020 - 2024).

- Institutional engagement in each of the agendas to guarantee the mechanism´s possible results.

- Prototyping process with community leaders balancing their needs with the institutional capacities and planning /public policy priorities of the administration (Bottom-up and Top-down approach).

| Items | Challenges | Strategy |
|---|---|---|
| Communication and diffusion | - Clarity in communication<br>- Wide diffusion<br>- Incorporate people's own communication channels | - Work with leaders of the territory |
| Training | - Low use of technologies<br>- Low levels of digital literacy<br>- Commitment to the entire training process<br>- Availability to replicate knowledge<br>- Low quality of digital services provided by the Mayor's Office<br>- Difficult access to digital service spaces<br>Simple and intuitive navigability | - Achieve simple navigability<br>- Create an innovative tool<br>- Readiness of entities to cooperate<br>- Low quantity of this type of tools in the territory |
| Effectiveness | - To compromise public institutions<br>- Involve private actors<br>- Facilitate spaces for balanced dialogue that reach points of consensus on which build and manage effective solutions | - Strengthen networks of existing organizations and leaders<br>- Strengthen bi-directional communication between public and private actors<br>- Interest of the institutions in the alliance and the advocacy exercise |

# 3.21. Open Manifesto Project

**Irene Martín**                                    **Pablo Martín Muñoz**
CIECODE                                                    CIECODE
Madrid, Spain                                          Madrid, Spain
*irene.martin@ciecode.es*                    *pablo.martin@ciecode.es*

## 3.21.1. Introduction

In response to the generalized lack of political accountability and availability of clear public information, we developed an intelligent assistant that provides an open and easy access to electoral programmes. The Personal Political Analyst (PPA) is a project led by the Civic Technology and Empowerment Chair (Delegació de Transparència i Govern Obert de la Diputació de València with the Polytechnical University of Valencia) and Political Watch.

The PPA is a voice assistant built on top of Google Platform that allows queries and consultations of the electoral promises made by the different political parties in the General, European, regional and local Spanish elections. The objective of this project is to promote public accountability of the political class, as well as citizen participation at the different governmental levels.

In Spain, the reality is that each political party develops their electoral programmes following a different structure. And not only that, also electoral programmes tend to be available only a few days before the elections – sometimes less than a week before –, they are extremely long – almost 300 pages in some cases – and confusing. And, very little after the elections have passed they tend to disappear from the web pages of the parties.

All these elements make it almost impossible for citizens to have any kind of control over the information published during the electoral process, so nobody really knows what are the real promises made by the different parties. A big added problem is that electoral programmes disappear once the elections are over. There is no way of monitoring because, as you don't know what parties have promised, you can't demand anything from them during the next 4 years of government.

Taking this situation into account, the Civic Technology and Empowerment Chair (Delegació de Transparència i Govern Obert de la Diputació de València with the Polytechnical University of Valencia) and Political Watch started the Open Manifesto Project, and then developed the PPA.

## 3.21.2. Project objectives

This project had two main goals. Firstly to provide citizens with easy access to the content of electoral programmes, and secondly to enhance their capacity to monitor those electoral promises during the elections but also once the electoral process is over. To achieve this goals, two different process where needed:

- The development (through a collaborative process) of an open and public standard to simplify the structure of electoral programmes, focusing on the main elements any programme should have in order to make them easy to access and compare by the citizenry, media and CSO.

- Once the standard was established, the development of an easy-to-use tool directed to citizens, so they could access the programmes, and do all the consultations needed to make their voting decisions and also to monitor the execution of those promises once the elections were over.

## 3.21.3. Main results

The project and both processes needed were focused mainly on serving citizens. That is why we held as many public consultations as we could, in order to know the needs of citizens, media, CSO and even political representatives. Also, the standard was developed following the open source principles; therefore all the information and processes generated were and are open to the citizenry or any interested parties. The process of creating the standard followed an iterative development model composed by the following phases:

- **Phase I, Study and Analysis:** By way of work sessions and face-to-face or online interviews with political parties, social agents and citizens, information needs were determined and catalogued in relation to the electoral manifestos and proposals, and a series of use cases were developed to serve as guidelines for the creation and adoption of the standard.

- **Phase II, Proposal and Feasibility:** Based on the information generated, the structure and the attributes to be included in the open standard were defined. Each proposal was evaluated by the technical team to determine the appropriate data type to be used and the feasibility of automatically generating the necessary information. Likewise, the incremental adaptation over the previous version was resolved.

- **Phase III, Validation:** Once the proposal and feasibility stage had passed, the new version of the open standard was generated and documented in JSON Schema format. A process of validation by the parties involved was then initiated.

- **Phase IV, Publication:** After validation, the new version was published together with the documentation produced. All the information produced was published under an open license, without dependencies to other standards that are not open and free of legal or technical clauses that may limit its use in any area.

The result of the process was a standard for electoral manifestos, a structured format that any political party (not only in Spain but anywhere) can follow when publishing their manifestos in order to enhance citizens' understanding and capacity to demand accountability. The standard was achieved by:

- Applying the same format and structure to the electoral manifestos of all political parties, making it easier to compare them.

- Organizing the proposals of political parties in an explicit and concrete manner, classified by topics. In this way it becomes easier to know the positions and proposals of different political parties concerning any given issue, thus allowing users to make more informed decisions.

- Indicating the level of importance political parties give to each of their proposals. This way users would find out, for instance, the "red lines" parties are not willing to cross and the sequence of implementation of their proposals.

- Publishing electoral manifestos in an open format that is easy to access. The information contained in the manifestos is accessible to everyone, at any time (even after elections), for consultation or to be reused for other purposes.

- Transforming the electoral manifestos into a manageable and accessible instrument for citizens and other interested parties (such as journalists) not only for political parties.

- Improving the capacity for cooperation and dialogue between political parties, public administrations, social agents and citizens.

- Allowing the interconnection of electoral manifestos with other political information, through civic tech tools.

Once the first version of the standard was achieved we started a dialogue process with political parties with the intention of making them adopt it before the elections. Due to the lack of reply of most political parties, our team started to adapt by ourselves all the european, national and regional programmes to the standard. Once this was done, we were in possession of a database with every electoral programme, all of them with the same structure and categories, making it simpler to compare and contrast. Also, once we had a homogenized database with all the electoral programmes, we decided to develop a tool that could make it easy for users to access the database and, most importantly, to convert the data into valuable and useful information. To face this challenge in an innovative way, because probably a web page with a comparison table between programmes would have been easier, we developed the PPA, a voice assistant built using Google Platform, available for every person with an Android phone. The PPA was developed using DialogFlow, a natural language understanding platform to create an app that could offer interactive voice response to users. The idea was to offer a simple and easy way to access the information regarding the electoral promises of every party, and to compare the different promises made by each of them.

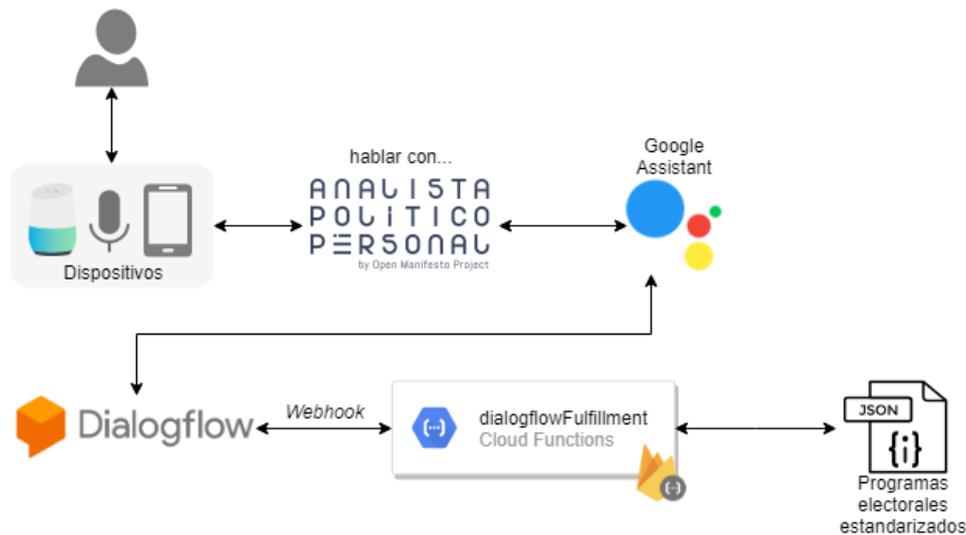

Figure 3.33: Components flow

We developed a conversational diagram, including different types of questions that could be made by users (mainly declarative and comparative in the first version). We then trained the PPA and with the help of volunteers we improved the quality of the responses. The first version was launched by October 2019, a month before the elections.

One important element that we wanted to focus on was interoperability; that is why we followed Schema.org's recommendations when developing the Open Manifesto's format. Also, our voice assistant is built on top of Google Platform and we use DialogFlow as a way to interpret natural language. From the beginning of the project, the team developed a communication strategy that implied meetings with several CSOs, citizenship and media as well as participatory workshops with citizenship and media. These alliances and supports will ensure the participation at long term level. Moreover, a communication online strategy developed specifically for this project will guarantee the engagement of new users and promote the participation of citizenship.

### 3.21.4. Challenges

One of the main challenges we found at the beginning of the project was to transform the political manifestos from PDF format to structured and semantic JSON data (Open Manifesto format). So, our mission was to publish all those data in open data format, allowing anyone to build new apps with them, like we did by building the PPA. It's important to remember that if electoral programmes are built in origin using this (or any similar) standard, we wouldn't have had to transform all those PDF into a structured format, so it would be one less step to deal with. That is why we set meetings with representatives from every major political party, to explain the standard and encourage them to build their programmes using it.

The second main challenge we had to face was to set meetings with the major political parties, partly because of the complex political situation in Spain during 2019 (two general elections (April and November) plus autonomic and european elections. In this context it was difficult to contact the parties, as the people we wanted to deal with were 'in the eye of the storm' during those months. Even though we were able to meet with 2 out of the 4 main national political parties (in the first stage we tried to contact with Ciudadanos, PSOE, PP and Unidas Podemos), only Unidas Podemos agreed to apply the standard for the November election (and, in the end, they could not manage to do it).

On the positive side, every political representative we met with was very interested in the standard, and gave us many useful suggestions to make it more accessible for political parties. We truly believe that under other circumstances we would have achieved better results regarding the adoption of the standard by political parties.

All the problems related to the adoption of the standard caused many delays in the implementation of the PPA, mainly because we needed all the programmes to be already homogeneous before we could add them to the app (and many programmes were not even published until one week before the elections). Also the political context, with people already tired of so many electoral appointments, made it more difficult to get people to know and use our app, especially without a bigger budget to spend on human resources and advertising.

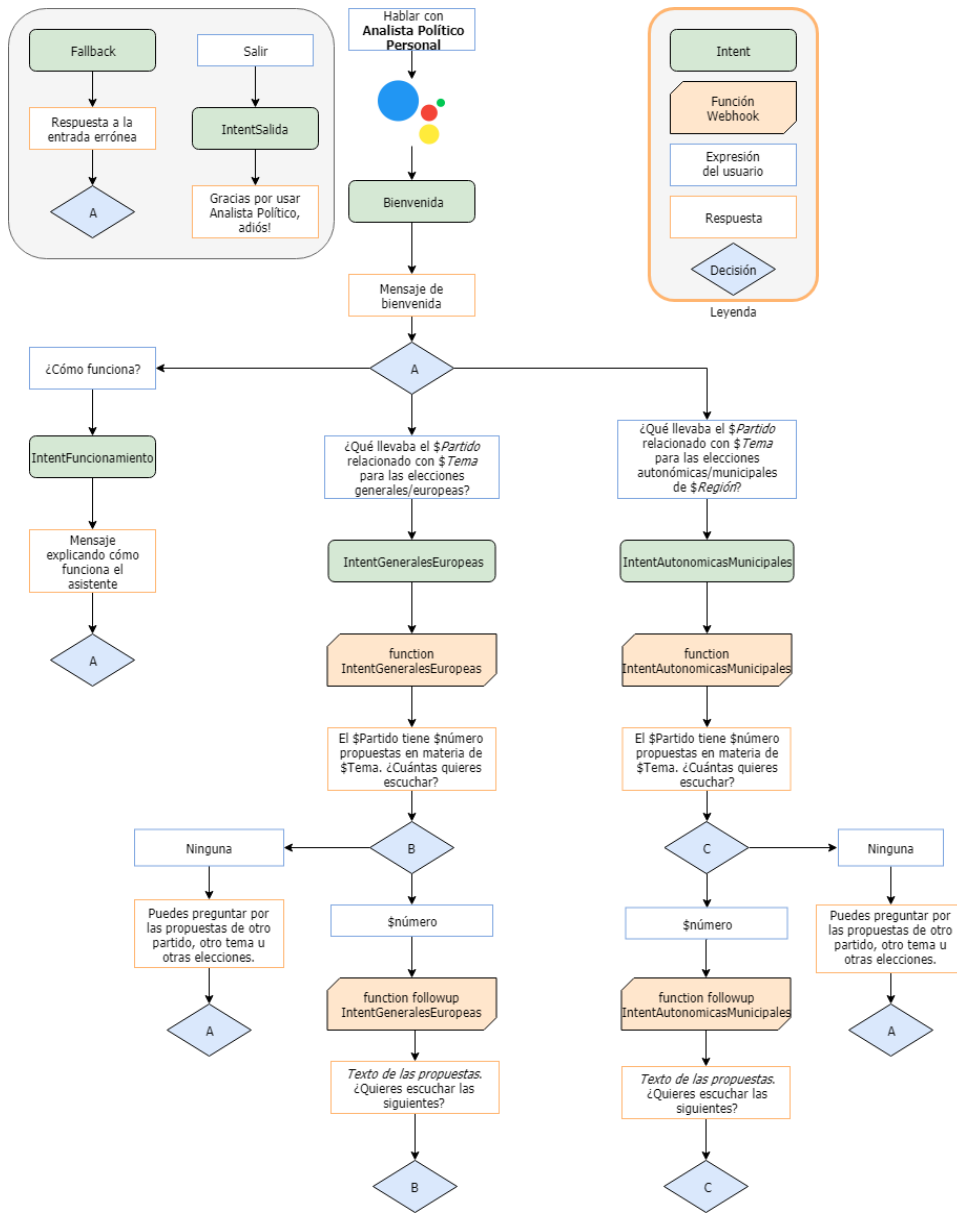

Figure 3.34: Conversational Diagram used for the PPA

Finally, once the elections had passed the challenge was, and still is, to keep the conversations with the parties alive in order to achieve bigger commitments to the standard so in the following elections they can adopt it straightforwardly, without the need for a team like us to translate the documents into a more structured format. The other challenge is to keep track of the promises the parties made, especially now that the COVID pandemic has changed completely the social and economic context, and that most people do not remember the promises of the parties they voted for.

# 3.22. Parlamento 2030

**Belén Agüero**
CIECODE
Madrid, Spain
*belen.aguero@ciecode.es*

**Pablo Martín Muñoz**
CIECODE
Madrid, Spain
*pablo.martin@ciecode.es*

## 3.22.1. Introduction

In response to the generalized lack of political accountability and availability of clear public information related to the political measures adopted to implement the Agenda 2030, this online tool provides an open and qualified access to the Spanish Parliament's activity from the Sustainable Development Goals (SDGs) perspective. It promotes a transparent and participative implementation of this Agenda and allows CSOs, policy makers and the media to monitor political discussion, proposals and decisions. Parlamento 2030 is being implemented in the Spanish Congress in the framework of a collaboration agreement between Political Watch and the Secretariat of State for International Cooperation (SECI), responsible for international cooperation policy in the Spanish Ministry of Foreign Affairs, European Union and Cooperation (MAUC). It was included in the Spanish Action Plan for the Implementation of the 2030 Agenda as a public tool for monitoring the SDGs.

Parlamento 2030 is an innovative online tool created to track, compile, tag and provide (in a free, open, clear, reliable and easily understandable manner) all the information about the activity of national parliaments related with the 2030 Agenda. It gathers all the information published by national parliaments into a database and, through an automatic process of massive tagging, classifies it according to the different linkages to specific SDGs and targets. It then offers this information freely and openly through an online browser for users to search, find and download. This combination of advanced computer science and the traditional knowledge provided by policy makers, CSOs and academia makes Parlamento 2030 one of the most advanced tools at an international level to access parliamentary activity information and for public monitoring.

The team behind this tool consists of people with experience in technical engineering, political coherence, and sustainable development and advocacy. Parlamento 2030 is a project of Political Watch, a civil society platform of engineers, social researchers and journalists created under the umbrella of CIECODE, a think- and-do tank focused on policy coherence for sustainable development. Our mission is to generate, in Spain and also in every country or region we work, positive conditions for the designing, adoption and implementation of public policies and private practices that promote a more equal, fair and sustainable society. We develop innovative technologies and tools to automatically track sustainable development-related political activity and offer public, direct, and free access to this information to citizens and key stakeholders.

## 3.22.2. Project objectives

The ultimate objective of Parlamento 2030 is to promote ambitious, properly designed and adequately financed public policies oriented towards the implementation of the 2030 Agenda. The specific objectives are:

- Provide citizenry with an open and online tool to access information about the activity of the Parliament.

- Provide the media and CSO with a reliable tool to access to all the information of the Parliament in terms of the 2030 Agenda.

- Create an incentive system for deputies in order to publicly recognize the good practices in terms of 2030 Agenda parliamentary activity and point out the bad practices or initiatives that go against its principles.

## 3.22.3. Main results

Parlamento 2030 is an online tool that obtains and classifies thematically the parliamentary information through an advanced system of scrapping and automatic massive tagging. These technologies build a database and enable users to navigate parliamentary activity through an openly and freely online search browser. The information provided is key for civic monitoring and public accountability of parliamentary activity. The technology used by Parlamento 2030 is called TIPI which stands for Transparency + Information + Participation + Influence. It was first used in a previous project (TiPi Ciudadano, which is still active) that tags and offers parliamentary information classified by 22 social issues.

The process carried out by Parlamento 2030's technology is:

1. The computer system extracts the information directly from the website of the Parliament. It then registers and organizes the information in its own database [Scrapping phase].

2. The software compares all this information with the 17 thematic dictionaries (a knowledge base containing more than 3.000 keywords) and labels each parliamentary act with those thematic areas they are related with [Labeling phase].

3. Parlamento 2030 offers all this information through an open, user-friendly and free access online browser. Users can select their own search parameters to tailor their inquiries according to their specific needs.

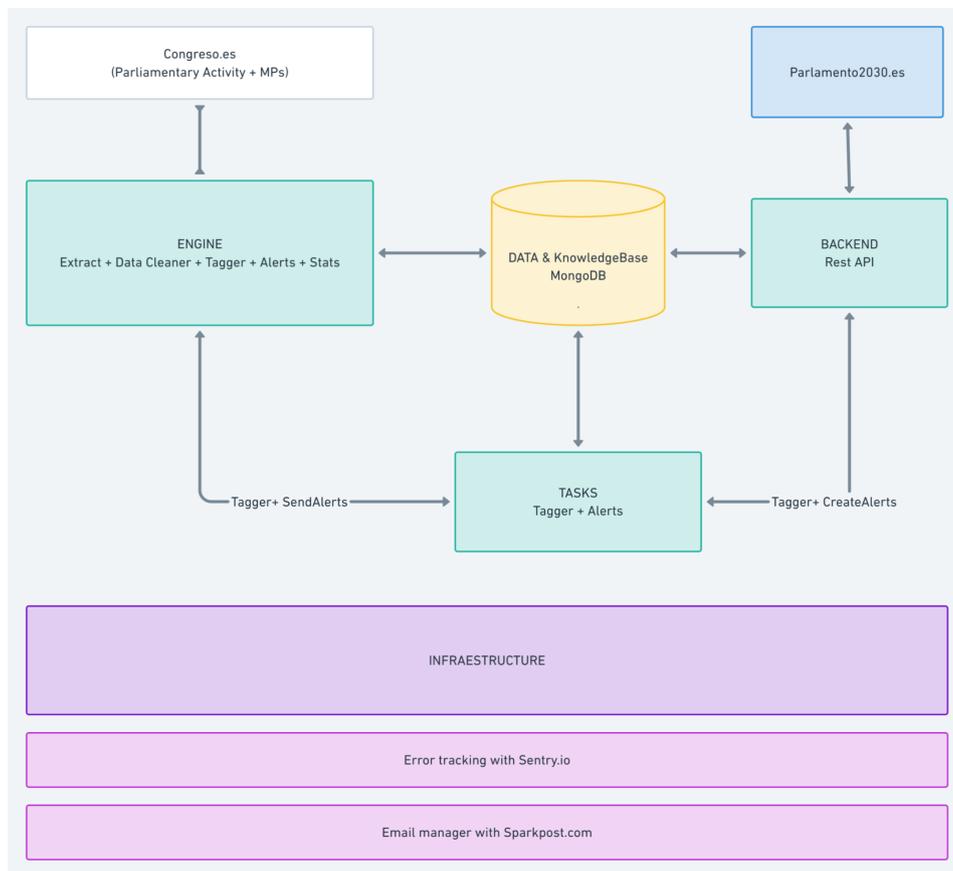

Figure 3.35: Components flow

Parlamento 2030 would not function properly without its 'dictionaries'. They represent the knowledge base of the tool, which is essential for the labeling phase, where each parliamentary initiative is associated with one or more specific topics of the 2030 Agenda. The software is capable of relating not only keywords but also goals and targets. This knowledge base was built using 'regular expressions' (see below) which were written by the Political Watch team and then submitted to a consultative process with experts from international CSO, government institutions, etc.

Table 3.9: Examples of regular expressions (please note that the originals are in Spanish).

| Tags | Regular expressions |
|---|---|
| Poor worker | Poor worker |
| Social protection | Social (protection\|program\|aid) |
| Access to vaccines in developing countries | Access to.*vaccines.*(developing countries\|least developed country\|LDC) |

As seen in the image above, our technology has different layers: a user layer, a backend layer, a data layer and an alerts layer. All of them are built using free software technologies and languages like Python, VueJS, D3js, MongoDB, Luigi, and Docker amongst many others. Our code is available for free on Political Watch's Github page. Our main data source is the Spanish Parliament website. Using webcrawler techniques, we transform HTML into structured data. We then publish all those data in an open data format, allowing anyone to build apps with them, and offering them through our API.

Responding to its main objectives, Parlamento 2030 technology is designed to cover the specific needs of its users (whether citizens, journalists, social movements or political representatives). To do so, after "processing" all the political activity of the Congress and "labeling" it according to its relation with the 17 SDG's, the technology provides the information through an online browser, with which the user can refine the search in terms of multiple criteria (author, date, subject, key word, etc.). The browser also allows users to download the result of their search in a reusable format (csv) and activate personalized alerts to be informed of any news about their subjects of interest.

### 3.22.4. Challenges

The implementation of Parlamento 2030 faced different obstacles in each phase. During the development process we faced several challenges as a consequence of the Agenda 2030's "essence":

- The cross-cutting nature of the SDGs is an extra challenge for monitoring, accountability and for the Policy Coherence implementation.

- Most connections between goals could be unseen and, therefore, difficult the coordination of measures and the creation of synergies.

- Pre-SDG institutions are still in charge of dealing with SDG challenges.

- Need of a 'whole of government' approach.

We knew that, in order to develop a truly useful tool that could answer all these challenges, we needed to make some changes in the technology (in the original coding), but also in the content (the knowledge base). As the SDGs are completely interconnected also the dictionaries of our knowledge base had to consider and embrace the same connections and relations.

Besides these adaptations, we knew we had a pending issue related to our communication capacities. In order to increase the usage of the tool by citizenry but also by public officials we needed them to know the tool and its utilities. With that purpose, we carried out an efficient communication strategy on the potential uses and the information provided by the tool.

Public administrations and CSOs from outside Spain have already shown their interest in the adaptation and implementation of the tool (especifically Andorra, Paraguay, Dominican Republic and Peru). This shows that with a little increase in the investment done to announce and communicate the tool, the demand on these types of tools grows immediately.

# 3.23. Sidewalk Accessibility in the US and Mexico: Policies, Tools, and A Preliminary Case Study

**Jon E. Froehlich**
University of Washington
Washington, USA
jonf@cs.uw.edu

**Edgar Martínez**
Liga Peatonal
Mexico City, Mexico
edgar.martinez@ligapeatonal.org

**Michael Saugstad**
University of Washington
Washington, USA
saugstad@cs.uw.edu

**Rebeca de Buen Kalman**
University of Washington
Evans School of Public Policy and Governance
Washington, USA
rdebuen@uw.edu

*We examine US and Mexico disability rights legislation and its relevance to accessible pedestrian infrastructure, provide an overview of civic tools for sidewalk mapping and assessment in Mexico, and describe the initial deployment of one such tool, Project Sidewalk, into two Mexican cities.*

## 3.23.1. Introduction

In the US, sidewalks have been shown to offer public health, economic, environmental, and accessibility benefits [9,17]; however, few studies have examined sidewalks in developing regions. In this workshop paper, we provide an overview of US and Mexico disability rights legislation and its relevance to accessible pedestrian infrastructure. Within this context, we highlight efforts in Mexico to map and assess sidewalks and describe the initial deployment of one such tool, Project Sidewalk, into two Mexican cities (Fig. 3.36, 3.37, 3.38, 3.39). We close by enumerating key challenges.

## 3.23.2. Background

Sidewalks play a crucial role in a city's urban mobility and quality of life. They can provide a safe space for pedestrians, help interconnect mass transit services, and also serve as unique public spaces for food, commerce, and leisure [15]. In Mexico, 55% of school commuters and 23% of workers travel by walking or rolling [32]; however, an alarming 44% of traffic-related deaths are pedestrian—often due to poor or non-existent pedestrian infrastructure. For people with disabilities, safe and accessible sidewalks can be even more important, affecting independence [25], quality of life [16], and overall physical activity [2]. The quality of sidewalks is, therefore, crucial to ensuring people's accessibility and mobility rights. Below, we provide brief historical overviews of disability rights legislation and sidewalk policies in the US and Mexico.

US Policy and Accessible Sidewalks

Spurred by returning WWI veterans with injuries, initial US disability policies focused on employment and rehabilitation services [11]. By the 1960s, however, the disability rights movement helped reframe disability not as a problem of mind or body but as a socially constructed form of societal oppression [7]. Bolstered by these efforts, the Rehabilitation Act of 1973 was passed, stating that no qualified individual with a disability should be excluded from or denied benefits of any program receiving federal assistance (Section 504). It was not until the landmark Americans with Disabilities Act (ADA) in 1990, however, that protections were extended beyond the government sector. Critically, the ADA recognized the minority status of Americans with disabilities—modelled after the Civil Rights Act of 1964—and required places of "public accommodation", including sidewalks, to provide people with disabilities appropriate aids or services [1]. Together, the Rehabilitation Act and the ADA regulate the accessibility of public rights-of-way and facilities in the US [36]; however, they do not define the specific accessible design requirements themselves. For this, the US employs the US Access Board—an independent federal agency [34,35]. For pedestrian infrastructure specifically, the Access Board specifies technical requirements for sidewalks, including a minimum 1.5m (5ft) passing width, a maximum 5% grade, and curb ramps at intersections—requirements that have helped inform the design of our interactive civic tools.

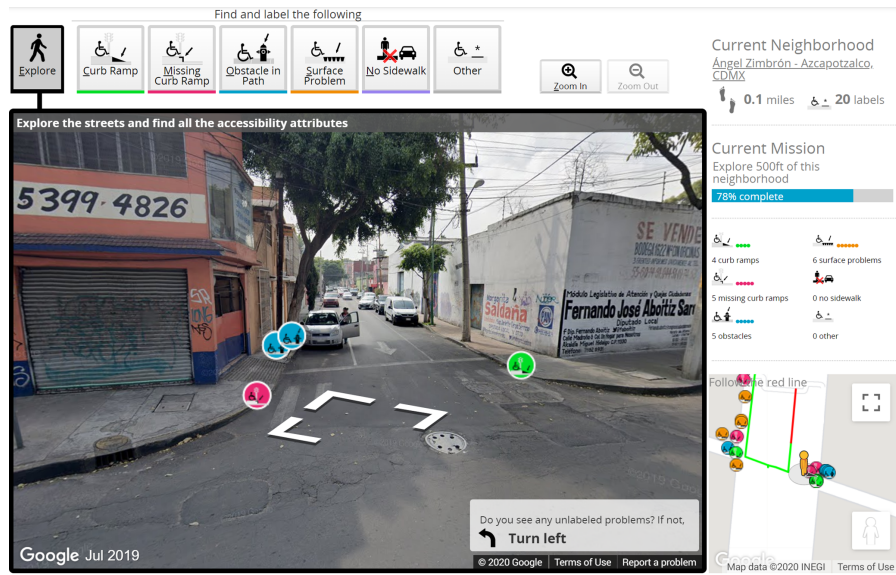

Figure 3.36: The Project Sidewalk labeling interface (in English) showing a labeled street scene in Azcapotzalco, MX with a curb ramp (green label), a missing curb ramp (red), and a tree and pole in the middle of a narrow sidewalk (blue).

Mexico Policy and Accessible Sidewalks

Like the US, disability-related policies in Mexico began in the mid-20th century. In 1944, *Ley del Seguro Social* [23] was passed to protect workers injured by occupational hazards followed by broader social assistance programs in 1977 (DIF) [24]. It was not until 2005, however, that the first national law was signed to specifically protect the rights of individuals with disabilities, called *Ley General de las Personas con Discapacidad* [25] [23]. This law mandated the creation of a federal agency, the *Consejo Nacional para las Personas con Discapacidades* (CONADIS) [26], charged with guiding and monitoring disability related programs throughout other government agencies. In 2011, Mexico passed more comprehensive disability rights legislation, called the Ley General para la Inclusión de las Personas con Discapacidad (with significant revisions in 2018) [27] [24]. Like the ADA, this law guarantees the rights of people with disabilities and promotes, protects, and ensures the inclusion of disabled citizens in society. While sidewalks ("banquetas") are not mentioned, the law stipulates that public spaces and urban environments should be accessible and those that are not should be progressively updated. Similar to US law, Mexican legislation does not specifically enumerate technical design requirements; however, in 2019, design guidelines were published for street-level projects [28][31]. Some city governments have gone beyond federal policy. For example, since 2016, Mexico City has passed a series of local legislation and guidelines to improve public transit and sidewalks for safety and accessibility [3–5].

Important Policies but a Lack of Tools and Accountability

While the disability rights legislation and accessible design requirements in both countries demonstrate commendable progress, there remains a lack of tools, data, and open standards for tracking sidewalks, their topology, and their accessibility [8,29]. Consequently, it is difficult to assess sidewalk development and ensure compliance with recent legislation. Typically, in the US, large-scale sidewalk accessibility renovations occur only in response to civil rights litigation such as in New York [12], Seattle [10], and Los Angeles [27]. For example, in response to a lawsuit, LA recently pledged $1.3 billion to fix broken sidewalks and address accessibility problems—estimating that over 40% of the city was affected [27]. Accessibility audits are also expensive: Seattle paid $700,000 to survey just the curb ramps in the city

---

[23]Social Security Law
[24]Sistema Nacional para el Desarrollo Integral de la Familia
[25]General Law of Persons with Disabilities
[26]National Council for Persons with Disabilities
[27]General Law for the Inclusion of Persons with Disabilities
[28]It is not clear if these are enforceable standards like in the US

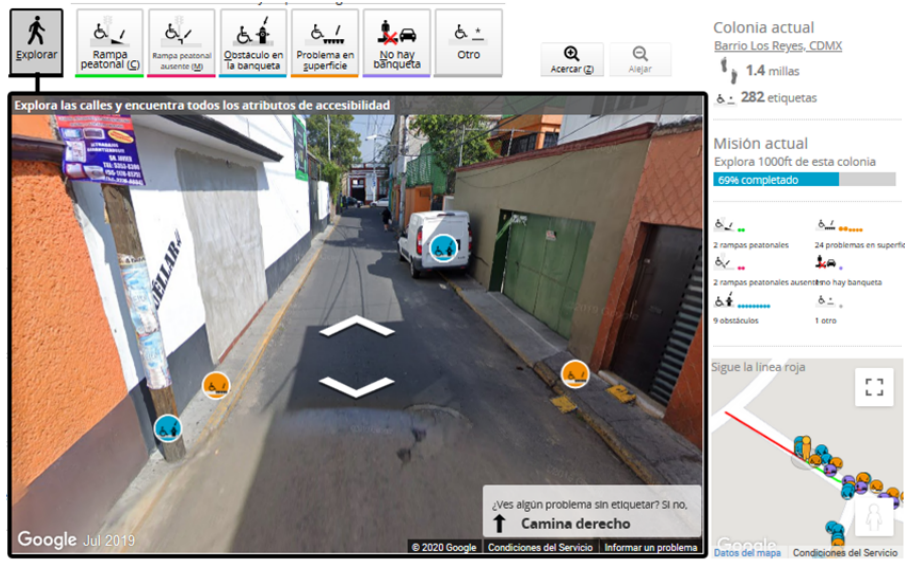

Figure 3.37: The Project Sidewalk interface (in Spanish) showing narrow sidewalks with surface problems (orange labels) and physical obstacles (blue labels) in Azcapotzalco, MX. Screenshot from Twitter @Gari01234.

[10]—again in response to a lawsuit.

### 3.23.3. Sidewalk Data and Civic Tools in Mexico

Given the recency of pedestrian and sidewalk accessibility policies in Mexico, there is a growing movement of activists and tools. In 2010, Mexico conducted a national housing inventory, which included physically surveying sidewalks across 40 cities (1,129,728 blocks in total) and released the data publicly [13]. They found that 33% of blocks had full sidewalk coverage, 36% had partial coverage, and 27% were without sidewalks [22]; however, the survey did not collect data on nor analyze accessibility-related features such as sidewalk surfaces and curb ramps. In addition to government efforts, grassroots organizations and NGOs have conducted pedestrian infrastructure audits, typically using paper forms and on-site inspections. For example, *Caminito de La Escuela* provides paper forms for citizens to evaluate pedestrian environments around schools, including sidewalks, crosswalks, and vehicular traffic [18]. Similarly, *La Banqueta se Respeta* collects data on sidewalk quality via in-person, paper-based mapping exercises—and use the resulting data to influence public policies [19]. Others take a more direct activist approach, such as *Grupo Salvando Vidas Oaxaca* who document and fix sidewalk problems on their own [9]—reminiscent of the 1970s DIY ramp activists in the US [12]. Most relevant to our work are interactive civic tools that rely on crowdsourcing such as *Mapatón* for bus routes [26], *Supercívicos* for documenting public infrastructure as a citizen journalist [37], and *Mapeatón* for sidewalk accessibility [6]. *Mapeatón* is a community mapping project: volunteers upload geo-referenced photos/videos of sidewalk journeys to Mapillary, particularly when using wheelchairs or pushing strollers (similar to JourneyCam [28] in the UK). In general, while these on-site physical inspection techniques provide the "gold standard" for pedestrian infrastructure audits, they can be expensive, logistically difficult to manage, and limit both who can supply data and how much data each individual can supply.

### 3.23.4. Project Sidewalk in Mexico: A Case Study

In our work, we are exploring complementary sidewalk auditing approaches that are fast, low-cost, and scalable using a combination of remote crowdsourcing, machine learning, and online map imagery. Our most recent tool, called Project Sidewalk, enables online crowdworkers to remotely label sidewalks and find and identify accessibility problems by virtually walking through city streets in Google Street View (Fig. 3.36, 3.37, 3.38, 3.39). Rather than relying solely on local populations, our potential user pool scales to anyone with an Internet connection and a web browser. In a 2018 pilot deployment, 1,400 users from across the world virtually audited 2,934+ km of Washington DC streets, providing 255,000 sidewalk accessibility labels with 92% accuracy [30]. We have now expanded to three more US cities: Seattle, WA, Newberg, OR, and Columbus, OH. Project Sidewalk and all collected data is fully open and accessible at http://projectsidewalk.io/api.

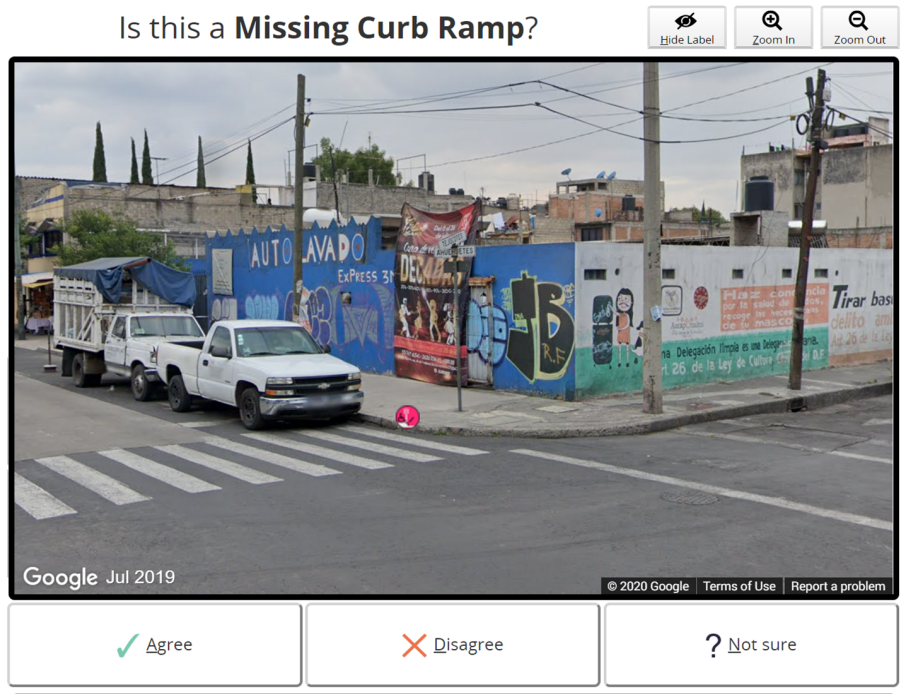

Figure 3.38: In addition to sidewalk labeling missions, Project Sidewalk users are also given validation missions to verify labels from other users. Shown above, a user is asked to verify a missing curb ramp label (which is correctly placed in this case).

| | Users | Total km Complete | Total Labels |
|---|---|---|---|
| Seattle, WA | 1,626 | 1618.1 | 92,591 |
| Columbus, OH | 278 | 195.8 | 17,590 |
| Newberg, OR | 204 | 224.6 | 16,076 |
| Azcapotzalco, MX | 255 | 80.7 | 5,864 |
| SPGG, MX | 269 | 35.2 | 4,442 |

Table 3.10: The number of users, total km audited, and sidewalk accessibility labels collected across five Project Sidewalk sites.

In early 2020, we were contacted by Liga Peatonal ("Pedestrian League")—an NGO focused on pedestrian improvements to increase the safety and accessibility of public spaces in Mexico—to explore deploying Project Sidewalk in Mexico. Working closely with their staff, we have been translating Project Sidewalk's interfaces into Spanish, adding locale-specific label tags, and co-brainstorming Mexico-dependent features. As a start, we deployed Project Sidewalk into two major metropolitan areas: the Azcapotzalco municipality in Mexico City (population 400,000; 33.6 km2) and San Pedro Garza García (SPGG) in Monterrey (population 122,000; 69.4 km2). While the Azcapotzalco deployment is grassroots, we are working directly with the local government in SPGG—the mayor helped launch the site in August 2020 [33].

Thus far, we have collected 10,313 sidewalk accessibility labels across 115.9km (72 miles) of streets in Azcapotzalco and SPGG (Table 3.10). While our deployments are ongoing and our analysis preliminary, we found that sidewalk accessibility issues in both Mexican cities were both more common and rated worse, on average, than labels of the same type in our three most recent US deployments (Table 3.40). For example, there are 2.5 surface problems per 100m in Azcapotzalco vs. 0.6/100m in Seattle and 1.4/100m in Columbus, and the average curb ramp was rated as a 2.8 severity vs. 1.5 in the other cities (higher is worse).

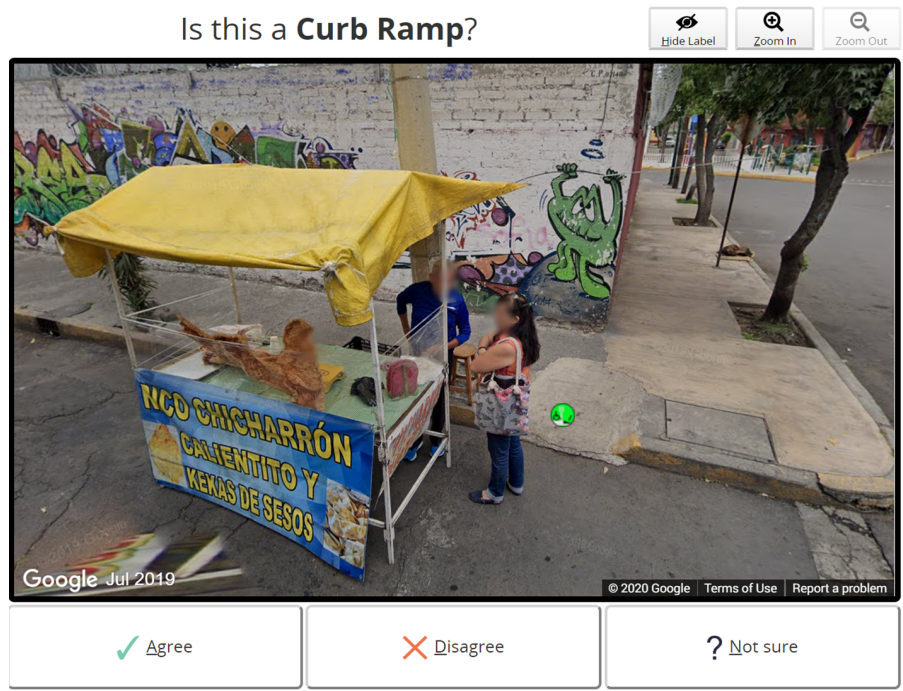

Figure 3.39: Another validation interface example. Here, a user validates a curb ramp label, which again is correct, but the pathway is partially obstructed by a food stand. Pop-up markets are a strong part of Mexican culture and provide an important informal economy; however, they can present unique accessibility challenges on sidewalks and streets.

### 3.23.5. Discussion and Conclusion

In this workshop paper, we described civic tools for sidewalks in Mexico, including two initial Project Sidewalk deployments, and the surrounding socio-political context regulating accessible design. Below, we enumerate key challenges co-identified by our cross-regional team. **Data for advocacy and change.** While there is a recent push towards open data and evidenced-based decision making in Mexico (e.g., the Agencia de Innovación Pública in Mexico City), we believe our work with SPGG is the first collaboration between a local government and a pedestrian-related crowdsourcing tool in Mexico. This collaboration is similar to Project Sidewalk's Newberg deployment where we worked with community activists and the city government. Here, Project Sidewalk's data was used to successfully lobby for new sidewalk-related funding programs (see visualization). Similarly, SPGG plans to use Project Sidewalk data to inform public policy, assess and triage problems, and as an outreach tool to involve citizens. However, as we describe in our recent CSCW paper [29], the design, availability, use, and maintenance of sidewalks are shaped not just by policy but by socio-cultural influences and competing funding priorities.

**Who participates?** As a remote tool, anyone on the Internet can contribute to Project Sidewalk; however non-local users may lack cultural awareness and miss or misidentify problems. Moreover, the reliance on technology itself can be exclusionary. In Mexico, 43% of the population uses a computer and 66% have a smartphone [14]. *Liga Peatonal* recently suggested creating paper "audit" forms for some members in their community (to be manually filled out and entered into the Project Sidewalk database). To reduce the need for a computer, we have also been increasingly adding smartphone-related features.

**Why participate?** To engage users, Liga Peatonal hosts virtual mapathons and advertises on social media [20,21] but sustaining participation is challenging. From informal conversations, most registered users are committed pedestrian or disability rights enthusiasts; however, Project Sidewalk itself currently has no specific features to support community building or a shared sense of social participation—which both our partners in the US and Mexico have requested (e.g., discussion forums, leaderboards). In some cities (DC and Seattle), we have employed paid crowdworkers from Mechanical Turk—which allows us to quickly audit large areas—but requires external funding and does not draw upon local

| | Curb Ramp | Missing Curb Ramp | Missing Side. | Obstacle | Surface Problem |
|---|---|---|---|---|---|
| Seattle, WA | 1.5 (0.7) | 3.8 (1.0) | 4 (0.8) | 3.2 (1.1) | 2.9 (0.9) |
| Columbus, OH | 1.4 (0.7) | 3.8 (1.2) | 4.1 (1.1) | 2.2 (1.4) | 2.1 (1.0) |
| Newberg, OR | 1.5 (0.7) | 3.9 (1.0) | 3.9 (0.9) | 3.1 (1.1) | 2.7 (1.0) |
| Azcapotzalco, MX | 2.8 (1.4) | 4.7 (0.6) | 4.6 (0.8) | 4.1 (1.0) | 3.6 (1.2) |
| SPGG., MX | 2.8 (1.4) | 4.4 (0.9) | 4.5 (0.9) | 4 (0.9) | 3.6 (1.1) |

Figure 3.40: The average severity ratings for curb ramps, missing curb ramps, missing sidewalks, obstacles, and surface problems. Ratings are 1-5 (5 is worst). Standard deviations are in parentheses.

citizens.

**The need for mixed-methods.** To fully assess sidewalk infrastructure, its use, and barriers to change, we suggest a mixed-methods approach, including ethnographic observation of sidewalk usage, interviews and surveys of key stakeholders, and an examination of federal and local legislation and policy. As a complement, Project Sidewalk's data affords both geo-spatial quantitative analyses—similar to that presented in Table 2—as well as qualitative examinations of the image-based labels—for example, to discover and taxonomize the types of problems in a city. We have found that the images themselves are powerful and help contextualize the numerical analyses (Fig. 5).

**In closing,** our overarching aim is to develop new low-cost and scalable sidewalk tracking tools that support evidence-based advocacy and policy making, provide government accountability, and enable new pedestrian tools. With the Spanish-based version of Project Sidewalk and our collaborators at *Liga Peatonal*, we have an opportunity to explore deployments in other regions in Central and South America. At the workshop, we look forward to discussing this possibility and pedestrian-related civic tools more generally in the US, Mexico, and beyond.

# Bibliography

[1] Spectra, 7vit y ava: aplicaciones para dispositivos que controlan el dengue, parkinson y el consumo de alimentos, 2013. `https://bit.ly/364YN0x`.

[2] Gravitreks, Jun 2017. `https://play.google.com/store/apps/details?id=com.gravitreks.www`.

[3] Public Meeting Requirements In The Age Of COVID-19. `https://www.law360.com/articles/1258008/public-meeting-requirements-in-the-age-of-covid-19`, 2020.

[4] Metadel Adane, Bezatu Mengistie, Worku Mulat, Girmay Medhin, and Helmut Kloos. The most important recommended times of hand washing with soap and water in preventing the occurrence of acute diarrhea among children under five years of age in slums of addis ababa, ethiopia. *Journal of community health*, 43(2):400–405, 2018.

[5] Hunt Allcott and Matthew Gentzkow. Social media and fake news in the 2016 election. *Journal of Economic Perspectives*, 31(2):211–236, 2017. doi: 10.1257/jep.31.2.211.

[6] Adriana Alvarado Garcia, Alyson L. Young, and Lynn Dombrowski. On making data actionable: How activists use imperfect data to foster social change for human rights violations in mexico. *Proc. ACM Hum.-Comput. Interact.*, 1(CSCW), December 2017. doi: 10.1145/3134654. URL `https://doi-org.prx.library.gatech.edu/10.1145/3134654`.

[7] Neil Andersson, Elizabeth Nava-Aguilera, Jorge Arosteguí, Arcadio Morales-Perez, Harold Suazo-Laguna, José Legorreta-Soberanis, Carlos Hernandez-Alvarez, Ildefonso Fernandez-Salas, Sergio Paredes-Solís, Angel Balmaseda, et al. Evidence based community mobilization for dengue prevention in nicaragua and mexico (camino verde, the green way): cluster randomized controlled trial. *Bmj*, 351:h3267, 2015.

[8] Neil Andersson, Jorge Arostegui, Elizabeth Nava-Aguilera, Eva Harris, and Robert J Ledogar. Camino verde (the green way): evidence-based community mobilisation for dengue control in nicaragua and mexico: feasibility study and study protocol for a randomised controlled trial. *BMC public health*, 17(1):407, 2017.

[9] Katy Appleton and Andrew Lovett. Gis-based visualisation of development proposals: reactions from planning and related professionals. *Computers, Environment and Urban Systems*, 29(3): 321–339, 2005.

[10] Pablo Aragón. *Characterizing online participation in civic technologies*. PhD thesis, Universitat Pompeu Fabra, 2019.

[11] Pablo Aragon, Adriana Alvarado Garcia, Christopher A. Le Dantec, Claudia Flores-Saviaga, and Jorge Saldivar. Civic technologies: Research, practice, and open challenges. In *Conference Companion Publication of the 2020 on Computer Supported Cooperative Work and Social Computing*, CSCW '20 Companion, page 537–545, New York, NY, USA, 2020. Association for Computing Machinery. ISBN 9781450380591. doi: 10.1145/3406865.3430888. URL `https://doi.org/10.1145/3406865.3430888`.

[12] Sherry R Arnstein. A Ladder of Citizen Participation. *Journal of the American Institute of Planner*, 35(4):216–224, 1969.

[13] Rebeca Arteta, Jorge Saldivar, Cristhian Parra, and Luca Cernuzzi. Participaaware: Fomentando la participación ciudadana por medio de notificaciones móviles vinculadas a lugares de interés del ciudadano. In *CIbSE*, pages 665–672, 2019.

[14] Mariam Asad and Christopher A. Le Dantec. Illegitimate civic participation: Supporting community activists on the ground. In *Proceedings of the 18th ACM Conference on Computer Supported Cooperative Work & Social Computing*, CSCW '15, page 1694–1703, New York, NY, USA, 2015. Association for Computing Machinery. ISBN 9781450329224. doi: 10.1145/2675133.2675156. URL `https://doi-org.prx.library.gatech.edu/10.1145/2675133.2675156`.

[15] Lori Bowen Ayre and Jim Craner. Technology column: public libraries as civic technology hubs. *Public Library Quarterly*, 36(4):367–374, 2017.

[16] AN Babu, E Niehaus, S Shah, and et.al. Unnithan. Smartphone geospatial apps for dengue control, prevention, prediction, and education: Mosapp, disapp, and the mosquito perception index (mpi). *Environmental monitoring and assessment*, 2019.

[17] Vian Bakir and Andrew McStay. Fake news and the economy of emotions. *Digital Journalism*, 6 (2):154–175, 2018. doi: 10.1080/21670811.2017.1345645. URL `https://doi.org/10.1080/21670811.2017.1345645`.

[18] Shaowen Bardzell. Feminist hci: Taking stock and outlining an agenda for design. In *Proceedings of the SIGCHI Conference on Human Factors in Computing Systems*, CHI '10, page 1301–1310, New York, NY, USA, 2010. Association for Computing Machinery. ISBN 9781605589299. doi: 10.1145/1753326.1753521. URL `https://doi.org/10.1145/1753326.1753521`.

[19] Victor Barreiro Jr. Data.gov.ph launches: Open data for good governance, 01 2014. URL `https://www.rappler.com/nation/48101-open-data-philippines-data-gov-ph`.

[20] Michel Bauwens and Vasilis Kostakis. Towards a new reconfiguration among the state, civil society and the market. *The Journal of Peer Production*, 2015.

[21] Jorn Berends, Wendy Carrara, and Heleen Vollers. Analytical report n6: Open data in cities 2, 2017.

[22] Tony Bergstrom and Karrie Karahalios. Vote and be heard: Adding back-channel signals to social mirrors. In *IFIP Conference on Human-Computer Interaction*, pages 546–559. Springer, 2009.

[23] Laura H Berry, Jessica Koski, Cleo Verkuijl, Claudia Strambo, and Georgia Piggot. *Making space: how public participation shapes environmental decision-making*. JSTOR, 2019.

[24] Kirsten Boehner and Carl DiSalvo. Data, design and civics: An exploratory study of civic tech. In *Proceedings of the 2016 CHI Conference on Human Factors in Computing Systems*, CHI '16, page 2970–2981, New York, NY, USA, 2016. Association for Computing Machinery. ISBN 9781450333627.

[25] James Boyle. The second enclosure movement and the construction of the public domain. *Law and contemporary problems*, 66(1/2):33–74, 2003.

[26] Daren C Brabham. Crowdsourcing the public participation process for planning projects. *Planning Theory*, 8(3):242–262, 2009.

[27] Frank M Bryan. *Real democracy: The New England town meeting and how it works*. University of Chicago Press, 2010.

[28] Michael Canares and Satyarupa Shekhar. Open data and subnational governments: Lessons from developing countries. *The Journal of Community Informatics*, 12(2), Jun. 2016. URL `http://www.ci-journal.net/index.php/ciej/article/view/1260`.

[29] Michael Canares, Dave Marcial, and Marijoe Narca. Enhancing citizen engagement with open government data. *Open Data Research Symposium*, 2015. URL `http://www.opendataresearch.org/dl/symposium2015/odrs2015-paper15.pdf`.

[30] Manuel Castells. The New Public Sphere: Global Civil Society, Communication Networks, and Global Governance. *The ANNALS of the American Academy of Political and Social Science*, 616 (1):78–93, 2008. doi: 10.1177/0002716207311877. URL https://doi.org/10.1177/0002716207311877.

[31] Andrew Chadwick and Christopher May. Interaction between States and Citizens in the Age of the Internet: "e-Government" in the United States, Britain, and the European Union. *Governance*, 16(2):271–300, 2003.

[32] Kathy Charmaz. *Grounded Theory*. American Cancer Society, 2007. ISBN 9781405165518. doi: 10.1002/9781405165518.wbeosg070. URL https://onlinelibrary.wiley.com/doi/abs/10.1002/9781405165518.wbeosg070.

[33] Peter Checkland. *Systems thinking, systems practice*. John Wiley and Sons, 1981.

[34] Peter Checkland and Sue Holwell. Action research: its nature and validity. *Systemic practice and action research*, 11(1):9–21, 1998.

[35] Xinran Chen, Sei-Ching Joanna Sin, Yin-Leng Theng, and Chei Sian Lee. Why students share misinformation on social media: Motivation, gender, and study-level differences. *The Journal of Academic Librarianship*, 41(5):583 – 592, 2015. ISSN 0099-1333. doi: 10.1016/j.acalib.2015.07.003. URL http://www.sciencedirect.com/science/article/pii/S0099133315001494.

[36] David Cheruiyot, Stefan Baack, and Raul Ferrer-Conill. Data journalism beyond legacy media: The case of african and european civic technology organizations. *Digital Journalism*, 7(9):1215–1229, 2019.

[37] Chun-Wei Chiang, Eber Betanzos, and Saiph Savage. Exploring blockchain for trustful collaborations between immigrants and governments. In *Extended Abstracts of the 2018 CHI Conference on Human Factors in Computing Systems*, pages 1–6, 2018.

[38] Joohee Choi and Yla Tausczik. Characteristics of collaboration in the emerging practice of open data analysis. In *Proceedings of the 2017 ACM Conference on Computer Supported Cooperative Work and Social Computing*, CSCW '17, pages 835–846, New York, NY, USA, 2017. ACM. ISBN 978-1-4503-4335-0. doi: 10.1145/2998181.2998265. URL http://doi.acm.org/10.1145/2998181.2998265.

[39] Ian Nicolas P. Cigaral. Eo on freedom of information takes effect today: Palace, November 2016. URL http://www.bworldonline.com/content.php?section=Nation&title=eo-on-freedom-of-information-takes-effect-today-palace&id=136876.

[40] Benjamin Y Clark, Jeffrey L Brudney, and Sung-Gheel Jang. Coproduction of government services and the new information technology: Investigating the distributional biases. *Public Administration Review*, 73(5):687–701, 2013.

[41] Adele E Clarke and Susan Leigh Star. The social worlds framework: A theory/methods package. *The handbook of science and technology studies*, 3:113–137, 2008.

[42] RY Clarke. Civic tech fuels us state and local government transformation, 2014.

[43] Codeforall.org. *Code for All*, 2020 (accessed October 19, 2020). URL https://codeforall.org.

[44] Cooke. *The Relational State*. Institute for Public Policy Research, 2012.

[45] Eric Corbett and Christopher A. Le Dantec. Going the distance: Trust work for citizen participation. In *Proceedings of the 2018 CHI Conference on Human Factors in Computing Systems*, CHI '18, page 1–13, New York, NY, USA, 2018. Association for Computing Machinery. ISBN 9781450356206. doi: 10.1145/3173574.3173886. URL https://doi-org.prx.library.gatech.edu/10.1145/3173574.3173886.

[46] Eric Corbett and Christopher A. Le Dantec. Exploring trust in digital civics. In *Proceedings of the 2018 Designing Interactive Systems Conference*, DIS '18, page 9–20, New York, NY, USA, 2018. Association for Computing Machinery. ISBN 9781450351980.

[47] Sasha Costanza-Chock, Taya Wagoner, Berhan Taye, Caroline Rivas, Chris Schweidler, Georgia Bullen, and the T4SJ Project. #MoreThanCode: Practitioners reimagine the landscape of technology for justice and equity., 2018.

[48] Costis Dallas. Digital curation beyond the "wild frontier": A pragmatic approach. *Archival Science*, 16(4):421–457, 2016.

[49] Anna De Liddo, Brian Plüss, and Paul Wilson. A novel method to gauge audience engagement with televised election debates through instant, nuanced feedback elicitation. In *Proceedings of the 8th International Conference on Communities and Technologies*, pages 68–77, 2017.

[50] Anna De Liddo, Nieves Pedreira Souto, and Brian Plüss. Let's replay the political debate: Hypervideo technology for visual sensemaking of televised election debates. *International Journal of Human-Computer Studies*, 145:102537, 2020.

[51] United Nations DESA. United nations e-government survey 2020: Digital government in the decade of action for sustainable development. `https://www.un.org/development/desa/publications/publication/2020-united-nations-e-government-survey`. (Accessed on 09/04/2020).

[52] Michael A Devito, Ashley Marie Walker, Jeremy Birnholtz, Kathryn Ringland, Kathryn Macapagal, Ashley Kraus, Sean Munson, Calvin Liang, and Herman Saksono. Social technologies for digital wellbeing among marginalized communities. In *Conference Companion Publication of the 2019 on Computer Supported Cooperative Work and Social Computing*, pages 449–454, 2019.

[53] Jessa Dickinson, Mark Díaz, Christopher A Le Dantec, and Sheena Erete. " the cavalry ain't coming in to save us" supporting capacities and relationships through civic tech. *Proceedings of the ACM on Human-Computer Interaction*, 3(CSCW):1–21, 2019.

[54] Daniel Dietrich. The role of civic tech communities in psi reuse and open data policies. *European Public Sector Information Platform Topic Report*, 5, 2015.

[55] Shubhra Dixit. The hindu. 2017. URL `https://www.thehindu.com/entertainment/movies/the-curious-case-of-indian-animation/article17646834.ece`.

[56] Stacy Donohue. Engines of change: What civic tech can learn from social movements. 2016.

[57] Robert Dreibelbis, Peter J Winch, Elli Leontsini, Kristyna RS Hulland, Pavani K Ram, Leanne Unicomb, and Stephen P Luby. The integrated behavioural model for water, sanitation, and hygiene: a systematic review of behavioural models and a framework for designing and evaluating behaviour change interventions in infrastructure-restricted settings. *BMC public health*, 13(1): 1015, 2013.

[58] Veena B Dubal. Wage slave or entrepreneur: Contesting the dualism of legal worker identities. *Calif. L. Rev.*, 105:65, 2017.

[59] Andrew Duffy, Edson Tandoc, and Rich Ling. Too good to be true, too good not to share: the social utility of fake news. *Information, Communication & Society*, pages 1–15, 2019. doi: 10.1080/1369118X.2019.1623904. URL `https://doi.org/10.1080/1369118X.2019.1623904`.

[60] Amir Ehsaei, Thomas Sweet, Raphael Garcia, Laura Adleman, and Jean M Walsh. Successful public outreach programs for green infrastructure projects. In *International Low Impact Development Conference 2015: LID: It Works in All Climates and Soils*, pages 74–92, 2015.

[61] Sheena Erete, Emily Ryou, Geoff Smith, Khristina Marie Fassett, and Sarah Duda. Storytelling with data: Examining the use of data by non-profit organizations. In *Proceedings of the 19th ACM Conference on Computer-Supported Cooperative Work & Social Computing*, CSCW '16, pages 1273–1283, New York, NY, USA, 2016. ACM. ISBN 978-1-4503-3592-8. doi: 10.1145/2818048.2820068. URL `http://doi.acm.org/10.1145/2818048.2820068`.

[62] Margarita Escobar de Morel. La participación ciudadana en paraguay. análisis a partir de la transición democrática. *Revista Internacional de Investigación en Ciencias Sociales*, 8(1):119–140, 2012.

[63] Andrew Feenberg. *Technosystem: The Social Life of Reason*. Harvard University Press, 2017.

[64] Code for Pakistan. *Impact Report 2019*, 2019 (accessed October 19, 2020). URL `https://codeforpakistan.org/impact-report-2019`.

[65] Code for Pakistan. *Code for Pakistan*, 2020 (accessed October 19, 2020). URL `https://codeforpakistan.org`.

[66] Code for Pakistan. *CrowdDurshal: A lifeline for the startup ecosystem in KP*, 2020 (accessed October 19, 2020). URL `https://codeforpakistan.org/blog/portfolio-item/crowddurshal`.

[67] Code for Pakistan. *Pakistan, D. DocSeek*, 2020 (accessed October 19, 2020). URL `https://codeforpakistan.org/blog/portfolio-item/docseek`.

[68] Code for Pakistan. *Pakistan, K. KP Gov Innovation Fellowship Program*, 2020 (accessed October 19, 2020). URL `https://codeforpakistan.org/programs/fellowship`.

[69] Code for Pakistan. *Grants Management and Disbursement System*, 2020 (accessed October 19, 2020). URL `https://codeforpakistan.org/blog/portfolio-item/grants-management-and-disbursement-system`.

[70] Code for Pakistan. *KP Excise Motor Vehicle Seizure and Confiscation System*, 2020 (accessed October 19, 2020). URL `https://codeforpakistan.org/blog/portfolio-item/kp-emvscs/`.

[71] Code for Pakistan. *Report Electricity Theft, N. NoKunda*, 2020 (accessed October 19, 2020). URL `https://codeforpakistan.org/blog/portfolio-item/nokunda/`.

[72] Code for Pakistan. *Open Data Portal for PMRU*, 2020 (accessed October 19, 2020). URL `https://codeforpakistan.org/blog/portfolio-item/open-data-portal-for-pmru`.

[73] Code for Pakistan. *Reimagining the KP Government Innovation Fellowship Program Amidst the COVID-19 Pandemic*, 2020 (accessed October 19, 2020). URL `https://medium.com/@CodeforPakistan/reimagining-the-kp-government-innovation-fellowship-program-amidst-the-covid-19-pa`

[74] Oscar Franco-Bedoya, David Ameller, Dolors Costal, and Xavier Franch. Open source software ecosystems: A systematic mapping. *Information and software technology*, 91:160–185, 2017.

[75] Deen Freelon. Discourse architecture, ideology, and democratic norms in online political discussion. *New Media & Society*, 17(5):772–791, 2015.

[76] Sonia Frías. Violación e intento de violación de mujeres, patrones de búsqueda de ayuda y denuncia. un análisis a partir de la endireh 2016. *Papeles de población*, 24(95):237–272, 2018.

[77] Archon Fung. Varieties of Participation in Complex Governance. *Public Administration Review*, 66(s1):66–75, 12 2006. ISSN 0033-3352. doi: 10.1111/j.1540-6210.2006.00667.x. URL `http://doi.wiley.com/10.1111/j.1540-6210.2006.00667.x`.

[78] Ana Martha Galindes and Marco Angelo S. Zaplan. Open research in the philippines: The lessons and challenges, Apr 2018. URL `https://blog.okfn.org/2018/04/24/open-research-in-the-philippines-the-lessons-and-challenges/`.

[79] John Gastil. *Political communication and deliberation*. Sage, 2008.

[80] Hollie Russon Gilman. Civic tech for urban collaborative governance. *PS, Political Science & Politics*, 50(3):744, 2017.

[81] Eric Gordon and Rogelio Alejandro Lopez. The practice of civic tech: Tensions in the adoption and use of new technologies in community based organizations. *Media and Communication*, 7 (3):57–68, 2019.

[82] Eric Gordon and Paul Mihailidis. *Civic media: Technology, design, practice*. MIT Press, 2016.

[83] Eric Gordon, Steven Schirra, and Justin Hollander. Immersive planning: a conceptual model for designing public participation with new technologies. *Environment and Planning B: Planning and Design*, 38(3):505–519, 2011.

[84] Mary Gray. Databite No. 119: Mary L. Gray, 2019. URL `https://www.youtube.com/watch?v=zj2DEQCOTh0&t=784s`.

[85] Pedro Gómez Silgueira. El parque caballero debe resurgir de sus ruinas, January 2018. `https://vivapy.wordpress.com/2018/01/05/el-parque-caballero-debe-resurgir-de-sus-ruinas/`.

[86] Civic Hall. Making the uk's political debates more responsive to public needs | civic hall. `https://civichall.org/civicist/political-debates-more-responsive-public-needs/`, 2016.

[87] Benjamin V Hanrahan, Ning F Ma, Eber Betanzos, and Saiph Savage. Reciprocal research: Providing value in design research from the outset in the rural united states. In *Proceedings of the 2020 International Conference on Information and Communication Technologies and Development*, pages 1–5, 2020.

[88] Garrett Hardin. The tragedy of the commons. *Science*, 162(3859):1243–1248, 1968. ISSN 0036-8075.

[89] Michael Hardt and Antonio Negri. *Assembly*. Oxford University Press, 2017.

[90] Abdiwahab Hashi, Abera Kumie, and Janvier Gasana. Hand washing with soap and wash educational intervention reduces under-five childhood diarrhoea incidence in jigjiga district, eastern ethiopia: a community-based cluster randomized controlled trial. *Preventive medicine reports*, 6:361–368, 2017.

[91] Anne Helmond. The Platformization of the Web: Making Web Data Platform Ready. *Social Media + Society*, 1:1–11, 2015.

[92] Gabriel Bodin Hetland. *Making Democracy Real: Participatory Governance in Urban Latin America*. PhD thesis, University of California, Berkeley, 2015.

[93] Sarah Holder. This Uber Driver Started a Legal Battle That Could Upend the Gig Economy, 2019. URL `https://www.bloomberg.com/news/articles/2019-08-22/why-uber-drivers-are-fighting-for-their-data`.

[94] Youyang Hou. *Understanding the Design and Implementation of Civic Technologies in Resource-Limited Public Organizations*. PhD thesis, University of Michigan, 2018.

[95] Yu-Tang Hsiao, Shu-Yang Lin, Audrey Tang, Darshana Narayanan, and Claudina Sarahe. vtaiwan: An empirical study of open consultation process in taiwan. 2018.

[96] Seth Hunter, Pattie Maes, Stacey Scott, and Henry Kaufman. Memtable: an integrated system for capture and recall of shared histories in group workspaces. In *Proceedings of the SIGCHI Conference on Human Factors in Computing Systems*, pages 3305–3314. ACM, 2011.

[97] Ronald F Inglehart and Pippa Norris. Trump, brexit, and the rise of populism: Economic have-nots and cultural backlash. 2016.

[98] Judith E Innes and David E Booher. Consensus building and complex adaptive systems: A framework for evaluating collaborative planning. *Journal of the American planning association*, 65(4):412–423, 1999.

[99] Judith E Innes and David E Booher. Reframing public participation: strategies for the 21st century. *Planning theory & practice*, 5(4):419–436, 2004.

[100] Lilly Irani. Difference and Dependence among Digital Workers: The Case of Amazon Mechanical Turk. *South Atlantic Quarterly*, 114:225–234, 2015.

[101] Michael C Jackson. *Systems thinking: Creative holism for managers*. Citeseer, 2003.

[102] Slinger Jansen, Michael A Cusumano, and Sjaak Brinkkemper. *Software ecosystems: analyzing and managing business networks in the software industry*. Edward Elgar Publishing, 2013.

[103] Mahmood Jasim, Pooya Khaloo, Somin Wadhwa, Amy X. Zhang, Ali Sarvghad, and Narges Mahyar. Communityclick: Capturing and reporting community feedback from town halls to improve inclusivity. *The 23rd ACM Conference on Computer-Supported Cooperative Work and Social Computing*, 2020. 29 pages. To Appear.

[104] Asitha DL Jayawardena, Sarah Romano, Kevin Callans, M Shannon Fracchia, and Christopher J Hartnick. Family-centered information dissemination: A multidisciplinary virtual covid-19 "town hall". *Otolaryngology–Head and Neck Surgery*, page 0194599820935419, 2020.

[105] Henry Jenkins, Mizuko Ito, et al. *Participatory culture in a networked era: A conversation on youth, learning, commerce, and politics*. John Wiley & Sons, 2015.

[106] Ian G Johnson, John Vines, Nick Taylor, Edward Jenkins, and Justin Marshall. Reflections on deploying distributed consultation technologies with community organisations. In *Proceedings of the 2016 CHI Conference on Human Factors in Computing Systems*, pages 2945–2957, 2016.

[107] Harry Hochheis Jonathan Lazar, Jinjuan Heidi Feng. *Research Methods in Human-Computer Interaction*. 2017.

[108] Emilia Kacprzak, Laura Koesten, Luis-Daniel Ibáñez, Tom Blount, Jeni Tennison, and Elena Simperl. Characterising dataset search - an analysis of search logs and data requests. *Journal of Web Semantics*, 55:37–55, 2019.

[109] Anne Marie Kanstrup and Ellen Christiansen. Model power: still an issue? In *Proceedings of the 4th decennial conference on Critical computing: between sense and sensibility*, pages 165–168, 2005.

[110] Amit Katwala. The tricks johnson and corbyn will use to win the itv leaders debate | wired uk. `https://www.wired.co.uk/article/itv-leaders-debate-general-election-2019`, 2019.

[111] Vera Khovanskaya, Lynn Dombrowski, Jeffrey Rzeszotarski, and Phoebe Sengers. The Tools of Management: Adapting Historical Union Tactics to Platform-Mediated Labor. *Proceedings of the ACM on Human-Computer Interaction*, 3:1–22, 2019.

[112] Soomin Kim, Jinsu Eun, Changhoon Oh, Bongwon Suh, and Joonhwan Lee. Bot in the bunch: Facilitating group chat discussion by improving efficiency and participation with a chatbot. In *Proceedings of the 2020 CHI Conference on Human Factors in Computing Systems*, pages 1–13, 2020.

[113] Knight Foundation. The Emergence of Civic Tech : Investments in a Growing Field, 2013. URL `http://www.knightfoundation.org/media/uploads/publication_pdfs/knight-civic-tech.pdf`.

[114] Antti Knutas, Victoria Palacin, Giovanni Maccani, and Markus Helfert. Software engineering in civic tech a case study about code for ireland. In *2019 IEEE/ACM 41st International Conference on Software Engineering: Software Engineering in Society (ICSE-SEIS)*, pages 41–50. IEEE, 2019.

[115] Laura M. Koesten, Emilia Kacprzak, Jenifer F. A. Tennison, and Elena Simperl. The trials and tribulations of working with structured data: -a study on information seeking behaviour. In *Proceedings of the 2017 CHI Conference on Human Factors in Computing Systems*, CHI '17, pages 1277–1289, New York, NY, USA, 2017. ACM. ISBN 978-1-4503-4655-9. doi: 10.1145/3025453.3025838. URL http://doi.acm.org/10.1145/3025453.3025838.

[116] Peter Kollock. Social dilemmas: The anatomy of cooperation. *Annual review of sociology*, 24 (1):183–214, 1998.

[117] Constantine Kontokosta, Christopher Tull, David Marulli, Maha Yaqub, and Renate Pinggera. Web-based visualization and prediction of urban energy use from building benchmarking data. 09 2015.

[118] Christopher A. Le Dantec, Mariam Asad, Aditi Misra, and Kari E. Watkins. Planning with crowdsourced data: Rhetoric and representation in transportation planning. In *Proceedings of the 18th ACM Conference on Computer Supported Cooperative Work & Social Computing*, CSCW '15, page 1717–1727, New York, NY, USA, 2015. Association for Computing Machinery. ISBN 9781450329224. doi: 10.1145/2675133.2675212. URL https://doi-org.prx. library.gatech.edu/10.1145/2675133.2675212.

[119] Myeong Lee and Brian S Butler. How are information deserts created? a theory of local information landscapes. *Journal of the Association for Information Science and Technology*, 70(2): 101–116, 2019.

[120] Sung-Chul Lee, Jaeyoon Song, Eun-Young Ko, Seongho Park, Jihee Kim, and Juho Kim. Solutionchat: Real-time moderator support for chat-based structured discussion. In *Proceedings of the 2020 CHI Conference on Human Factors in Computing Systems*, pages 1–12, 2020.

[121] Jeremy R Levine and Carl Gershenson. From political to material inequality: Race, immigration, and requests for public goods. In *Sociological Forum*, volume 29, pages 607–627. Wiley Online Library, 2014.

[122] Peter Levine, Archon Fung, and John Gastil. Future directions for public deliberation. *Journal of Public Deliberation*, 1(1), 2005.

[123] Rensis Likert. A technique for the measurement of attitudes. *Archives of psychology*, 1932.

[124] Dennis Linders. From e-government to we-government: Defining a typology for citizen co-production in the age of social media. *Government Information Quarterly*, 29(4):446 – 454, 2012. ISSN 0740-624X. doi: https://doi.org/10.1016/j.giq.2012.06.003. URL http: //www.sciencedirect.com/science/article/pii/S0740624X12000883. Social Media in Government - 12th Annual International Conference on Digital Government Research (dg.o2011).

[125] Lucía Liste and Knut H Sørensen. Consumer, client or citizen? How Norwegian local governments domesticate website technology and configure their users. *Information, Communication & Society*, 18(7):733–746, 7 2015. ISSN 1369-118X. doi: 10.1080/1369118X.2014.993678. URL http://www.tandfonline.com/doi/abs/10.1080/1369118X.2014.993678.

[126] Can Liu, Mara Balestrini, and Giovanna Nunes Vilaza. From social to civic: Public engagement with iot in places and communities. In *Social Internet of Things*, pages 185–210. Springer, 2019.

[127] Derk Loorbach, Niki Frantzeskaki, and Roebin Lijnis Huffenreuter. Transition management: taking stock from governance experimentation. *Journal of Corporate Citizenship*, (58):48–66, 2015.

[128] Russanne Low. Citizen scientists as community agents of change: Globe observer mosquito habitat mapper. *AGUFM*, 2018.

[129] Carolyn Lukensmeyer, Susanna Haas Lyons, and America Speaks. 21 st century town meeting®, 2020.

[130] Christoph Lutz and Christian Pieter Hoffmann. The dark side of online participation: exploring non-, passive and negative participation. *Information Communication and Society*, 20(6):876–897, 2017. ISSN 14684462. doi: 10.1080/1369118X.2017.1293129. URL `http://dx.doi.org/10.1080/1369118X.2017.1293129`.

[131] May O Lwin, Santosh Vijaykumar, and Owen Noel Newton et.al. Fernando. A 21st century approach to tackling dengue: Crowdsourced surveillance, predictive mapping and tailored communication. *Acta tropica*, 2014.

[132] May O Lwin, Santosh Vijaykumar, and Gentatsu et.al. Lim. Baseline evaluation of a participatory mobile health intervention for dengue prevention in sri lanka. *Health Education & Behavior*, 2016.

[133] Monika Mačiulienė and Aelita Skaržauskienė. Building the capacities of civic tech communities through digital data analytics. *Journal of Innovation & Knowledge*, 2019.

[134] A Mahmood. *The dirty truth: 41 million Pakistanis without toilets*, 2020 (accessed October 19, 2020). URL `https://www.dawn.com/news/1168630`.

[135] Narges Mahyar, Kelly J Burke, Jialiang Ernest Xiang, Siyi Cathy Meng, Kellogg S Booth, Cynthia L Girling, and Ronald W Kellett. Ud co-spaces: A table-centred multi-display environment for public engagement in urban design charrettes. In *Proceedings of the 2016 ACM on Interactive Surfaces and Spaces*, pages 109–118. ACM, 2016.

[136] Narges Mahyar, Diana V Nguyen, Maggie Chan, Jiayi Zheng, and Steven Dow. The civic data deluge: Understanding the challenges of analyzing large-scale community input. ACM Designing Interactive Systems (DIS) (to appear), 2019.

[137] Jane Mansbridge, Janette Hartz-Karp, Matthew Amengual, and John Gastil. Norms of deliberation: An inductive study. 2006.

[138] Helen Margetts and Patrick Dunleavy. The second wave of digital-era governance: a quasi-paradigm for government on the Web. *Philosophical Transactions of the Royal Society A: Mathematical, Physical and Engineering Sciences*, 371(1987):20120382, 2013. URL `https://royalsocietypublishing.org/doi/abs/10.1098/rsta.2012.0382`.

[139] Michael Margolis and David Resnick. *Politics as usual: The Cyberspace 'Revolution'*, volume 6. SAGE Publications Ltd, 2000.

[140] Estanislao Gacitúa Marió, Annika Silva-Leander, and Miguel Carter. Paraguay: Social development issues for poverty alleviation. *Social Development Papers*, 63, 2004.

[141] Andrew May and Tracy Ross. The design of civic technology: factors that influence public participation and impact. *Ergonomics*, 61(2):214–225, 2018.

[142] Lorraine Green Mazerolle. *Managing calls to the police with 911/311 systems*. US Department of Justice, Ofifce of Justice Programs, National Institute of Justice, Washington, D.C., 2005.

[143] Curtis McCord. Popular technologies: Expertise, organization and production in civic tech toronto. In *EASST-4S virPrague 2020*.

[144] Curtis McCord and Christoph Becker. Sidewalk and toronto: Critical systems heuristics and the smart city. In *ICT4S '19*, 2019.

[145] Michael McGuire. Collaborative public management: Assessing what we know and how we know it.

[146] Brian McInnis, Alissa Centivany, Juho Kim, Marta Poblet, Karen Levy, and Gilly Leshed. Crowdsourcing law and policy: a design-thinking approach to crowd-civic systems. In *Companion of the 2017 ACM conference on Computer Supported Cooperative Work and Social Computing*, pages 355–361, 2017.

[147] Spencer McKay and Chris Tenove. Disinformation as a threat to deliberative democracy. *Political Research Quarterly*, 2020. doi: 10.1177/1065912920938143. URL https://doi.org/10.1177/1065912920938143.

[148] Amanda Meng, Carl DiSalvo, and Ellen Zegura. Collaborative data work towards a caring democracy. *Proc. ACM Hum.-Comput. Interact.*, 3(CSCW), November 2019. doi: 10.1145/3359144. URL https://doi-org.prx.library.gatech.edu/10.1145/3359144.

[149] Sanju Menon, Weiyu Zhang, and Simon T. Perrault. Nudge for deliberativeness: How interface features influence online discourse. In *Proceedings of the 2020 CHI Conference on Human Factors in Computing Systems*, CHI '20, page 1–13, New York, NY, USA, 2020. Association for Computing Machinery. ISBN 9781450367080.

[150] David G Messerschmitt, Clemens Szyperski, et al. Software ecosystem: understanding an indispensable technology and industry. *MIT Press Books*, 1, 2005.

[151] Amina J. Mohammed. Participation, consultation and engagement: Critical elements for an effective implementation of the 2030 agenda, May 2017. (Accessed on 09/04/2020).

[152] David Moore. Unpacking civic tech–inside and outside of government. 2015. URL http://www.participatorypolitics.org/civic-tech-inside-and-outside-of-government/.

[153] Judith Morse, Margaret Ruggieri, and Karen Whelan-Berry. Clicking our way to class discussion. *American Journal of Business Education*, 3(3):99–108, 2010.

[154] Michael J. Muller. Participatory Design. In Andrew Sears and Julie A. Jacko, editors, *Human-Computer Interaction*. CRC Press, 3 2007. ISBN 9780429139390. doi: 10.1201/9781420088892. URL https://www.taylorfrancis.com/books/9781420088892.

[155] Stephen Peter Mwangungulu, Robert David Sumaye, Alex Julius Limwagu, Doreen Josen Siria, Emmanuel Wilson Kaindoa, and Fredros Oketch Okumu. Crowdsourcing vector surveillance: using community knowledge and experiences to predict densities and distribution of outdoor-biting mosquitoes in rural tanzania. *PLoS One*, 2016.

[156] Seungahn Nah and Masahiro Yamamoto. civic technology and community building: interaction effects between integrated connectedness to a storytelling network (icsn) and internet and mobile uses on civic participation. *Journal of Computer-Mediated Communication*, 22(4):179–195, 2017.

[157] United Nations. Sustainable development goals. united nations. *United Nations Sustainable Development*, 2015.

[158] Matti Nelimarkka. A review of research on participation in democratic decision-making presented at sigchi conferences. toward an improved trading zone between political science and hci. *Proceedings of the ACM on Human-Computer Interaction*, 3(CSCW):1–29, 2019.

[159] Jakob Nielsen and Rolf Molich. Heuristic evaluation of user interfaces. In *SIGCHI conference on Human factors in computing systems*, 1990.

[160] Beth Simone Noveck. *Wiki government: how technology can make government better, democracy stronger, and citizens more powerful*. Brookings Institution Press, 2009.

[161] Beth Simone Noveck. *Smarter Citizens, Smarter State.* Havard University Press, 2015.

[162] Patrick Olivier and Peter Wright. Digital civics: taking a local turn. *Interactions*, 22(4):61–63, 2015.

[163] Open Government Partnership.

[164] Tim O'Reilly. Government as a Platform. *Innovations: Technology, Governance, Globalization*, 6(1):13–40, 2011.

[165] Elinor Ostrom. *Governing the commons: The evolution of institutions for collective action*. Cambridge university press, 1990.

[166] Elinor Ostrom. How types of goods and property rights jointly affect collective action. *Journal of theoretical politics*, 15(3):239–270, 2003.

[167] Daniel Tumminelli O'Brien, Dietmar Offenhuber, Jessica Baldwin-Philippi, Melissa Sands, and Eric Gordon. Uncharted territoriality in coproduction: The motivations for 311 reporting. *Journal of Public Administration Research and Theory*, 27(2):320–335, 2017.

[168] Baar T. Grunewald P. Project website:training active listening by using digital technologies – isooko. `http://isooko.eu/2019/03/26/ training-active-listening-by-using-digital-technologies/`, 2019.

[169] Ugo Pagallo, Paola Aurucci, Pompeu Casanovas, Raja Chatila, Patrice Chazerand, Virginia Dignum, Christoph Luetge, Robert Madelin, Burkhard Schafer, and Peggy Valcke. Ai4people-on good ai governance: 14 priority actions, a smart model of governance, and a regulatory toolbox. 2019.

[170] Victoria Palacin, Matti Nelimarkka, Pedro Reynolds-Cuéllar, and Christoph Becker. The design of pseudo-participation. In *Proceedings of the 16th Participatory Design Conference 2020 - Participation(s) Otherwise - Volume 2*, PDC '20, page 40–44, New York, NY, USA, 2020. Association for Computing Machinery. ISBN 9781450376068. doi: 10.1145/3384772.3385141. URL `https://doi.org/10.1145/3384772.3385141`.

[171] C. Parra, L. Cernuzzi, and R. et.al. Rojas. [upcoming] synergies between technology, participation, and citizen science in a community-based dengue prevention program. *American behavioral scientist. Special Issue on "ICTs for Community Development: Bridging conceptual, theoretical and methodological boundaries"*, 2020.

[172] Mayur Patel, Jon Sotsky, Sean Gourley, and Daniel Houghton. The emergence of civic tech: Investments in a growing field. *Knight Foundation*, 2013.

[173] Evan M Peck, Sofia E Ayuso, and Omar El-Etr. Data is personal: Attitudes and perceptions of data visualization in rural pennsylvania. In *Proceedings of the 2019 CHI Conference on Human Factors in Computing Systems*, pages 1–12, 2019.

[174] Tiago Peixoto and Micah L Sifry. Civic tech in the global south, 2017.

[175] Karl Peltzer and Supa Pengpid. Oral and hand hygiene behaviour and risk factors among in-school adolescents in four southeast asian countries. *International Journal of Environmental Research and Public Health*, 11(3):2780–2792, 2014.

[176] Marta Poblet and Carles Sierra. Understanding help as a commons. *International Journal of the Commons*, 14(1), 2020.

[177] Marta Poblet, Esteban García-Cuesta, and Pompeu Casanovas. Crowdsourcing tools for disaster management: A review of platforms and methods. In *International Workshop on AI Approaches to the Complexity of Legal Systems*, pages 261–274. Springer, 2013.

[178] Marta Poblet, Esteban García-Cuesta, and Pompeu Casanovas. Crowdsourcing roles, methods and tools for data-intensive disaster management. *Information Systems Frontiers*, 20(6):1363–1379, 2018.

[179] Marta Poblet, Pompeu Casanovas, and Víctor Rodríguez-Doncel. *Linked Democracy: Foundations, tools, and applications*. Springer Nature, 2019.

[180] Thamy Pogrebinschi. Paraguay, 2017. `https://latinno.net/es/country/paraguay/`.

[181] Propakistani.pk. *3G/4G Users Cross the 61 Million Mark in Pakistan*, 2020 (accessed October 19, 2020). URL `https://propakistani.pk/2019/01/30/ 3g-4g-users-cross-the-61-million-mark-in-pakistan`.

[182] Annette Prüss-Ustün, Jennyfer Wolf, Jamie Bartram, Thomas Clasen, Oliver Cumming, Matthew C Freeman, Bruce Gordon, Paul R Hunter, Kate Medlicott, and Richard Johnston. Burden of disease from inadequate water, sanitation and hygiene for selected adverse health outcomes: An updated analysis with a focus on low-and middle-income countries. *International journal of hygiene and environmental health*, 222(5):765–777, 2019.

[183] Aare Puussaar, Ian G. Johnson, Kyle Montague, Philip James, and Peter Wright. Making open data work for civic advocacy. *Proc. ACM Hum.-Comput. Interact.*, 2(CSCW):143:1–143:20, November 2018. ISSN 2573-0142. doi: 10.1145/3274412. URL http://doi.acm.org/10.1145/3274412.

[184] Mark S Reed. Stakeholder participation for environmental management: a literature review. *Biological conservation*, 141(10):2417–2431, 2008.

[185] Juliana Rotich. Ushahidi: Empowering citizens through crowdsourcing and digital data collection. *Field Actions Science Reports. The journal of field actions*, (Special Issue 16):36–38, 2017.

[186] Gene Rowe and Lynn J Frever. A Typology of Public Engagement Mechanisms. *Science, Technology & Human Values*, 30(2):251–290, 2005. ISSN 0162-2439. doi: 10.1177/0162243904271724.

[187] Rebecca Rumbul. Novel online approaches to citizen engagement: Empowering citizens and facilitating civic participation through digital innovation in new zealand and australia. 2015.

[188] Rebecca Rumbul. Developing transparency through digital means? examining institutional responses to civic technology in latin america. *JeDEM-eJournal of eDemocracy and Open Government*, 8(3):12–31, 2016.

[189] Rebecca Rumbul. Tools for transparency? institutional barriers to effective civic technology in latin america. In *2016 Conference for E-Democracy and Open Government (CeDEM)*, pages 147–155. IEEE, 2016.

[190] Jathan Sadowski. The Internet of Landlords: Digital Platforms and New Mechanisms of Rentier Capitalism. *Antipode*, 52:562–580, 2020.

[191] Jorge Saldivar, Cristhian Parra, Carlos Rodríguez, Luca Cernuzzi, and Vincenzo D'Andrea. Participa: fostering civic participation for public services innovation. In *Participation for Development Workshop, Participatory Design Conference, PDC '14, Windhoek, Namibia, October 6-10, 2014.*, 2014.

[192] Jorge Saldivar, Cristhian Parra, Marcelo Alcaraz, Rebeca Arteta, and Luca Cernuzzi. Civic technology for social innovation. *Computer Supported Cooperative Work (CSCW)*, 28(1-2):169–207, 2019.

[193] Saiph Savage, Andres Monroy-Hernandez, and Tobias Höllerer. Botivist: Calling volunteers to action using online bots. In *Proceedings of the 19th ACM Conference on Computer-Supported Cooperative Work & Social Computing*, CSCW '16, page 813–822, New York, NY, USA, 2016. Association for Computing Machinery. ISBN 9781450335928. doi: 10.1145/2818048.2819985. URL https://doi-org.prx.library.gatech.edu/10.1145/2818048.2819985.

[194] Andrew Schrock. *Civic tech: Making technology work for people*. Rogue Academic Press, 2018.

[195] Richard W Schwester, Tony Carrizales, and Marc Holzer. An examination of the municipal 311 system. *International Journal of Organization Theory and Behavior*, 12(2):218, 2009.

[196] Dhavan V. Shah, Nojin Kwak, and R. Lance Holbert. "connecting" and "disconnecting" with civic life: Patterns of internet use and the production of social capital. *Political Communication*, 18(2): 141–162, 2001. doi: 10.1080/105846001750322952. URL https://doi.org/10.1080/105846001750322952.

[197] Emily Shaw. Why civic technologists should still care about e-gov, 2015.

[198] Yang Shi, Yang Wang, and John Chen. IdeaWall : Improving Creative Collaboration through Combinatorial Visual Stimuli. *Proceedings of the 2017 ACM Conference on Computer Supported Cooperative Work and Social Computing - CSCW '17*, pages 594–603, 2017.

[199] Shun Shiramatsu, Teemu Tossavainen, Tadachika Ozono, and Toramatsu Shintani. Towards continuous collaboration on civic tech projects: use cases of a goal sharing system based on linked open data. In *International Conference on Electronic Participation*, pages 81–92. Springer, 2015.

[200] Micah Sifry, Matt Stempeck, and Erin Simpson. Civic tech field guide. *URL: https://civictech.guide*, 2017.

[201] Micah L Sifry. Civic tech and engagement: In search of a common language. 5, 2014. URL `http://techpresident.com/news/25261/civic-tech-and-engagement-search-common-language`.

[202] Jesper Simonsen and Toni Robertson. *Routledge International Handbook of Participatory Design*. Routledge, 2012.

[203] Jesper Simonsen, Jørgen Ole Bærenholdt, Monika Büscher, and John Damm Scheuer. *Design research: Synergies from interdisciplinary perspectives*. Routledge, 2010.

[204] Aelita Skarzauskiene. Monitoring collective intelligence: A survey of lithuanian civic tech. In *Companion of the 2018 ACM Conference on Computer Supported Cooperative Work and Social Computing*, pages 277–280, 2018.

[205] Aelita Skarzauskiene and Monika Maciuliene. Mapping international civic technologies platforms, 2020. Paper presented at the 2020 annual conference of International Communication Association. Virtual.

[206] Adrian Smith and Pedro Prieto Martín. Going beyond the smart city? implementing technopolitical platforms for urban democracy in madrid and barcelona. *Journal of Urban Technology*, pages 1–20, 2020.

[207] Graham Smith. *Democratic innovations: Designing institutions for citizen participation*. Cambridge University Press, 2009.

[208] Tom Steinberg. 'civic tech' has won the name-game. but what does it mean?, 2014. URL `https://www.mysociety.org/2014/09/08/civic-tech-has-won-the-name-game-but-what-does-it-mean/`. (Accessed on 09/12/2018).

[209] Nathalie Stembert and Ingrid J Mulder. Love your city! an interactive platform empowering citizens to turn the public domain into a participatory domain. In *International Conference Using ICT, Social Media and Mobile Technologies to Foster Self-Organisation in Urban and Neighbourhood Governance, Delft, The Netherlands, 16-17 May 2013*, 2013.

[210] Matt Stempeck. Towards a taxonomy of civic technology. 7:2016, 2016. URL `https://blogs.microsoft.com/on-the-issues/2016/04/27/towards-taxonomy-civic-technology/`.

[211] Anselm Strauss. A social world perspective. *Studies in symbolic interaction*, 1(1):119–128, 1978.

[212] Anselm Strauss. Social worlds and legitimation processes. *Studies in symbolic interaction*, 4 (17):1, 1982.

[213] Yu Sun and Wenjie Yan. The power of data from the global south: environmental civic tech and data activism in china. *International Journal of Communication*, 14:19, 2020.

[214] Edson Tandoc, Darren Lim, and Rich Ling. Diffusion of disinformation: How social media users respond to fake news and why. *Journalism*, 21, 2019. doi: `10.1177/1464884919868325`.

[215] Barbee Teasley, Laura Leventhal, Brad Blumenthal, Keith Instone, and Daryl Stone. Cultural diversity in user interface design: Are intuitions enough? *SIGCHI Bull.*, 26(1):36–40, January 1994. ISSN 0736-6906. doi: 10.1145/181526.181533. URL https://doi.org/10.1145/ 181526.181533.

[216] Tom Tonthat. Virtual town halls keep communication flowing during ever-changing conditions. 2020.

[217] Lars Hasselblad Torres. Citizen sourcing in the public interest. *Knowledge Management for Development Journal*, 3(1):134–145, 2007.

[218] Karen Tracy and Margaret Durfy. Speaking out in public: Citizen participation in contentious school board meetings. *Discourse & Communication*, 1(2):223–249, 2007.

[219] UNICEF et al. Fact sheet: Handwashing with soap, critical in the fight against coronavirus, is 'out of reach'for billions, 2020. URL http://uni.cf/33HVzC1.

[220] Niels van Doorn and Adam Badger. Platform capitalism's hidden abode: Producing data assets in the gig economy. *Antipode*, 52(5):1475–1495, 2020.

[221] Vasillis Vlachokyriakos, Clara Crivellaro, Christopher A Le Dantec, Eric Gordon, Pete Wright, and Patrick Olivier. Digital civics: Citizen empowerment with and through technology. In *Proceedings of the 2016 CHI conference extended abstracts on human factors in computing systems*, pages 1096–1099, 2016.

[222] Lingjing Wang, Cheng Qian, Philipp Kats, Constantine Kontokosta, and Stanislav Sobolevsky. Structure of 311 service requests as a signature of urban location. *PloS one*, 12(10):e0186314, 2017.

[223] Mara Warwick. Philippines: Open data launch, March 2017. URL https://www.worldbank. org/en/news/speech/2017/03/02/open-data-launch.

[224] Sebastian Weise, Paul Coulton, and Mike Chiasson. Designing in between local government and the public–using institutional analysis in interventions on civic infrastructures. *Computer Supported Cooperative Work (CSCW)*, 26(4-6):927–958, 2017.

[225] Christopher Whitehead and Lydia Ray. Using the iclicker classroom response system to enhance student involvement and learning. *Journal of Education, Informatics and Cybernetics*, 2(1):18–23, 2010.

[226] UNICEF WHO, unicef, et al. Joint monitoring programme for water supply, sanitation and hygiene. *Program Data [online database]. https://washdata.org/data*, 10:2018, 2017.

[227] Lu Xiao, Weiyu Zhang, Anna Przybylska, Anna De Liddo, Gregorio Convertino, Todd Davies, and Mark Klein. Design for online deliberative processes and technologies. *Proceedings of the 33rd Annual ACM Conference Extended Abstracts on Human Factors in Computing Systems - CHI EA '15*, 2015.

[228] Harlan Yu and David Robinson. The new ambiguity of 'open government'. *UCLA law review discourse*, 59, 02 2012. doi: 10.2139/ssrn.2012489.

[229] Amy X Zhang and Justin Cranshaw. Making sense of group chat through collaborative tagging and summarization. *Proceedings of the ACM on Human-Computer Interaction*, 2(CSCW):196, 2018.

[230] Ethan Zuckerman. New Media, New Civics? *Policy & Internet*, 6(2):151–168, 2014. doi: 10.1002/1944-2866.POI360. URL https://onlinelibrary.wiley.com/doi/abs/ 10.1002/1944-2866.POI360.